\documentclass[iop]{emulateapj}

\usepackage{natbib}
\usepackage{graphics}

\shorttitle{LMC INTEGRATED LIGHT  ABUNDANCES}
\shortauthors{COLUCCI ET AL.}
\begin{document}
\newcommand{\msol}{M_\odot}
\newcommand{\etal}{et al.\ }
\newcommand{\kms}{km~s$^{-1}$ }
\newcommand{\rAA}{{\AA \enskip}}
\newcommand{\ew}{W_\lambda}
\newcommand{\mv}{$M_{V}^{\rm{tot}}$}
\newcommand{\bvo}{($B-V$)$_{\rm{o}}$} 

\defcitealias{mb08}{Paper I}
\defcitealias{milkyway}{Paper II}
\defcitealias{paper3}{Paper III}
\defcitealias{stars}{CB11}

 \defcitealias{pompeia08}{P08}
    \defcitealias{johnson06}{J06}
     \defcitealias{mucc08int}{M08}
      \defcitealias{mucc10old}{M10}
       \defcitealias{mucc1866}{M11}
        \defcitealias{hill00}{H00}
        \defcitealias{HILL95}{H95}
 \defcitealias{1992ApJS...79..303L} {LL92}
 \defcitealias{1989ApJS...70..865R} {RB89}

\title{ Globular Cluster Abundances from High-Resolution,
  Integrated-Light Spectroscopy. IV.  The Large Magellanic Cloud: $\alpha$, Fe-peak, Light, and Heavy Elements\footnotemark[1]}

\footnotetext[1]{This paper includes data gathered with the 6.5 meter Magellan 
Telescopes located at Las Campanas Observatory, Chile.}

\author{Janet E. Colucci}
\affil{Department of Astronomy and Astrophysics, 1156  High Street, UCO/Lick Observatory, \\ University of California, Santa Cruz, CA 95064; jcolucci@ucolick.org}

\author{Rebecca A. Bernstein}
\affil{Department of Astronomy and Astrophysics, 1156  High Street, UCO/Lick Observatory,\\
 University of California, Santa Cruz, CA 95064; rab@ucolick.org}

\author{Scott  A. Cameron}
\affil{Science Department, 3000 College Heights Blvd., Cerro Coso Community College, Ridgecrest, CA 93555; scameron@cerrocoso.edu}

\and

\author{Andrew McWilliam}
\affil{The Observatories of the Carnegie Institute of Washington, 813 Santa Barbara Street, Pasadena, CA 91101-1292; andy@ociw.edu}

\begin{abstract}
We present detailed chemical abundances in 8 clusters in the Large Magellanic Cloud (LMC). 
We measure abundances of 22 elements for clusters spanning a range in age of 0.05 to 12 Gyr, providing  a comprehensive picture  of the chemical enrichment and star formation history of the LMC. 
 The abundances were obtained from individual absorption lines  using a new method for analysis of high resolution ($R\sim$25,000)  integrated light spectra  of star clusters. This method was developed and  presented in Papers I, II, and III of this series. In this paper, we develop an additional integrated light  $\chi^2$-minimization spectral synthesis technique to facilitate measurement of weak ($\sim$15 m\AA) spectral lines and abundances in low signal-to-noise ratio data (S/N$\sim$30).  
Additionally, we supplement the integrated light abundance measurements with detailed abundances that we measure for individual stars in the youngest clusters (Age $<$2 Gyr) in our sample.
In both the integrated light and stellar abundances we find evolution of [$\alpha$/Fe] with [Fe/H] and age. Fe-peak abundance ratios are similar to those in the Milky Way, with the exception of [Cu/Fe] and [Mn/Fe], which are sub-solar at high metallicities.  The heavy elements Ba, La, Nd, Sm, and Eu are significantly enhanced in the youngest clusters. Also,  the heavy to light {\it s}-process ratio is elevated relative to the Milky Way ([Ba/Y] $>+0.5$)  and increases  with decreasing age, indicating a strong contribution of low-metallicity AGB star ejecta to the inter-stellar medium throughout the later history of the LMC. 
 We also find a correlation of integrated light  Na and Al abundances with cluster mass, in the sense that  more massive, older clusters,  are enriched in the light elements Na and Al with respect to Fe, which implies that these clusters harbor star-to-star abundance variations as is common in the Milky Way. Lower mass,  intermediate age and young clusters have Na and Al  abundances that are lower and  more consistent with LMC field stars. 
  Our results can be used  to constrain  both future chemical evolution models for the LMC and theories of GC formation.
\end{abstract}

\keywords{galaxies: individual (LMC) --- galaxies: star clusters: general --- galaxies: abundances --- globular clusters: individual(NGC 2005, NGC 2019, NGC 1916, NGC 1978, NGC 1718, NGC 1866, NGC 1711, NGC 2100) --- stars: abundances}

\section{Introduction}	
\label{sec:intro}
\setcounter{footnote}{1}

We are conducting an ongoing program to obtain detailed abundances from high resolution ($R\sim$25,000), integrated light (IL) spectra of extragalactic  star clusters.   In a series of papers, we have  developed and presented a method to obtain abundances of $\sim$20 elements from individual absorption lines in these spectra.  In \cite{mb08} and \cite{milkyway}, hereafter Paper I and Paper II, we developed this method on a training set of 7 globular clusters (GCs) in the Milky Way (MW), which have ages $>$10 Gyr.  In \cite{paper3} and \cite{stars}, hereafter Paper III and \citetalias{stars}, we extended this method to a training set of 8 massive, high surface brightness star clusters in   the Large Magellanic Cloud (LMC), for which we obtained measurements of [Fe/H] and age.    This sample of LMC clusters is critical for development of the IL spectra method on clusters with ages between 0.05 and 10 Gyrs, because the MW has few, if any,  massive high surface brightness clusters in this age regime.    In this paper, we present detailed chemical abundances of $\sim$20 additional elements  in the LMC training set clusters.

The IL method was designed to take advantage of the high luminosities and surface brightness of unresolved GCs,  as compared to individual stars, in order to study chemical evolution and cluster properties in distant galaxies  ($\sim$4 Mpc). Therefore, the purpose of developing this technique is to use GCs to constrain galaxy evolution in the same way that stars have been used to constrain formation of the MW.
 It has been demonstrated that star clusters trace the properties of major star formation events in their host galaxies, and therefore they are crucial tools for learning about more distant galaxies where individual stars cannot be studied in detail  \citep[e.g.][]{brodierev06}.

To this end, it is important to verify that our IL spectral analysis   provides accurate results for clusters of all ages.  The  study of young clusters, like those in our LMC training set, is particularly interesting for several reasons.
First, there are young massive star clusters currently forming in nearby star-bursting galaxies like M82, which can teach us about 
cluster formation in  conditions  of high star formation rate and intensity. These conditions are likely similar to those present at high redshift,  and are important  for understanding the formation of massive clusters in general  \citep[e.g.][]{2001ASPC..245..390L}. 
Young clusters are also interesting in light of the recent results that MW GCs have small age and abundance spreads, and thus had multiple, fast star formation episodes  at early times \citep[e.g.][]{2007ApJ...661L..53P,2008ApJ...681L..17M,2009A&A...497..755M,2011ApJ...737....3G,2008MNRAS.391..825D,2010A&A...516A..55C,2007A&A...464.1029D,2009A&A...507L...1D, 2011ApJ...726...36C, 2011arXiv1101.2208C}.
  By studying IL abundance trends with age and mass in these young clusters we can try and constrain  cluster formation timescales and internal chemical enrichment models. Additionally,  in  starbursting environments young clusters allow us to study the intrinsic cluster luminosity function and  cluster  disruption, which can be used to constrain the contribution of cluster stars to  field star populations.

Our training set of clusters in the LMC   not only provides a necessary test of the IL method on young clusters, but also comprehensive information on the chemical evolution of the LMC itself over a wide range in age and [Fe/H]. 
 While the star formation history (SFH) of the LMC field and cluster  populations has been well-studied photometrically and with low resolution spectra \citep[e.g.][]{hz2009, 1996ARA&A..34..511O, 2006A&A...448...77C,2008AJ....135..836C,1991AJ....101..515O}, detailed abundance information for the LMC is much more limited than for the MW.  This is simply due to the long integration times required to obtain high S/N, high resolution spectra of large samples of stars at the distance of the LMC (D$\sim$50 kpc).

 From prior work with high resolution photometry, we know that the LMC cluster 
system had its initial formation epic $>$10 Gyr ago, and at least one additional  burst 2-4 Gyr ago \citep[e.g.][]{1986A&A...156..261B,1991AJ....101..515O,hz2009}.  There is only one cluster that  formed in the ``age-gap'' between 3-10 Gyr. The disk field population seems to have had a nearly constant formation rate over most of the history of the LMC, with evidence of localized enhancement in star formation rate 1-4 Gyrs ago \citep[e.g.][]{1998AJ....115.1045G}.   The LMC also has a high surface brightness bar, which shows an underlying old population, and some evidence for a burst in star formation $\sim6$ Gyr ago \citep{2000AJ....120.1808C}.

With detailed chemical abundances, we can  further constrain the SFH of the LMC. There are several useful galaxy formation diagnostics that can be used when large samples of elements are available.
 One common  focus of detailed abundance studies  is to obtain $\alpha$-element abundances (e.g. O, Mg, Ca, Si, Ti), which  are produced mostly in Type II supernovae (SNe II), or massive stars, on short timescales.  Large over-abundances (with respect to Fe-peak elements) of $\alpha$-elements are therefore signatures of early, rapid or bursty star formation \citep[e.g][]{1979ApJ...229.1046T,1995ApJS..101..181W}.  
Another useful diagnostic   is the ratio of  [Ba/Y]. The high [Ba/Y] ratios observed in nearby dwarf galaxies \citep[e.g.][]{2001ApJ...548..592S,2003AJ....125..684S,2002astro.ph..5411S,2005ApJ...622L..29M,2007PASP..119..939G,2009ARA&A..47..371T} reflect the metallicity dependence of the {\it s}-process in asymptotic giant branch (AGB) stars \citep[e.g.][]{1999ARA&A..37..239B}. This metallicity dependence means that high [Ba/Y]  ratios   identify star formation that occurred    in long-lived low-metallicity environments, or low star formation rates.  
High [Ba/Y] ratios have  previously  been  observed in individual stars in the LMC \citep{pompeia08,mucc08int},  and we determine new [Ba/Y] ratios for the 8 clusters analyzed in this work.

In this paper,  our goals are  both to demonstrate the accuracy of our abundance analysis when applied to young clusters, and also to  obtain new abundance results for LMC clusters  in order to study the chemical evolution history of the LMC.
  In \textsection \ref{sec:obs} we present the observations and data reduction. In  \textsection \ref{sec:analysis} we review the IL abundance analysis, results from \citetalias{paper3}, and present a new, supplementary IL spectral synthesis technique.  In \textsection \ref{sec:young}, we present abundance results from the sample of individual stars in \citetalias{stars}, which are used to verify the precision of the IL method at young ages.
IL chemical abundance results are presented in  \textsection \ref{sec:results}.  In  \textsection \ref{sec:comparisons} we discuss our results  compared to abundance analyses from the literature. In \textsection \ref{sec:light-mass}, \textsection \ref{sec:gcheavy} and \textsection \ref{sec:model} we discuss the implications of our results for massive cluster formation and the formation and evolution of the LMC.   Throughout this work we will frequently refer to previous measurements of abundances in individual LMC stars by \cite{pompeia08},   \cite{johnson06}, \cite{mucc08int}, \cite{mucc10old}, \cite{mucc1866},  \cite{1992ApJS...79..303L}, \cite{1989ApJS...70..865R},  \cite{hill00} and  \cite{HILL95}, which will be abbreviated as P08, J06, M08, M10, M11, LL92, RB89, H00, and H95, respectively.

\section{Observations and Data Reduction}
\label{sec:obs}

\begin{figure*}
\centering
\includegraphics[scale=0.3,angle=90]{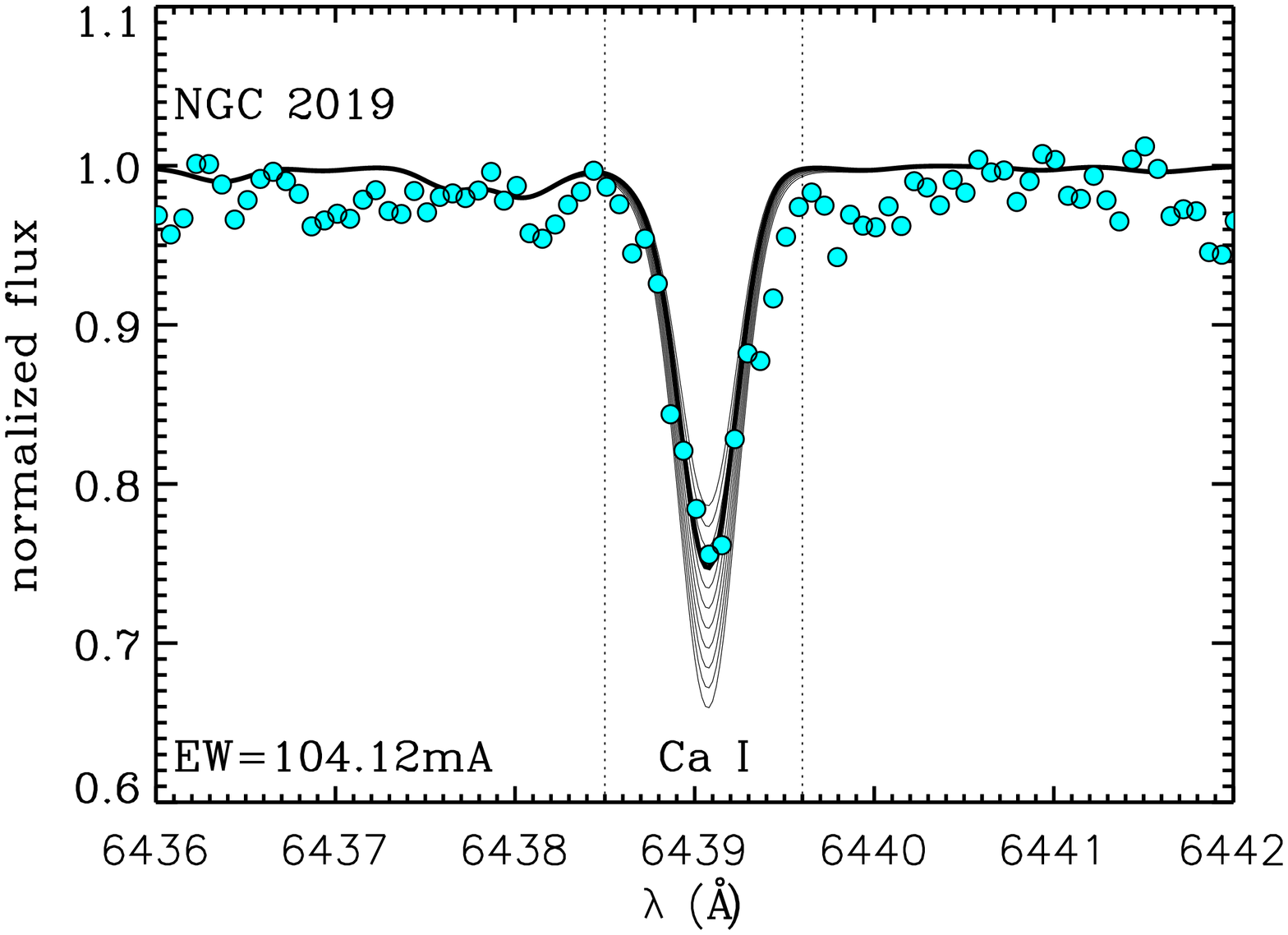}
\includegraphics[scale=0.3,angle=90]{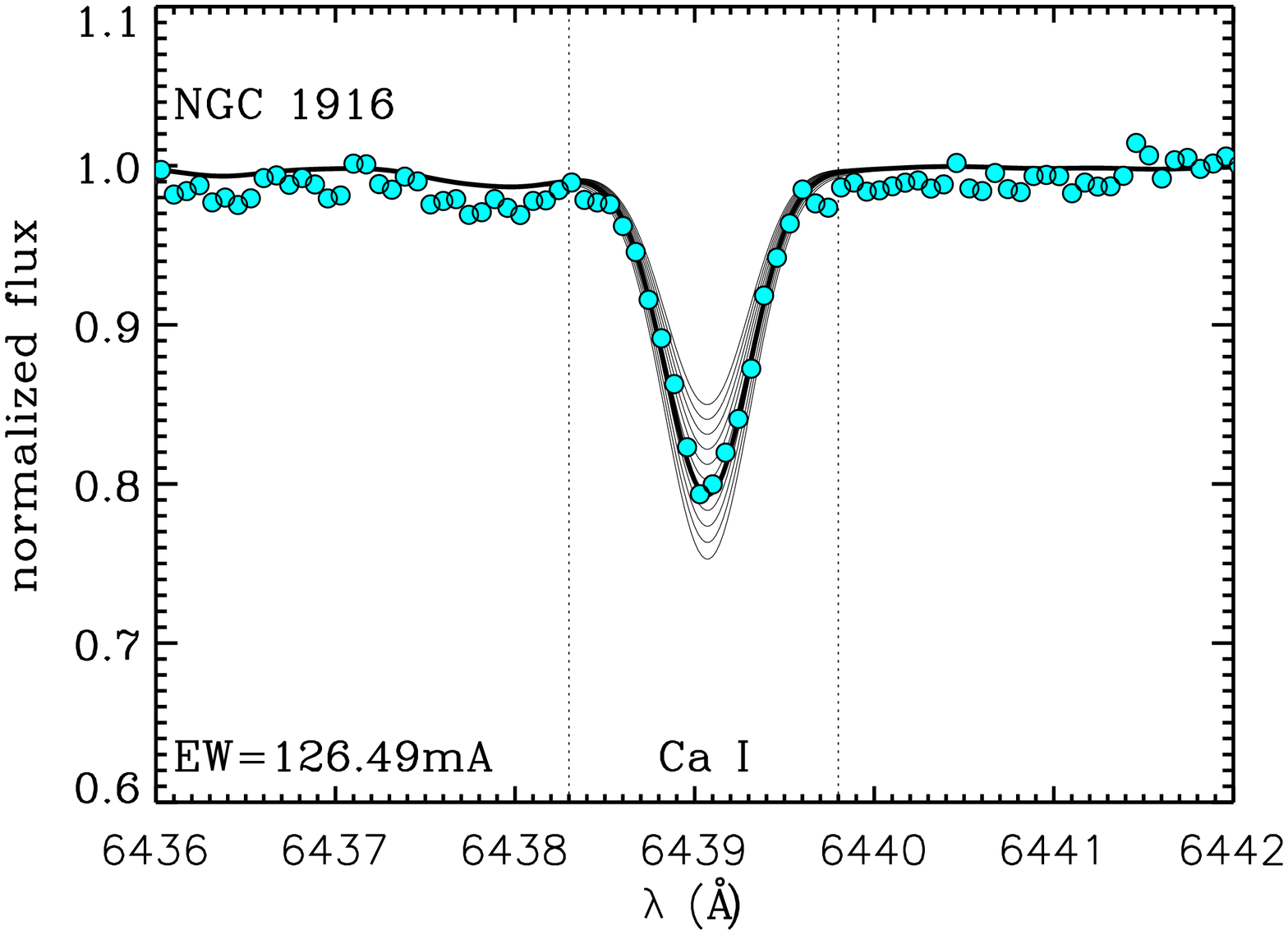}
\includegraphics[scale=0.3,angle=90]{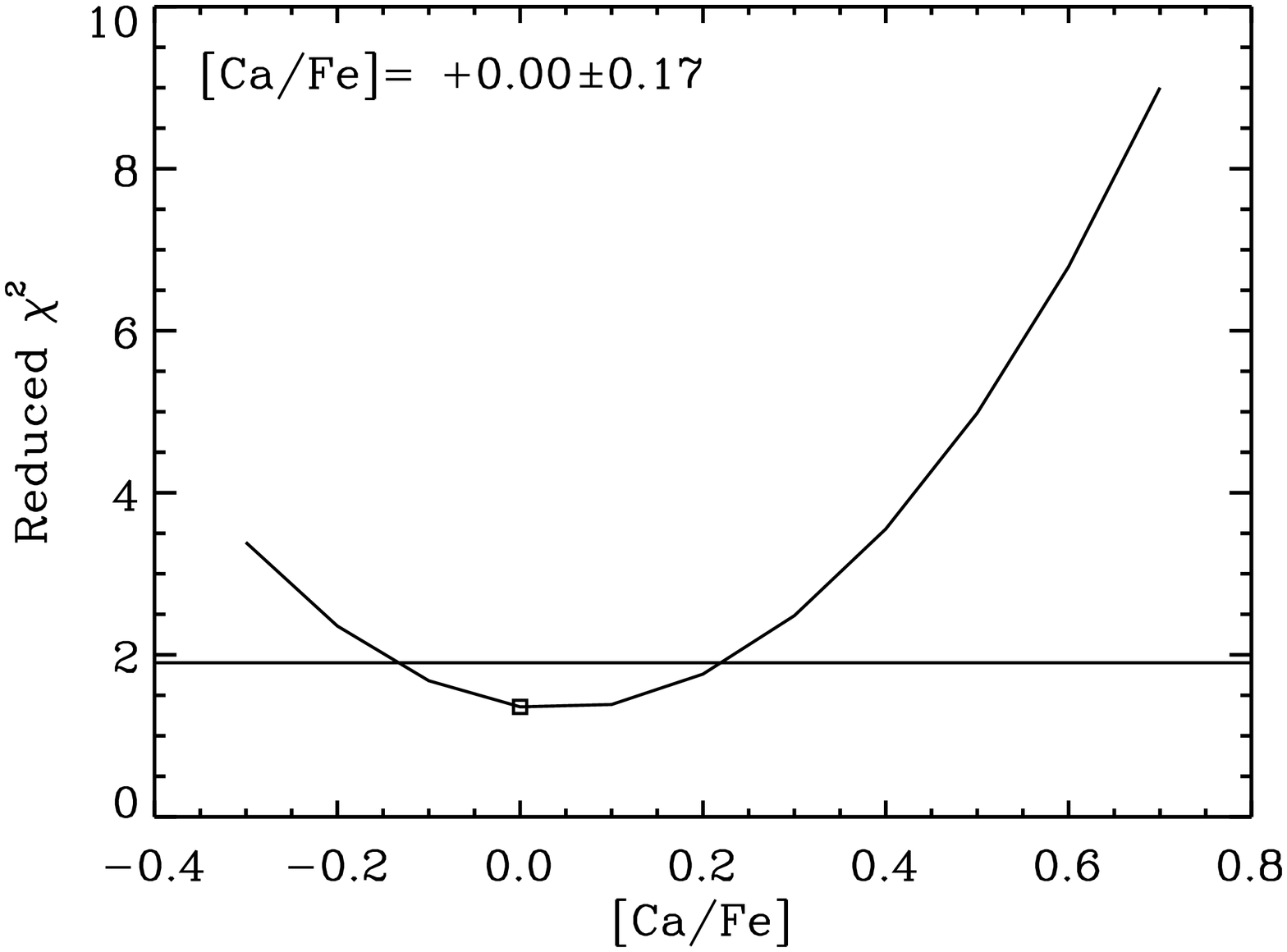}
\includegraphics[scale=0.3,angle=90]{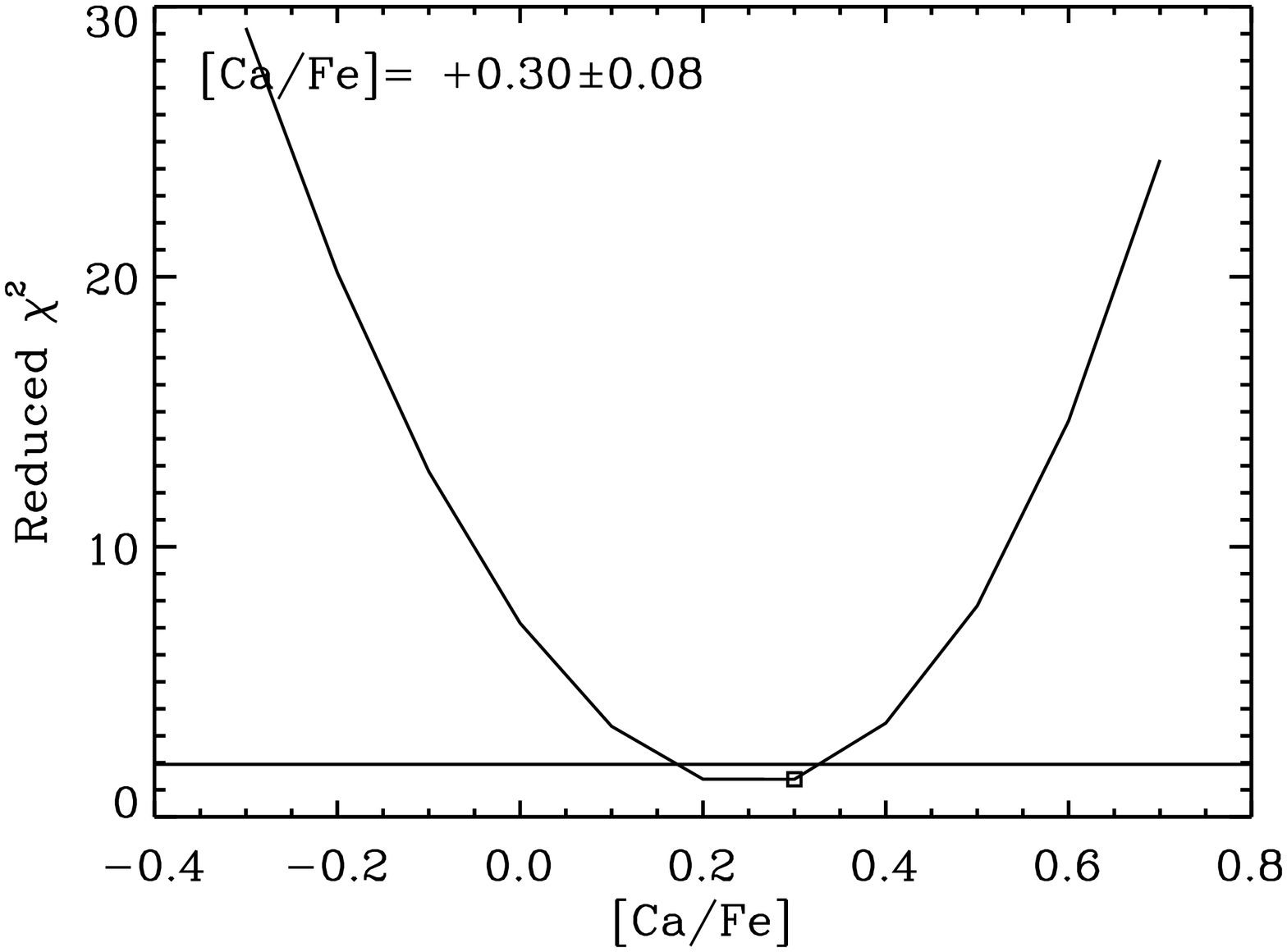}
\caption{Example of $\chi^2$ minimization technique for the Ca I 6439 \rAA line for NGC 2019 (left panels) and NGC 1916 (right panels).  The top panels show the observed data as cyan filled circles, and solid black lines show the IL  synthesized spectra corresponding to $-0.3 < $ [Ca/Fe] $ < +0.7$. The thick black line in each figure shows the IL spectrum corresponding to the minimum in the reduced $\chi^2$, which can be seen in the bottom panels as a function of [Ca/Fe]. }
\label{fig:syn_ca}
\end{figure*}

 Integrated light spectra for all of the LMC clusters were obtained by uniformly scanning the core regions of each cluster.  All spectra were obtained with the 2.5 m du Pont telescope at Las Campanas in 2000 December and 2001 January, as described in \citetalias{mb08}, \citetalias{milkyway} and \citetalias{paper3}, with the exception of NGC 1718 which was observed with the MIKE double echelle spectrograph  \citep{mike}  on the 6.5 m Magellan telescope in 2006 November.  The reduction of data  from the du Pont telescope is described fully in \citetalias{mb08}. The reduction of the data from the Magellan telescope is described fully in \citetalias{paper3}.

 The LMC training set includes 8 clusters that are well enough sampled to have well-constrained [Fe/H] and ages from our analysis in \citetalias{paper3} 
( NGC 1916, NGC 2005, NGC 1916, NGC 1718, NGC 1978, NGC 1866, NGC 1711, and NGC 2100).  These clusters were chosen to span the available range in age and [Fe/H] of high surface brightness clusters  found in the LMC.   In \citetalias{paper3} we divided the LMC clusters into three groups according to age. ``Old'' clusters are those with ages $>$5 Gyr, ``intermediate age'' clusters are those with ages 1$-$4 Gyr, and ``young'' clusters are those with ages $<$1 Gyr.   The clusters in each group are listed in Table~\ref{tab:previous_results}.

  In \citetalias{stars} we presented an analysis of the [Fe/H] of 10 cool giant stars in the intermediate and young clusters of our sample.  The stellar sample includes  2 stars in NGC 1978, 3 stars in NGC 1866, 3 stars in NGC 1711, and 2 stars in NGC 2100.    All of the stellar spectra were taken with the MIKE spectrograph on Magellan. The description of the target   selection, data reduction, and stellar parameter analysis is presented   in \citetalias{stars}.

\section{Integrated Light Abundance Analysis}
\label{sec:analysis}
The IL abundance analysis technique is described in detail in Paper I, II and III and \citetalias{stars}, and will not be repeated here.  In \textsection~\ref{sec:cmds}, we briefly summarize the ages and [Fe/H] derived for the LMC training set clusters in \citetalias{paper3} and \citetalias{stars}.  In \textsection~\ref{sec:lines}-\textsection~\ref{sec:synth}, we present the line lists we employ and discuss a new extension of the ILABUNDs code for determining abundances of individual lines with a $\chi^2$-minimization spectral synthesis technique.

\subsection{CMDs and Ages}        
\label{sec:cmds}

In \citetalias{paper3} and \citetalias{stars}  we presented  a detailed analysis of the [Fe/H] and age solutions for
 each cluster in the LMC training set.  The [Fe/H] and age for each cluster are used to create final, representative color magnitude diagrams (CMDs) from which we get the stellar parameters needed for spectral synthesis and abundance determination (see Papers I, II and III). Here we use these final representative CMDs  with the   age and [Fe/H] derived in \citetalias{paper3} and \citetalias{stars} for each cluster to calculate  abundances for the additional elements.

 For the final [Fe/H] and ages we use the results from \citetalias{paper3} for the old clusters, and the results from \citetalias{stars} for the intermediate age and young clusters.    These results are summarized in columns 3 and 4 in Table~\ref{tab:previous_results}.     
We constrained the age of old clusters to a range of $\sim$5 Gyrs,  intermediate age clusters to $\sim$1-2 Gyr, and young clusters to $\sim$0.4 Gyr.    The acceptable age range  for each cluster leads to  a range in derived [Fe/H], corresponding  to the different stellar populations  at each age.  For old clusters, the difference in [Fe/H] is generally small ($<$0.05 dex) because there are only subtle changes in the stellar populations at older ages.  Therefore, for the old clusters we use one best-fitting  CMD to report abundances calculated for other elements.    
For intermediate age and young clusters, which have more  rapidly evolving stellar populations,  the difference in derived abundances  between CMDs of different ages can  occasionally be larger than the line-to-line statistical scatter ($\sigma_{lines}$) of the individual Fe lines ($>$0.1 dex).   
For  intermediate age and young clusters,  we adopt the mean abundances calculated from two CMDs spanning the appropriate age range and  report an additional  uncertainty on the abundances due to the assumed age as $\sigma_{age}$.

\subsection{EWs and Line Lists}        
\label{sec:lines}

We use the semi-automated program GETJOB \citep{1995AJ....109.2736M} to measure absorption line equivalent widths (EWs)  for individual lines in both  the IL  and stellar spectra.  We interactively fit low order polynomials to continuum regions for each spectral order.  The line profiles are fit with single, double, or triple Gaussians as needed.  

 Line lists and log {\it gf} values were taken from \citetalias{mb08}, \citetalias{milkyway},  \citet{m31paper}, and \cite{2009A&A...497..611M},  and  supplemented with the Kurucz atomic and molecular line database \citep{1997IAUS..189..217K}. The same line list was used for analysis of both the cluster IL spectra and the individual stellar spectra.
 The lines and EWs  measured in the cluster IL spectra are listed in Table~\ref{tab:linetable_nf}, and the EWs measured in the stellar spectra are listed in Table~\ref{tab:linetable_stars}.  
 
In addition to the IL EW analysis used in our previous work, we have implemented an IL  line synthesis component to ILABUNDs, which we describe in detail in \textsection~\ref{sec:synth}.  Lines with abundances determined using the synthesis routine of ILABUNDs are labeled  ``SYN'' in column 13 of Table~\ref{tab:linetable_nf}, while abundances determined using our original EW analysis are labeled ``EW.''    We have calculated abundances with  hyperfine splitting (hfs) of energy levels taken into account  for the  elements Sc, V, Co, Cu, Mn,  Ba, Sr, La and  Eu.  These calculations are noted as ``HFS''  in  Table~\ref{tab:linetable_nf} and    Table~\ref{tab:linetable_stars}.

\subsection{ILABUNDs Line Synthesis}
\label{sec:synth}

\begin{figure}
\centering
\includegraphics[scale=0.35,angle=90]{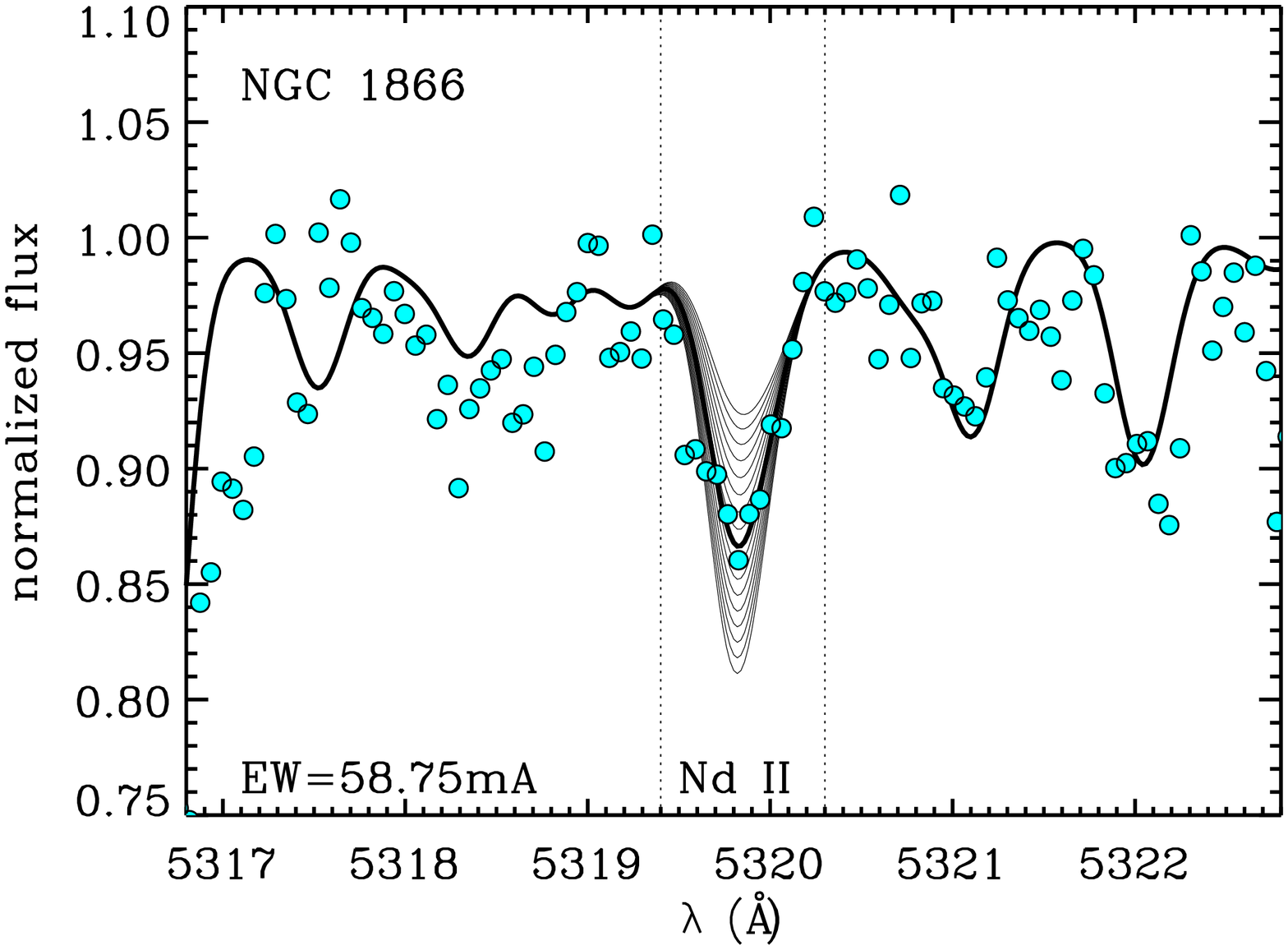}
\includegraphics[scale=0.35,angle=90]{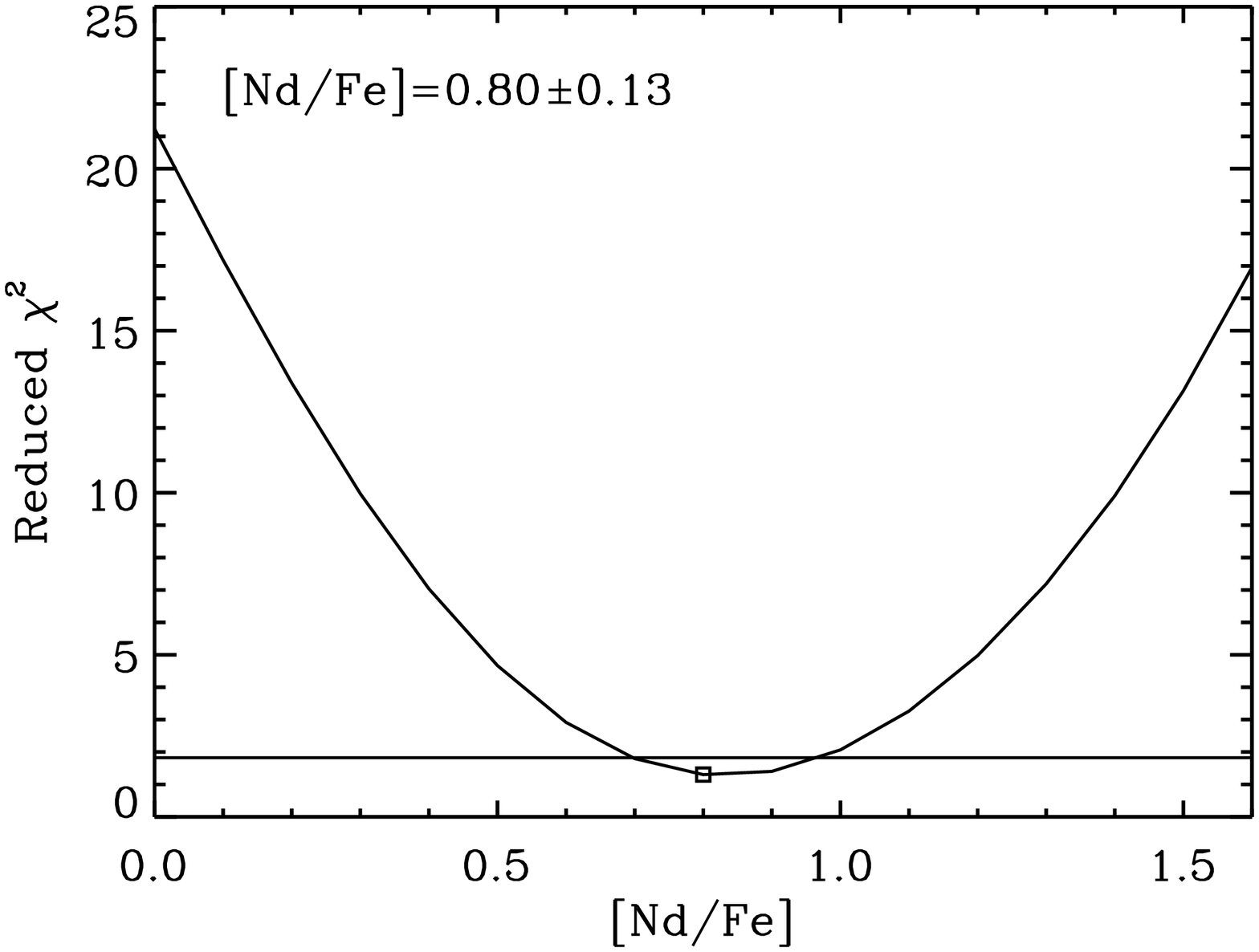}
\caption{Example of the $\chi^2$-minimization technique for the Nd II 5319 \rAA line for NGC 1866.  In the top panel, the observed data is shown as filled circles, and the solid black lines show the IL synthesized spectra corresponding to $+0.0 <$ [Nd/Fe] $< +1.6$. Bottom panel is the same as in Figure \ref{fig:syn_ca}, only for Nd. This example shows a region with lower S/N and weaker features than that of Figure \ref{fig:syn_ca}, and that the abundance is well determined with this technique. }
\label{fig:syn_nd}
\end{figure}

The development of the line synthesis routine for ILABUNDs was motivated by the desire to make additional measurements of poorly constrained elements by analyzing weak features for which EWs are not otherwise recoverable.  As for the analysis of all of the  elements discussed in this paper, the line synthesis routine is performed only with the best-fitting CMD already identified from the [Fe/H] analysis, as discussed above.
We note that the goal of our synthesis analysis is still to calculate abundances for a given species using individual lines, not to fit the entire spectrum at one time. 
Therefore, we limit the synthesis region to $\pm10$~\rAA from the line of interest, and supplement our standard line lists with the Kurucz database.

Like the EW calculation routine of ILABUNDs, the synthesis component calculates spectra for each stellar type in the CMDs using the 2010 version of MOOG \citep{moog}.  The spectra for the representative stars are then flux weighted according to the scheme described for the EWs in \citetalias{mb08}.  The  combined synthesized IL spectrum is then renormalized and convolved with the 1-dimensional velocity dispersion  ($\sigma_v$)  of the cluster. The  measurement of $\sigma_v$ for the LMC clusters follows the method described in \cite{m31paper}, and is presented in Colucci et al. (2012, in preparation).

A $\chi^2$-minimization scheme is used to determine the best-fitting abundance for a  given line.
First, the continuum level  is identified in a 30-40 \rAA window as the mean value with 3$\sigma$ lower rejection. Next, a 1 \rAA region used for the $\chi^2$ fitting must be identified by eye and adjusted on a cluster by cluster basis because of blending effects that vary due to the different $\sigma_{v}$ and metallicities of the clusters. An example of how this region can vary between clusters because of different $\sigma_{v}$ is shown
in  the top panels of Figure~\ref{fig:syn_ca}.  The left panel shows the Ca I 6439 \rAA line for NGC 2019 and the right panel shows the same for NGC 1916. The larger $\sigma_v$ of NGC 1916 makes a larger region necessary for calculating the $\chi^2$.  Note, the Ca I 6439 \rAA line is a strong, high S/N line for which EWs can be accurately measured using GETJOB, and is shown here only as a demonstration of the $\chi^2$ technique.

For a given line, synthesized spectra are created for a  range of abundances and the reduced $\chi^2$ is calculated for each.  The minimum in the $\chi^2$ can then be identified, as shown in the bottom panels of Figure~\ref{fig:syn_ca}.   We determine the approximate uncertainty in abundance using the range of values  that are within  $\sim$40$\%$ of the minimum  $\chi^2$, shown by the horizontal solid lines in the bottom panels of Figure~\ref{fig:syn_ca}.   EWs for synthesized lines  are calculated for the best fitting spectra and are listed in Table~\ref{tab:linetable_nf}.  We find excellent agreement between the EWs measured using GETJOB and the EWs measured with the $\chi^2$-minimization for the examples in Figure~\ref{fig:syn_ca}.  For both NGC 2019 and NGC 1916 the  EWs measured using the two techniques agree to  within 1$\%$.

As an example  of a case in which the line synthesis is particularly useful, we show
 the $\chi^2$-minimization for a weak (15$\%$ flux decrement) Nd II line at 5319 \rAA  in Figure~\ref{fig:syn_nd}.
  The Nd II line falls into the category of being in a low S/N region of the spectra.  The line synthesis is useful anywhere line depths are reduced.  This can occur in low S/N data, for clusters with high $\sigma_{v}$, and for weak transitions.  While the uncertainty on the derived abundance of these synthesized lines is higher even when using the $\chi^2$-minimization technique, we are able to meaningfully constrain abundances of lines with EWs as small as 20 m\AA, or in data with S/N$\sim$30. This additional technique makes it possible for us to use the IL abundance analysis method on lower S/N data of distant, unresolved clusters in future work.

\section{Accuracy of LMC IL Analysis: Comparison to Our Sample  of Individual Stars in Young Clusters}
\label{sec:young}

\begin{figure*}
\centering
\includegraphics[trim = 0mm 100mm 0mm 0mm, clip,angle=90,scale=0.50]{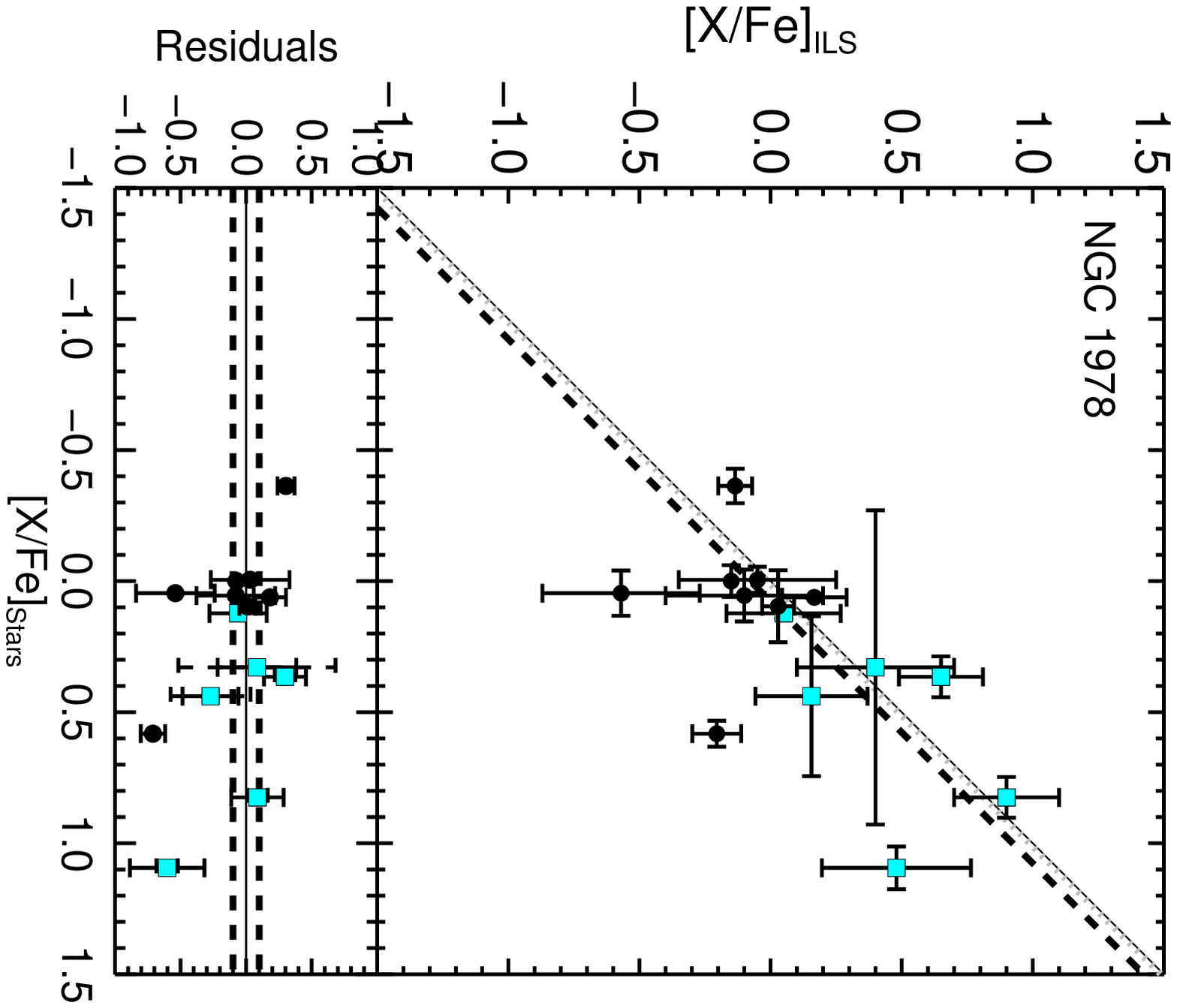}
\includegraphics[trim = 0mm 100mm 0mm 0mm, clip,angle=90,scale=0.50]{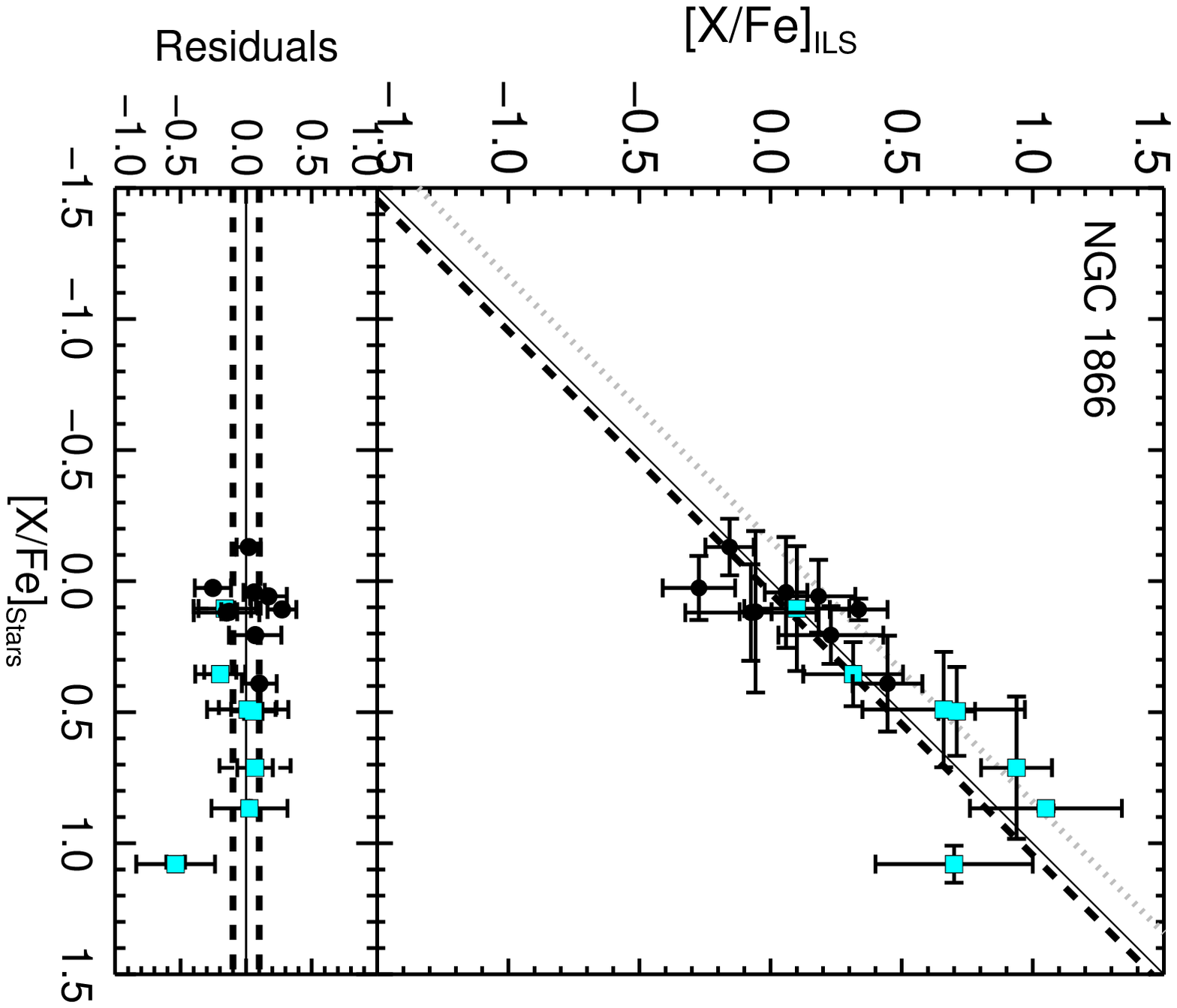}
\includegraphics[trim = 0mm 100mm 0mm 0mm, clip,angle=90,scale=0.50]{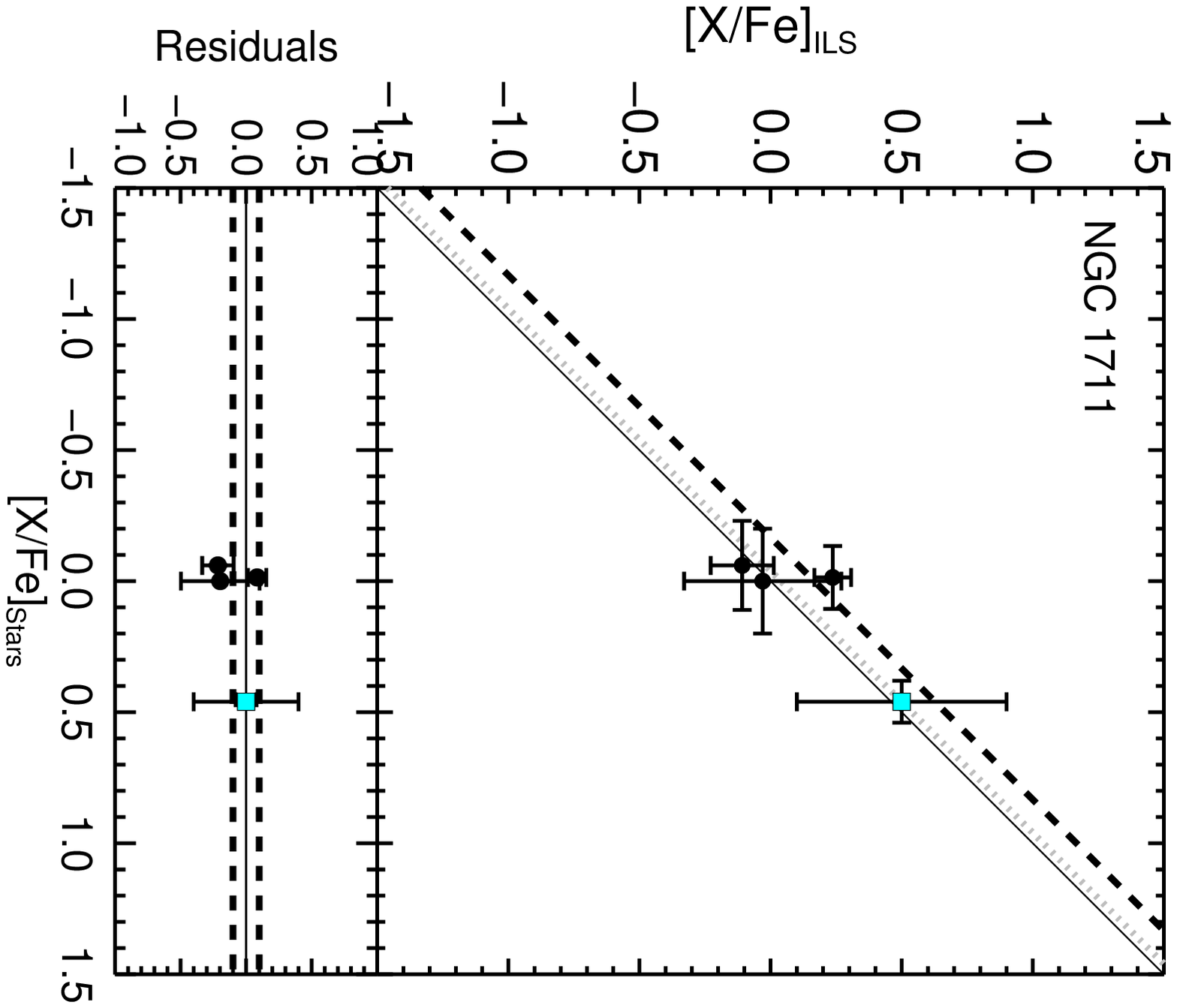}
\includegraphics[trim = 0mm 100mm 0mm 0mm, clip,angle=90,scale=0.50]{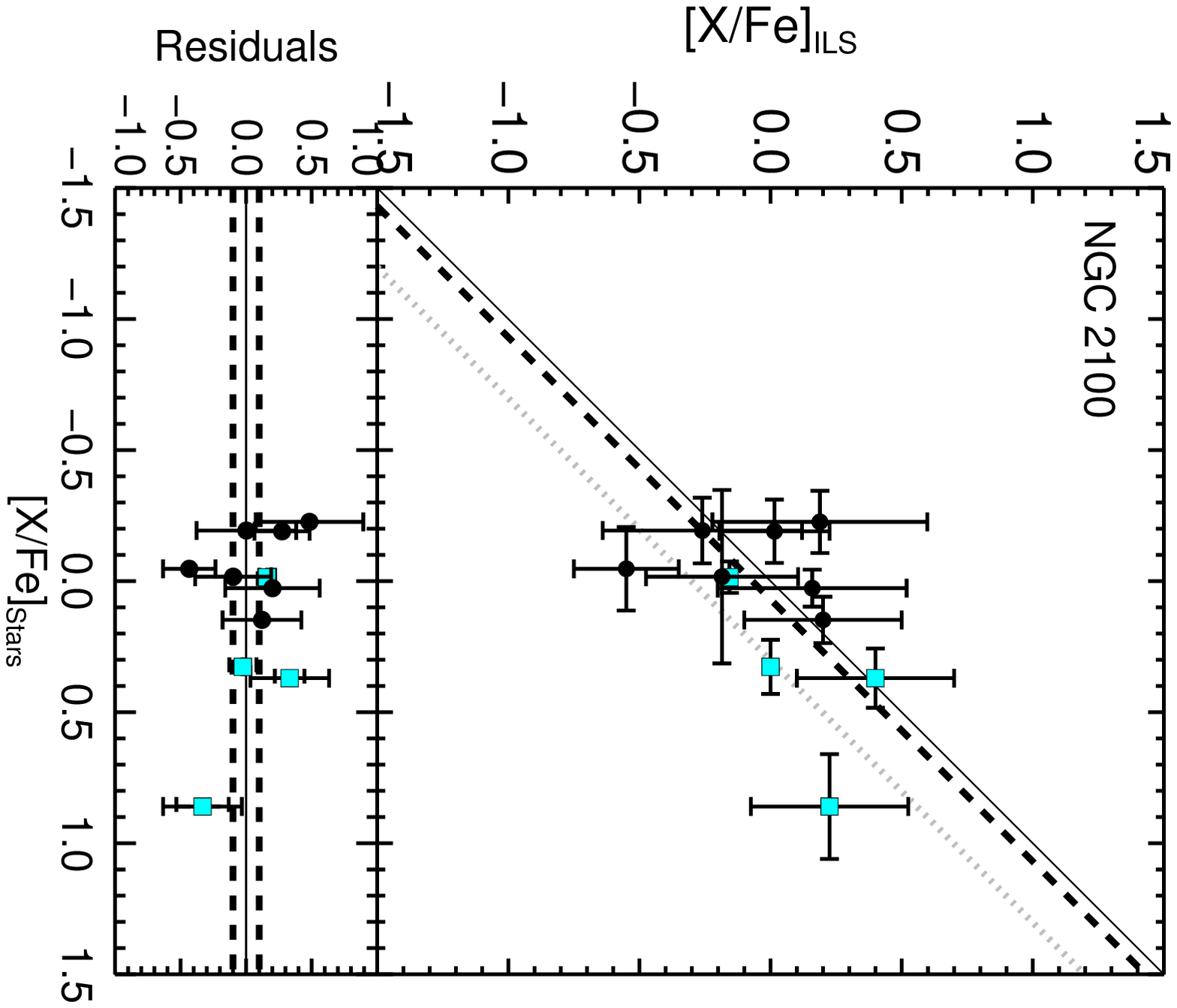}
\caption{Comparison of abundance ratios from IL and stellar analysis  for NGC 1978, NGC 1866, NGC 1711, and NGC 2100.  
 Black circles and cyan squares show abundances for neutral and ionized species, respectively.  The thick dashed line in the top panels of each plot shows a linear least squares fit to the neutral species, with the slope of the fit constrained to unity. The thin dotted line shows the same fit to the ionized species.  The bottom panels in each plot show the residuals of the neutral and ionized species around their respective trend lines.  The thick dashed lines in the bottom plots mark residuals of $\pm$0.1 dex, to guide the eye.  In general the two methods agree to the $\pm$0.1 dex level.  }
\label{fig:starcomparison}
\end{figure*}

To demonstrate that the IL method produces accurate abundances for clusters with ages of 0.05 to 2 Gyr, 
we first compare the abundances measured with our IL abundance analysis method to the abundances measured in individual stars.  These comparisons for NGC 1978, NGC 1866, NGC 1711, and NGC 2100 are shown in Figure \ref{fig:starcomparison}.  
Wherever possible we have used the same analysis techniques, i.e. identical line lists, line parameters, stellar atmospheres, and stellar spectral synthesis codes.  This eliminates many potential systematic offsets that can be present when  comparing our cluster  IL  results to abundance analyses  performed by other authors (see \textsection~\ref{sec:comparisons} for more details).

In Figure \ref{fig:starcomparison}  we show the comparison between the results obtained from individual stars in NGC 1978, NGC 1866, NGC 1711, and NGC 2100 to the results obtained from the IL spectra in those same clusters.   To evaluate systematic offsets between the two methods, we have performed linear least squares fits, with the slopes constrained to unity,  to both the neutral and ionized species in each cluster.   To show the scatter around the fits we also show the residuals in the bottom panels of the plots.  The derived offset and scatter for the neutral and ionized lines for each cluster are listed in Table~\ref{tab:starsvsil}.  	
We find that the systematic offset in neutral species is $<$0.1 dex in all cases except for NGC 1711, for which we only have three elements available for comparison.     The scatter about the fits for neutral species  is 0.40, 0.17, 0.18, and 0.29  dex for NGC 1978, 1866, 1711, and 2100, respectively.  For ionized species the average systematic offset is $<$0.16 dex in all cases, and the scatter is 0.33, 0.22, and 0.26 dex for NGC 1978, 1866, and 2100, respectively.   In both the neutral and ionized cases the scatter about the fits is comparable to the $\sigma_{\rm Tot}$ of the cluster measurements, and it is likely that with higher S/N IL spectra the scatter would decrease.  For example, the cluster for which we have the highest S/N data in the IL spectra, NGC 1866,  has the smallest scatter in both neutral and ionized species.  

For each cluster in Figure \ref{fig:starcomparison}  we can look for element species that are outliers in the fits.  It is interesting to note if the outliers for all of the clusters tend to be of the same species, which would mean that the IL analysis method has a large systematic error for these particular species.  For NGC 1978 the  outliers are Na I,  Ti I, Mn I and Nd II.  For NGC 1866 the only outlier is Sm II. For NGC 1711 we only have four elements available for comparison, with no obvious outliers. For NGC 2100 the outliers are Al I and  Ba II.
This comparison shows that the outliers in each cluster are different, so  it does not appear that the IL method has a large systematic offset for any particular element, with the caveat that we can only make this detailed comparison for the clusters in our sample that have ages $<$ 2 Gyr, due to the ages of our current sample of individual stars.

Finally, it is particularly interesting to look at the comparison for the youngest cluster in our sample, NGC 2100.  For this cluster we were only able to measure an upper limit in [Fe/H] \citepalias{paper3,stars}, which we found to be $\sim$0.4 dex more metal-rich than the mean  [Fe/H] we found from the two stars c12 and b22.  From inspection of Figure \ref{fig:starcomparison}, we find that even though the [Fe/H] is offset by $+$0.4 dex, the abundance ratios are extremely consistent with the abundance ratios that we derive for the individual stars. 
This is not unexpected, because the relative abundances of elements with respect to Fe change very slowly, even when the absolute changes are more significant.  This is also consistent with what we found in \citet{m31paper} for old clusters in M31.
Therefore,  the IL analysis method can provide reliable measurements of the abundance ratios even when we can only obtain a limit on the [Fe/H].

In conclusion, we find good agreement between the abundance ratios measured from  the cluster IL spectra and the abundance ratios measured from the individual stellar spectra.  The largest discrepancies are found for the cluster IL spectra with the lowest S/N.  In this sample, there are no elements with abundances that are always 
predictably inconsistent between  the IL and stellar analyses.  We conclude that in general the cluster IL analysis method results in abundances that are accurate to $\lesssim$0.1 dex over the age range 0.05$-$2 Gyr. 
The accuracy of the measured abundance for any species for any one
cluster naturally depends on the accuracy of the measured equivalent
widths, which in turn depend on the S/N of the data,  the velocity
dispersion of the cluster,  and the strength of the constraint on the CMD
solution. For the results in this paper, the abundances measured for
any one species in any one cluster have statistical uncertainties of
$0.05-0.3$ dex (see columns for $\sigma_{\rm lines}$, $\sigma_{\rm
  age}$, and $\sigma_{\rm Tot}$ in Tables~\ref{tab:abundance_table_2019} through
~\ref{tab:abundance_table_2100}).

\section{Results \& Discussion : Chemical Abundances}
\label{sec:results}

In the following sections we present abundance results in our LMC
clusters for $\sim$20 individual elements, including
$\alpha-$elements, light elements, Fe-peak elements, and neutron
capture elements.  We discuss the behavior of similar groups of
elements with respect to previous abundance work in LMC clusters and
field stars, as well as in comparison to abundance work in MW clusters
and field stars.  Many of the LMC abundances measured here are the
first such measurements for certain clusters, and will be discussed
further in \textsection \ref{sec:light-mass}, \textsection
\ref{sec:gcheavy} and \textsection \ref{sec:summary}.

For the eight LMC clusters, we show the final IL abundances, number of
available spectral lines and the uncertainty in the mean of
$N_{{lines}}$ ($\sigma_{lines}$=$\sigma$/$\sqrt{(N_{lines}-1)}$) for
each species in Tables~\ref{tab:abundance_table_2019} through
~\ref{tab:abundance_table_2100}.
For the young and intermediate age clusters we also report the
uncertainty in the [X/Fe] ratios that results from the uncertainty in
the age solution ($\sigma_{age}$). 
For the old clusters the uncertainty associated with the age of the
cluster is small---usually $<<$0.05 dex for [X/Fe]--- and so is not
included for the individual species.  Note that for the young and
intermediate age clusters, age uncertainties are the dominant
systematic uncertainty in the abundances, and the only systematic
uncertainty that does not have an analogy in any calculation of
abundances for individual stars.  As in any stellar abundance
analysis, systematic uncertainties due to the details of the
calculations (e.g., assumptions adopted in MOOG, the Kurucz model
atmospheres, $\alpha$-enhancement effects on H$-$ opacity,
{\it gf}-values) are unavoidable, difficult to quantify, and best
dealt with by making uniform comparison between results (see
\S\ref{sec:comparisons}).  See previous papers in this series for
further discussion.

\begin{figure*}[t]
\centering
\includegraphics[scale=0.4]{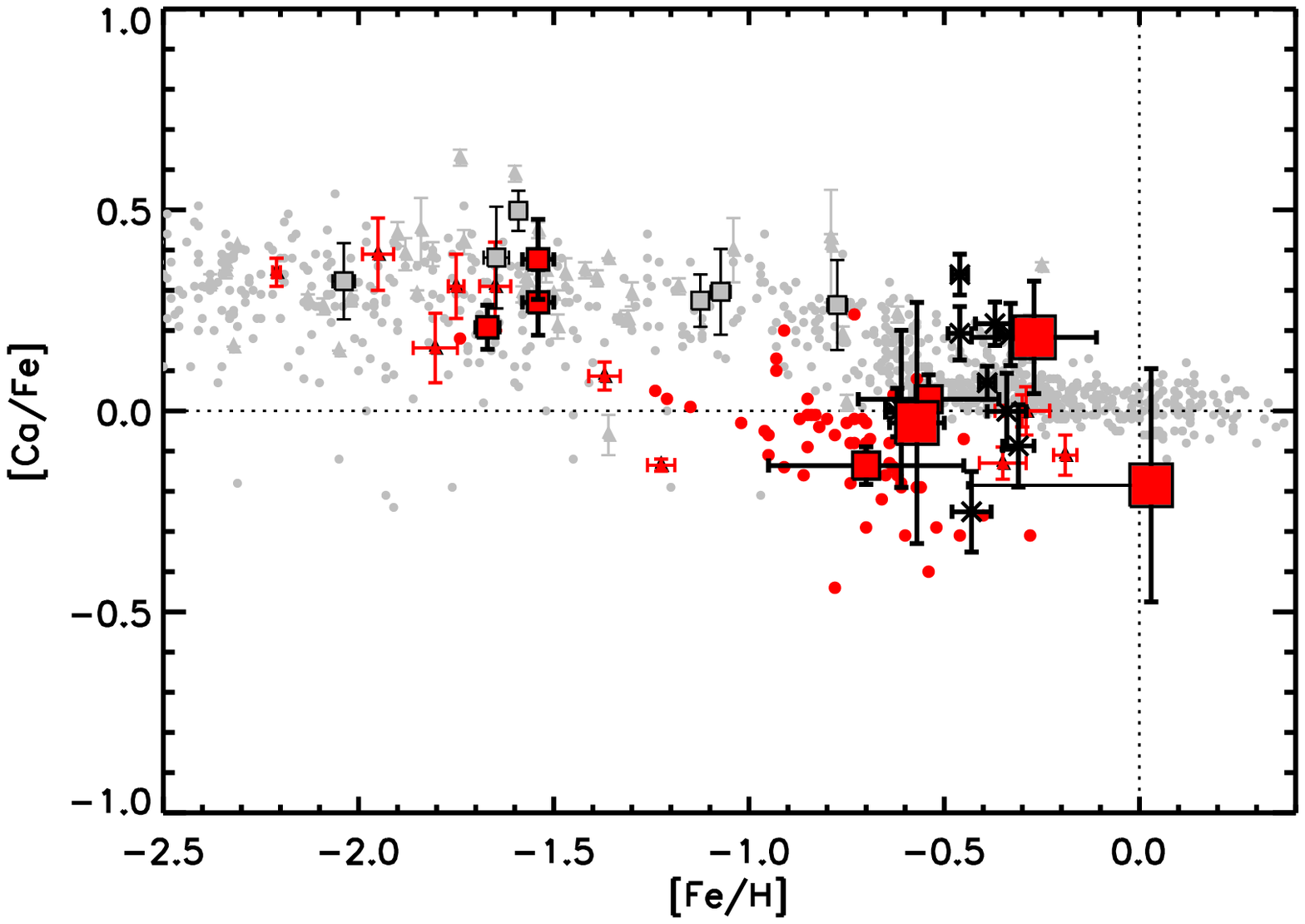}
\includegraphics[scale=0.4]{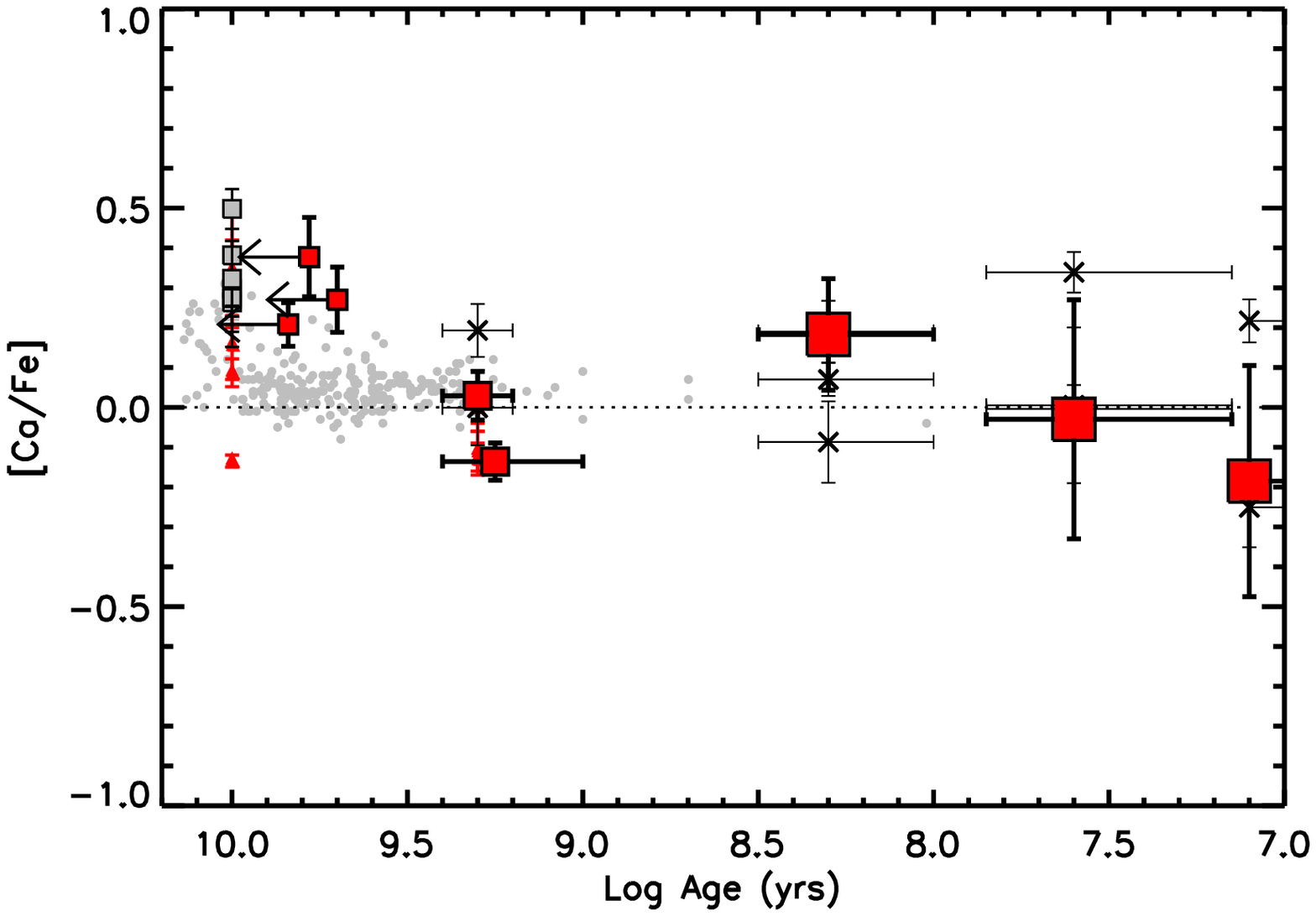}
\includegraphics[scale=0.4]{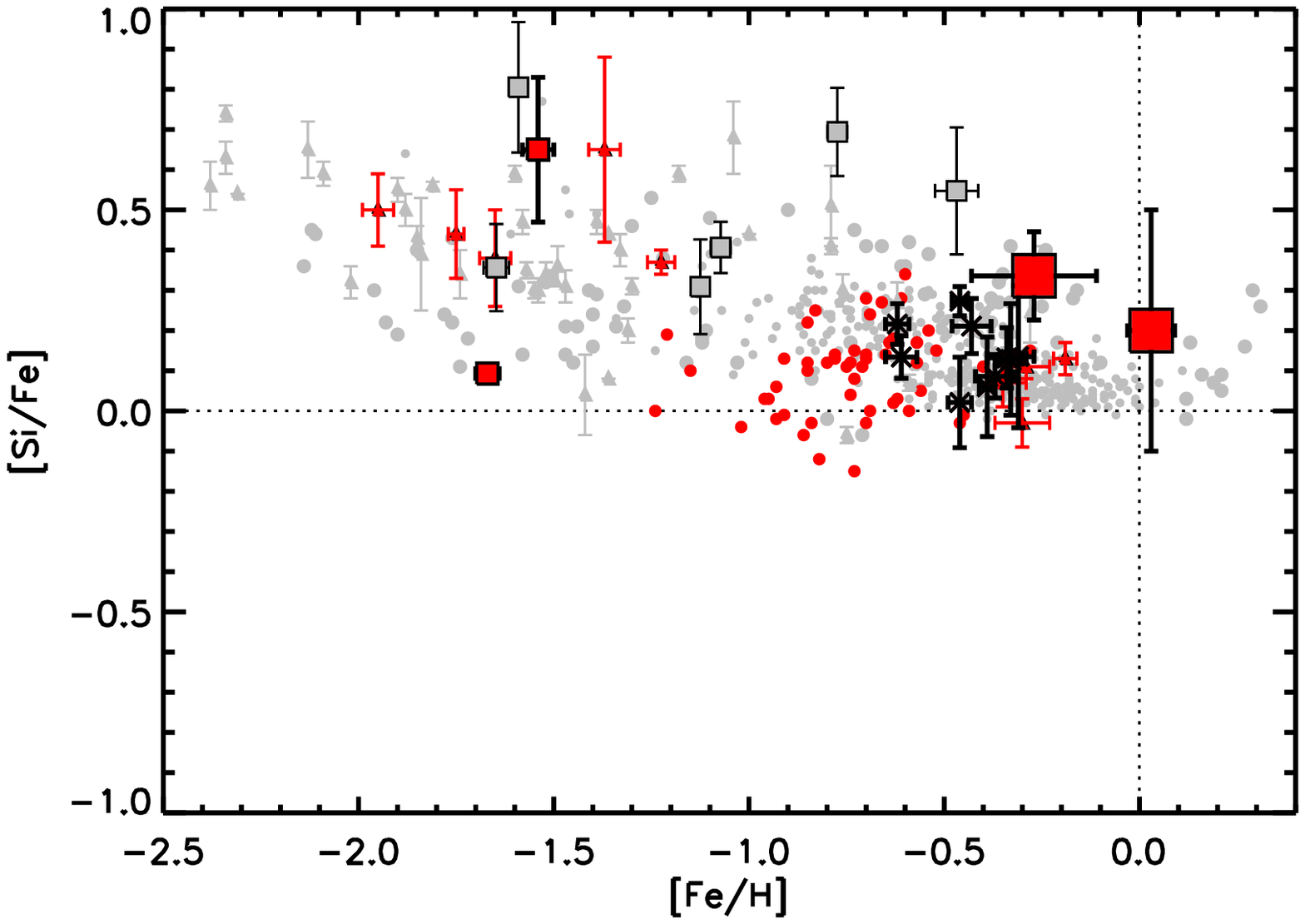}
\includegraphics[scale=0.4]{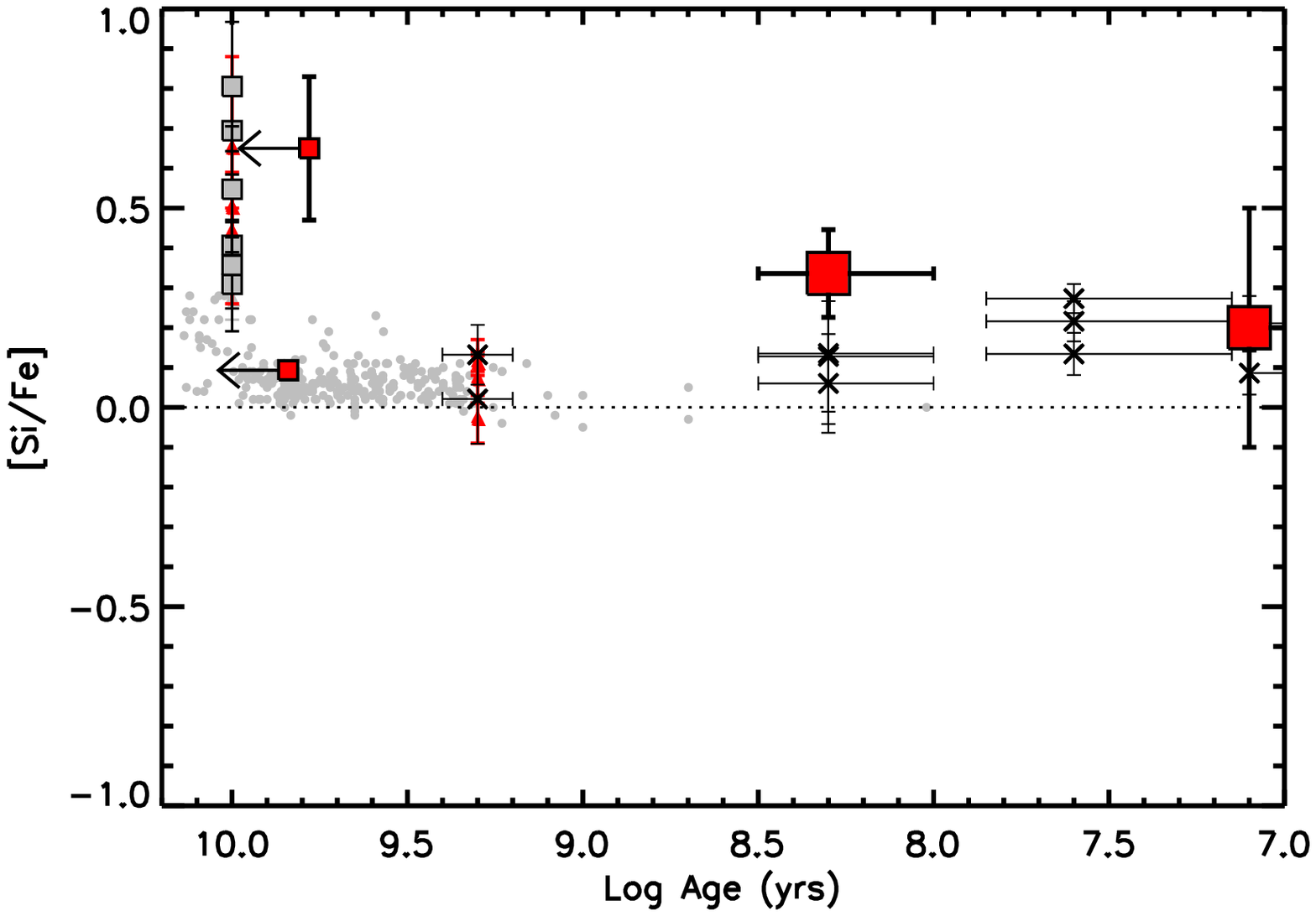}
\caption{Left panels show Ca I and Si I ratios as a function of [Fe/H], and right panels show the same as a function of age.  Small, medium and large red squares show the old, intermediate, and young cluster LMC IL abundances, respectively. Large grey squares show the MW abundances from \citetalias{milkyway}. Small black crosses show  the individual stars in the young LMC clusters. Small grey and red points show MW and  LMC stellar abundances from the literature.  Data for MW stars are from  \cite{bensby05}, \cite{2004AJ....128.1177V}, \cite{2005AJ....130.2140P}  and references therein. For the  LMC stars, triangles show abundances  from \citetalias{johnson06}, \citetalias{mucc08int}, \citetalias{mucc10old}, \citetalias{mucc1866} and  circles show abundances from \citetalias{pompeia08}. When possible the abundance ratios of other authors have been adjusted to be consistent with the solar abundance distribution of  \cite{2005ASPC..336...25A} that was used in our analysis. Ages for MW clusters and old LMC clusters are set to 10 Gyrs, and ages of LMC clusters of   \citetalias{mucc08int} are set to 2 Gyrs.} 
\label{fig:alpha}
\end{figure*}

\begin{figure*}
\centering
\includegraphics[scale=0.4]{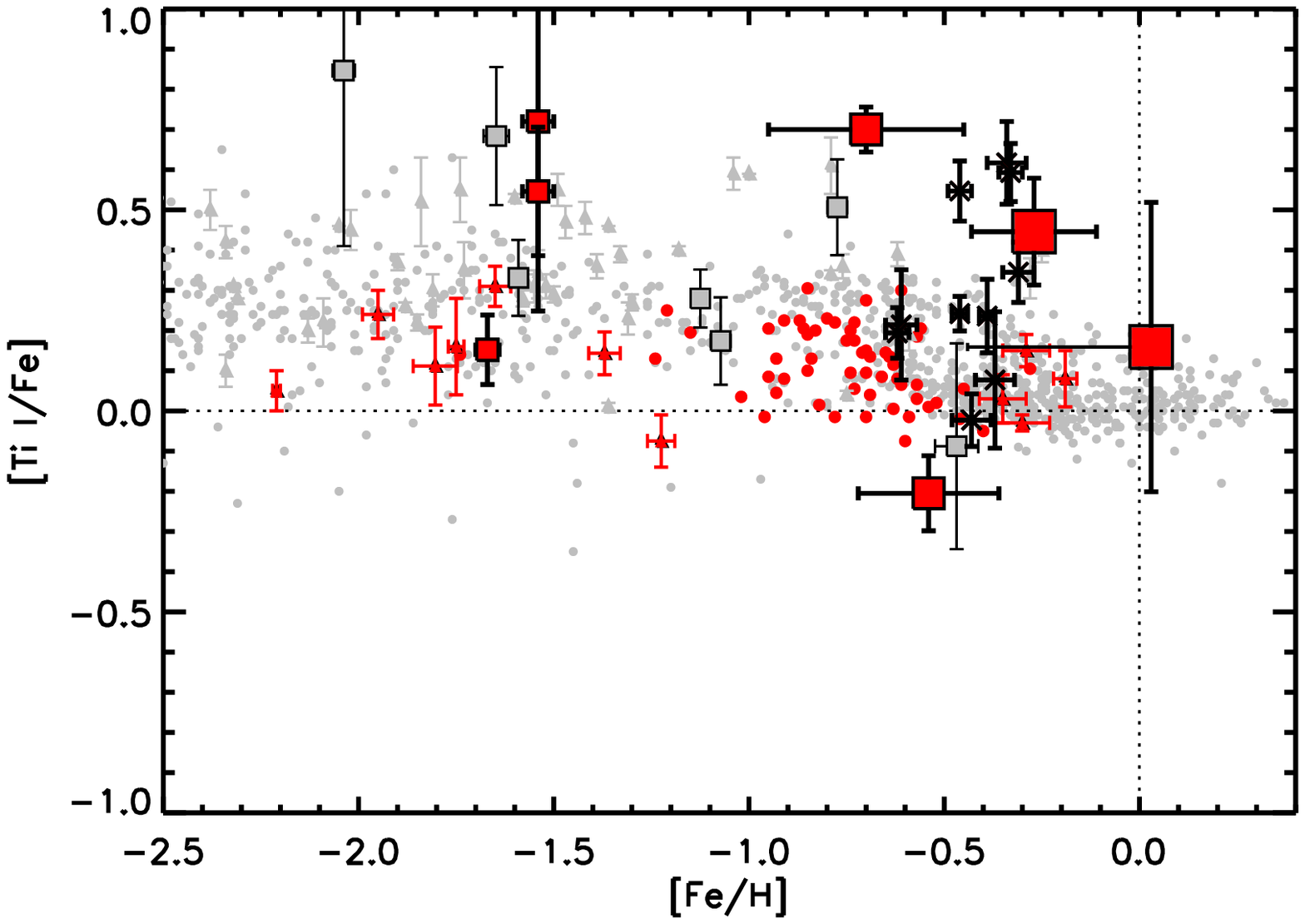}
\includegraphics[scale=0.4]{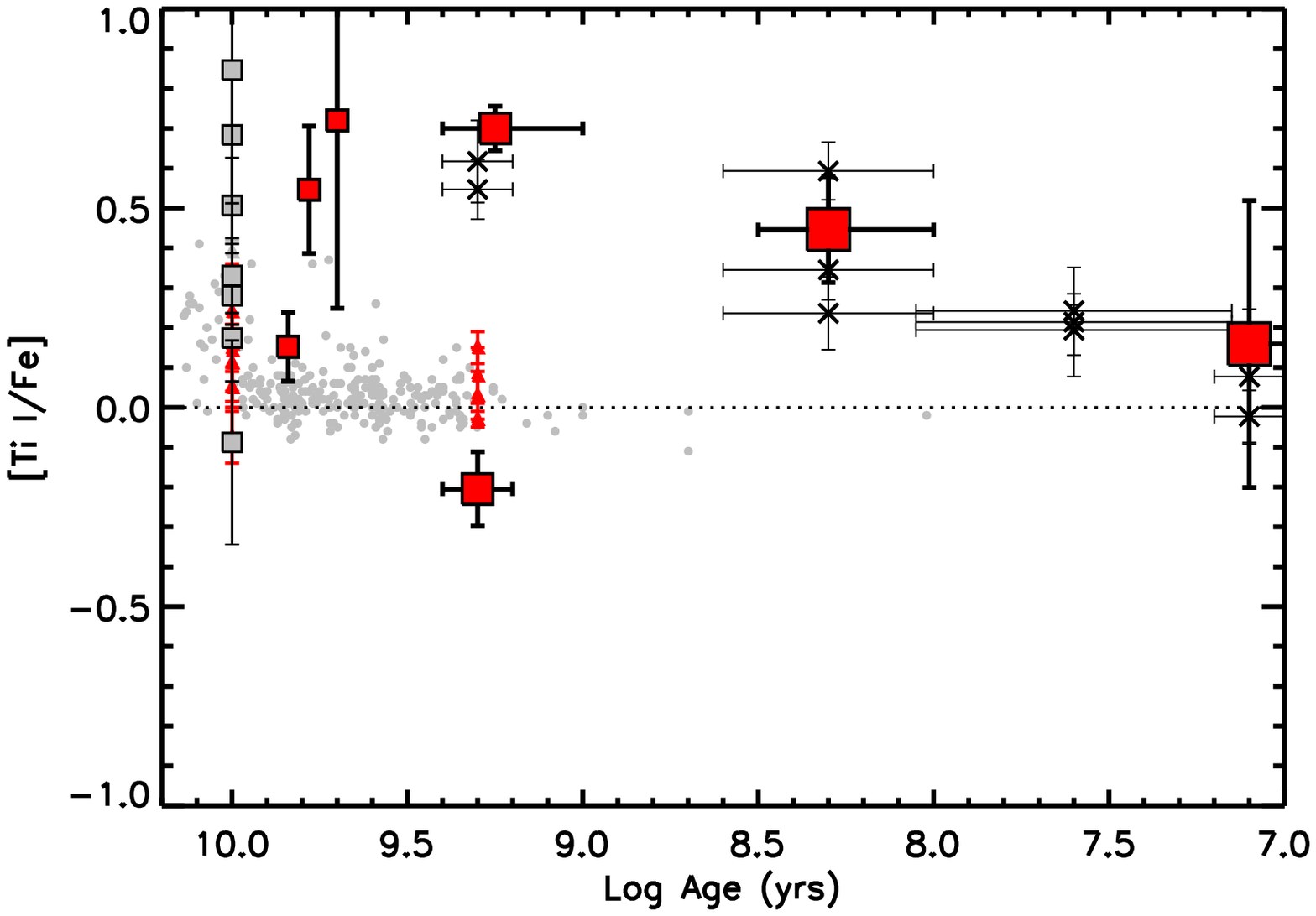}
\includegraphics[scale=0.4]{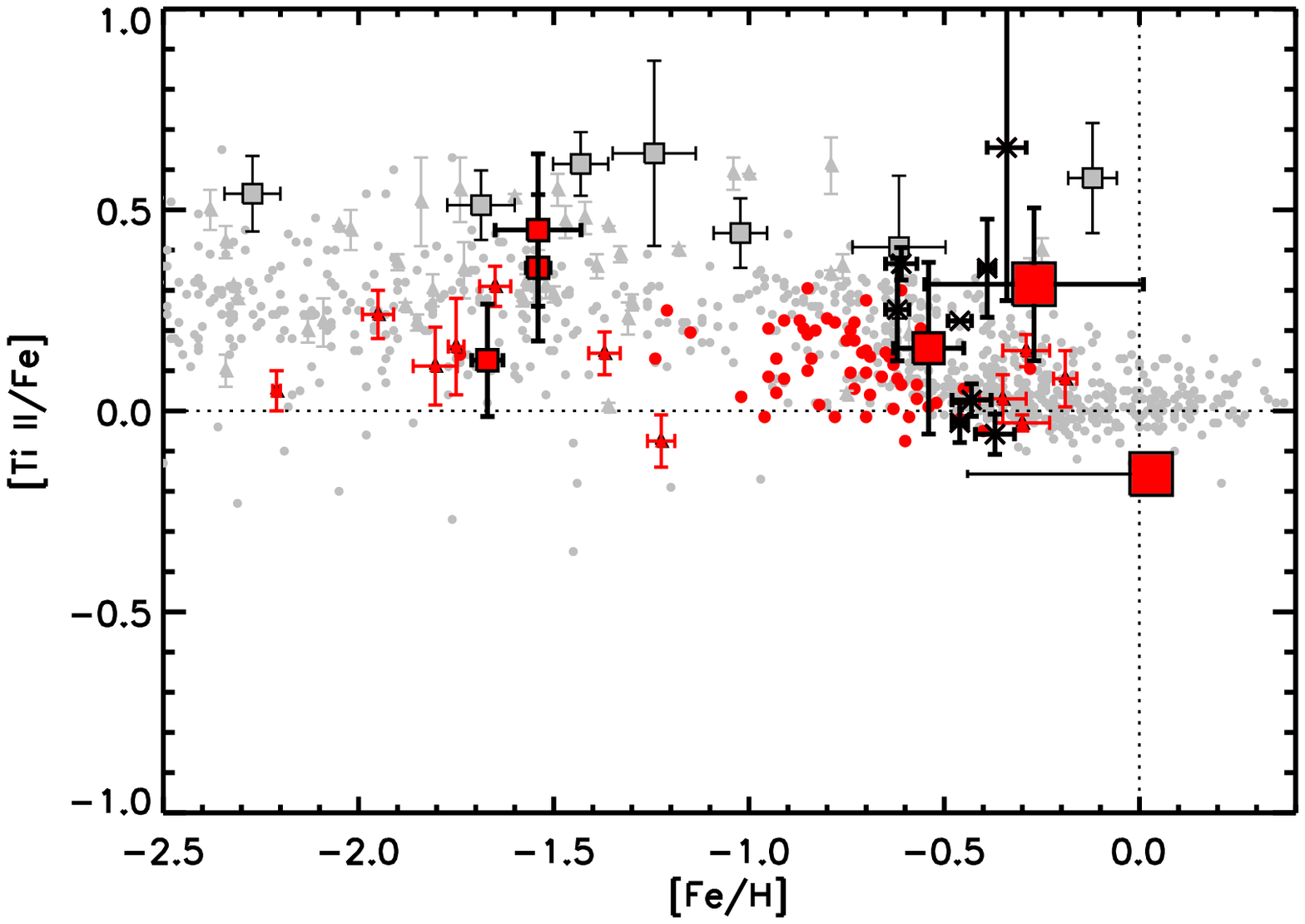}
\includegraphics[scale=0.4]{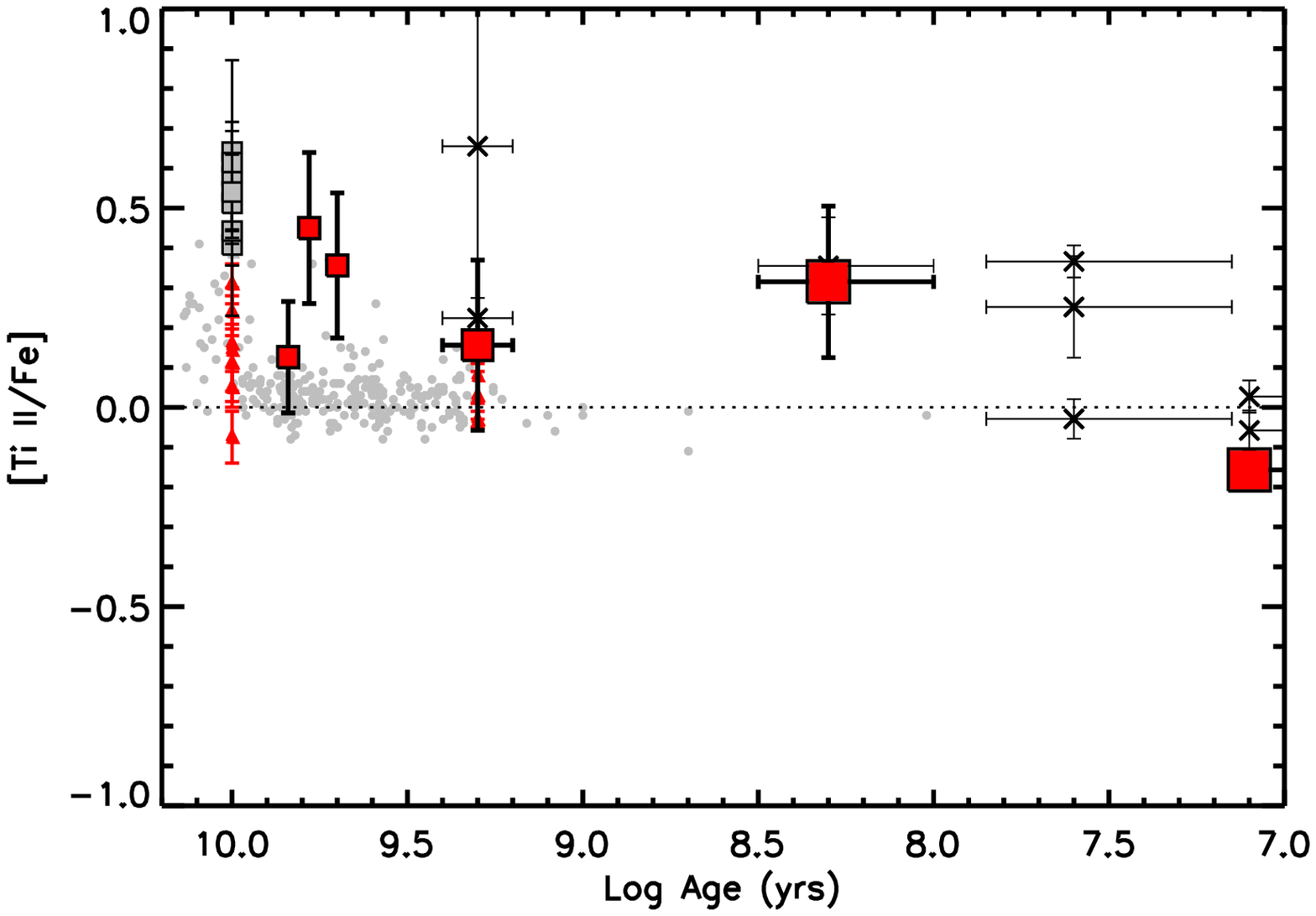}
\caption{The same as Figure \ref{fig:alpha} for Ti I and Ti II.}
\label{fig:ti}
\end{figure*}

The results for the 10 stars in the LMC clusters NGC 1978, NGC 1866,
NGC 1711 and NGC 2100 are reported in
Table~\ref{tab:abundance_table_stars}.  We also report the number of
lines of each species that are measured in each star, and the error in
the mean abundance.

For both the cluster IL and the individual stars,  abundances
relative to Fe are given using the solar abundance distribution of
\cite{2005ASPC..336...25A}, with a solar log$\epsilon$(Fe)=7.50. The only exception is that we use the 
solar log$\epsilon$(O)=8.93 of \cite{1989AIPC..183....9G}.  In the recent literature, we note that
there has been a convergence of the solar oxygen abundance to a lower
value, namely log$\epsilon$(O)=8.66 \citep{2005ASPC..336...25A}.  However, this
value has not yet propagated into the Kurucz model atmosphere grids
and so we are forced to adopt 8.93 dex in our calculations. Moreover,
the results in the literature to which we compare our work have also
 used a higher value for all calculations and normalizations.
For these reasons, we have elected to use the higher value for
internal consistency in our analysis and with the literature.  The
abundance ratios for neutral species are reported with respect to
[Fe/H]$_{\rm{I}}$, and ionized species with respect to
[Fe/H]$_{\rm{II}}$.

In Figures~\ref{fig:alpha} through \ref{fig:ncapture3}, we show the
LMC IL cluster abundances from our analysis (red squares) as a
function of [Fe/H] and age.  The abundances measured in the individual
stars are shown as black crosses.  LMC abundances are shown with the
MW IL GC training set abundances from \citetalias{milkyway} (gray
squares).  For reference, we show a compilation of abundances measured
for individual stars in the LMC as small red points and for the MW as
small gray points.  The small red and gray points that include error
bars correspond to stellar abundances from clusters in the LMC and MW,
respectively.  References for individual star abundances are located
in the captions of Figures~\ref{fig:alpha} through
\ref{fig:ncapture3}. For the old sample of LMC clusters the error bars
on the IL abundances in Figures~\ref{fig:alpha} through
\ref{fig:ncapture3} correspond to the standard error in the mean of
the lines for each species. For the intermediate age and young
clusters the error bar corresponds to $\sigma_{\rm{Tot}}$, which
includes the $\sigma_{lines}$ and $\sigma_{age}$ added in quadrature.
For elements where only one clean line was measured, $\sigma_{lines}$
was estimated using the $\chi^2$-minimization technique described in
\textsection~\ref{sec:synth}.

\subsection{Alpha Elements: Ca, Ti and Si}
\label{sec:alphas}

 In this section,  we present the abundances for $\alpha$-elements 
 Ca I, Ti I, Ti II and Si I.
These are the first-ever $\alpha$-element abundances for NGC 1916, NGC
1718, and NGC 1711. 
Abundances of O I and Mg I are discussed with the other light elements
in \textsection \ref{sec:light}. 
The abundances for all clusters  are plotted in Figure \ref{fig:alpha}
as a function of both [Fe/H] (left panels) and age (right panels).

The [Ca/Fe] for all clusters tends to decrease with decreasing
age and increasing [Fe/H], as one would expect.
The old clusters (NGC 1916, NGC 2019, and NGC 2005) have [Ca/Fe]  in
the range $+0.2$ to $+0.4$. 
The intermediate age clusters (NGC 1718 and NGC 1978) have 
[Ca/Fe] of $-0.14$ and $+0.03$, respectively, from IL. Individual
stars in NGC 1978 have [Ca/Fe]= $+0.10 \pm 0.13$.
From both IL and individual star spectra, the youngest clusters 
(age $<$ 1Gyr) have a spread in [Ca/Fe] between $-0.19$ and $+0.18$.

Our measurements, obtained by both IL and individual stellar spectra,
agree with previous results (see detailed comparison in \textsection \ref{sec:comparisons}).
While the abundance information available for the LMC is still
limited, our results add to increasing evidence that there is
generally a larger range in [$\alpha$/Fe] in LMC stars than in MW
stars of similar metallicity
\citep[e.g.][]{2004AJ....128.1177V,2005AJ....130.2140P,pompeia08,
  smith02}.  This spread could indicate incomplete mixing of the LMC
ISM, or possibly gas inflow or outflow.

 In the data for this sample, there are more individual lines
      available for Ca I than for any other $\alpha$-element.  Therefore the Ca I measurements given here are the most
      statistically significant of our IL $\alpha$-element results.  While Ca is not a ``pure'' $\alpha$-element
--- it is also produced in small quantities in Type Ia supernovae (SNe
Ia) --- it is the most accurate and consistent $\alpha$-element for
cluster comparisons.  We find that the most evolution in [Ca/Fe]
occurs between 10 and 2 Gyrs ago, with a decreasing mean value over
that range,  and remains more or less constant  since 2 Gyr ago.
This implies that the LMC was dominated by SNe II enrichment early on,
followed by significant SNe Ia enrichment before the second burst of
star formation ($\sim8$ Gyr later) formed the intermediate age
clusters \citep{hz2009}.  This is consistent with the evolutionary
timescale for both SNe-types.

Our [Si/Fe] measurements generally show the same behavior as our
[Ca/Fe] measurements: a spread in [Si/Fe] for older clusters and
decreasing [Si/Fe] with decreasing age.
Fewer Si I transitions make Si I more difficult to measure than Ca I.
Nonetheless previous measurements of [Si/Fe] in field and cluster
stars in the LMC are consistent with our results, including some
indication that mean values for [Si/Fe] are $\sim$0.1-0.2 dex higher
than those of [Ca/Fe], as can be seen in Figure \ref{fig:alpha}.

In Figure \ref{fig:ti}, we show the results for Ti I and Ti II,
separated by ionization state.  [Ti I/Fe] and [Ti II/Fe] generally
show both a larger line-to-line scatter and scatter between clusters
than [Ca/Fe] or [Si/Fe], which is likely due to the S/N of the data.
Despite the higher scatter, [Ti I/Fe] and [Ti II/Fe] are generally
consistent with [Ca/Fe] and [Si/Fe] in that Ti decreases with
decreasing age.  Our [Ti I/Fe] measurements are higher on average than
previous measurements obtained for field and cluster stars, but the
higher uncertainties in our measurements make it difficult to
determine if there is a systematic offset.  Our [Ti II/Fe]
measurements overlap more with the field and cluster star
measurements.

\begin{figure}
\centering
\includegraphics[scale=0.5]{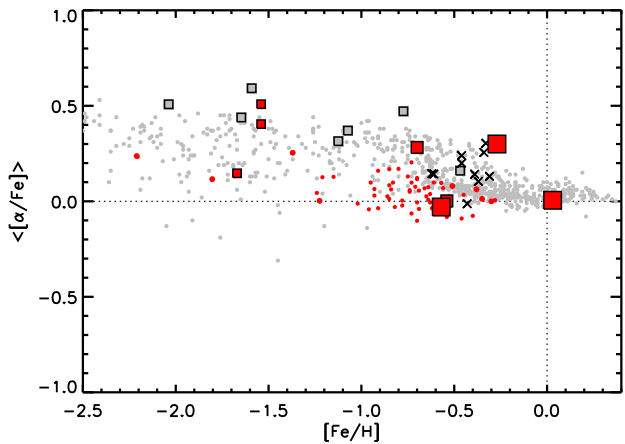}
\caption{Mean [$\alpha$/Fe] calculated from Ca I, Ti I, Ti II, and Si I.  Symbols and data for single stars are the same as in Figure~\ref{fig:alpha}.  }
\label{fig:alpha_avg}
\end{figure}

Although there is considerable scatter between clusters, we find that
the mean [Ca/Fe], [Si/Fe], and [Ti/Fe] for the old LMC clusters are
consistent with measurements for MW clusters
\cite[e.g.][]{2005AJ....130.2140P,milkyway}.  This is further evidence
that, like the MW, the LMC ISM was dominated by enrichment of SNe II
when the old clusters formed $\sim10$ Gyr ago.  
To clarify the overall trends, we have
tabulated the mean [$\alpha$/Fe] for the old LMC clusters in
Table~\ref{tab:alpha_table}, and the mean  [$\alpha$/Fe] for
the MW IL sample.  This is particularly useful for 
comparisons with results in the literature from other abundance
analysis techniques.  The mean [$\alpha$/Fe] for the old LMC clusters
is only $\sim$0.04 dex lower than for the MW training set GCs, and
consistent within the statistical uncertainty.  In
Figure~\ref{fig:alpha_avg} we show the mean [$\alpha$/Fe] as a
function of [Fe/H] for all of the LMC clusters, both old and young, as
well as field stars in the MW and LMC.

\subsection{Light Elements: O, Na, Mg, Al}
\label{sec:light}

It is well known that MW GCs exhibit star-to-star abundance variations
for light elements involved in high temperature proton-capture
nucleosynthesis \citep[see ][ for a recent
summary.]{2004ARA&A..42..385G}.  Recently, we found indirect evidence
for abundance variations in M31 GCs \citep{m31paper}, and
\cite{2009ApJ...695L.134M} confirmed that abundance variations for O,
Na, Mg, and Al are also present in three old, metal-poor GCs in the
LMC. This indicates that star-to-star abundance variations are likely to
be  ubiquitous and an integral part of massive cluster formation,
not just limited to the MW.  When star-to-star abundance variations
are present, a fraction of stars in the cluster can exhibit any of the
following to varying degrees: depleted O due to the ON-cycle, enriched
Na due to the NeNa-cycle, and depleted Mg and/or enriched Al due to
the MgAl-cycle.  Recently, several authors have tried to connect these
star-to-star abundance variations with the observations of multiple
populations of stars in MW globular clusters
\citep[e.g.][]{2008MNRAS.391..825D,2010A&A...516A..55C,2007A&A...464.1029D,2009A&A...507L...1D,2011ApJ...726...36C}.
 
As mentioned above, we have already shown that indirect evidence for
abundance variations can be measured in cluster IL spectra.
Specifically, the IL analysis shows large scatter in [Mg/Fe] when
compared to other $\alpha$-elements, as well as a lower mean [Mg/Fe]
and significantly elevated [Al/Fe] and [Na/Fe].  This was seen both in
our M31 GCs, mentioned above, and in  \citetalias{mb08} and
\citetalias{milkyway} for  the MW IL sample.
The same indications of star-to-star abundance variations are now also
clear in the IL of our old LMC clusters.

\begin{figure*}
\centering
\includegraphics[scale=0.4]{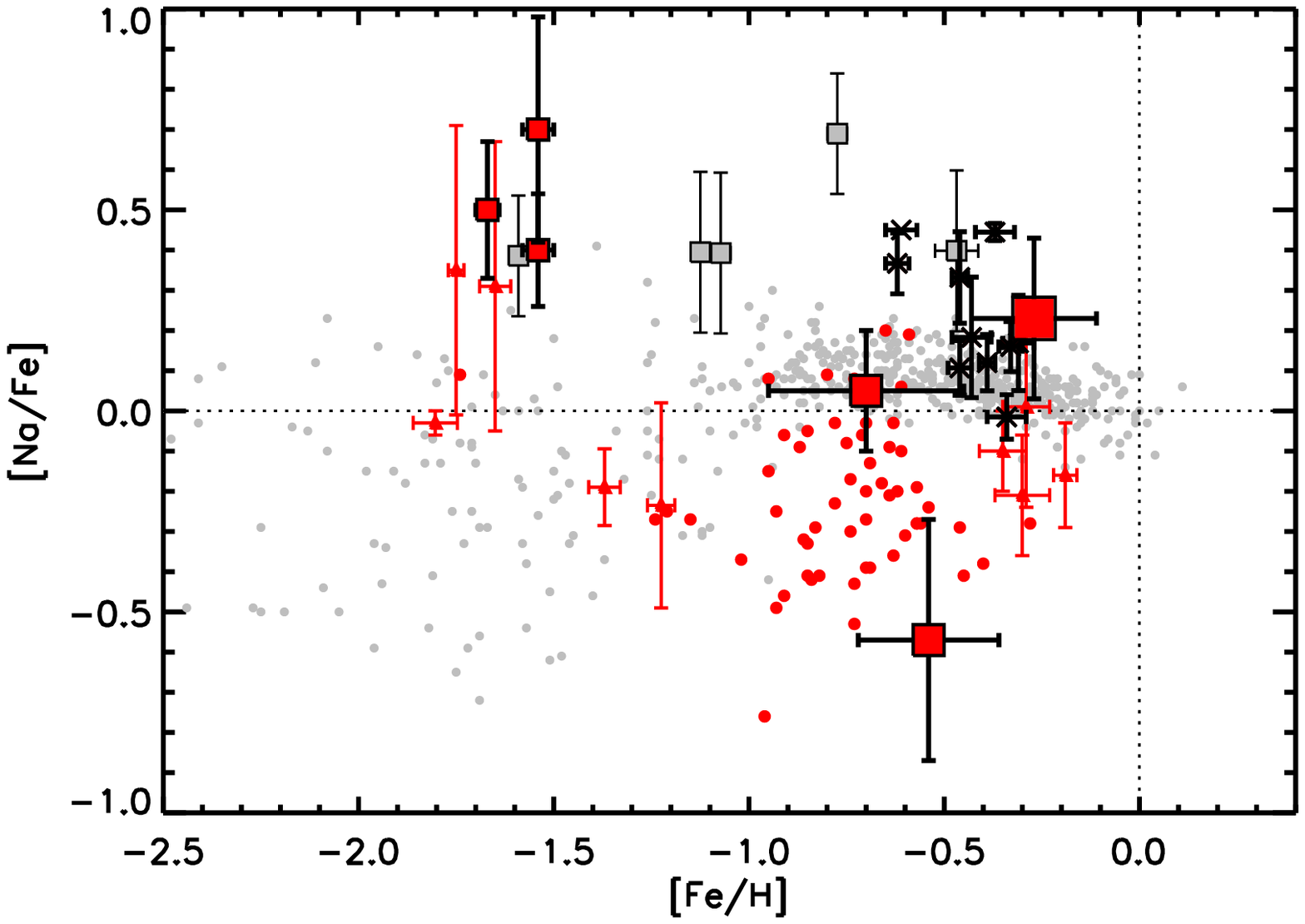}
\includegraphics[scale=0.4]{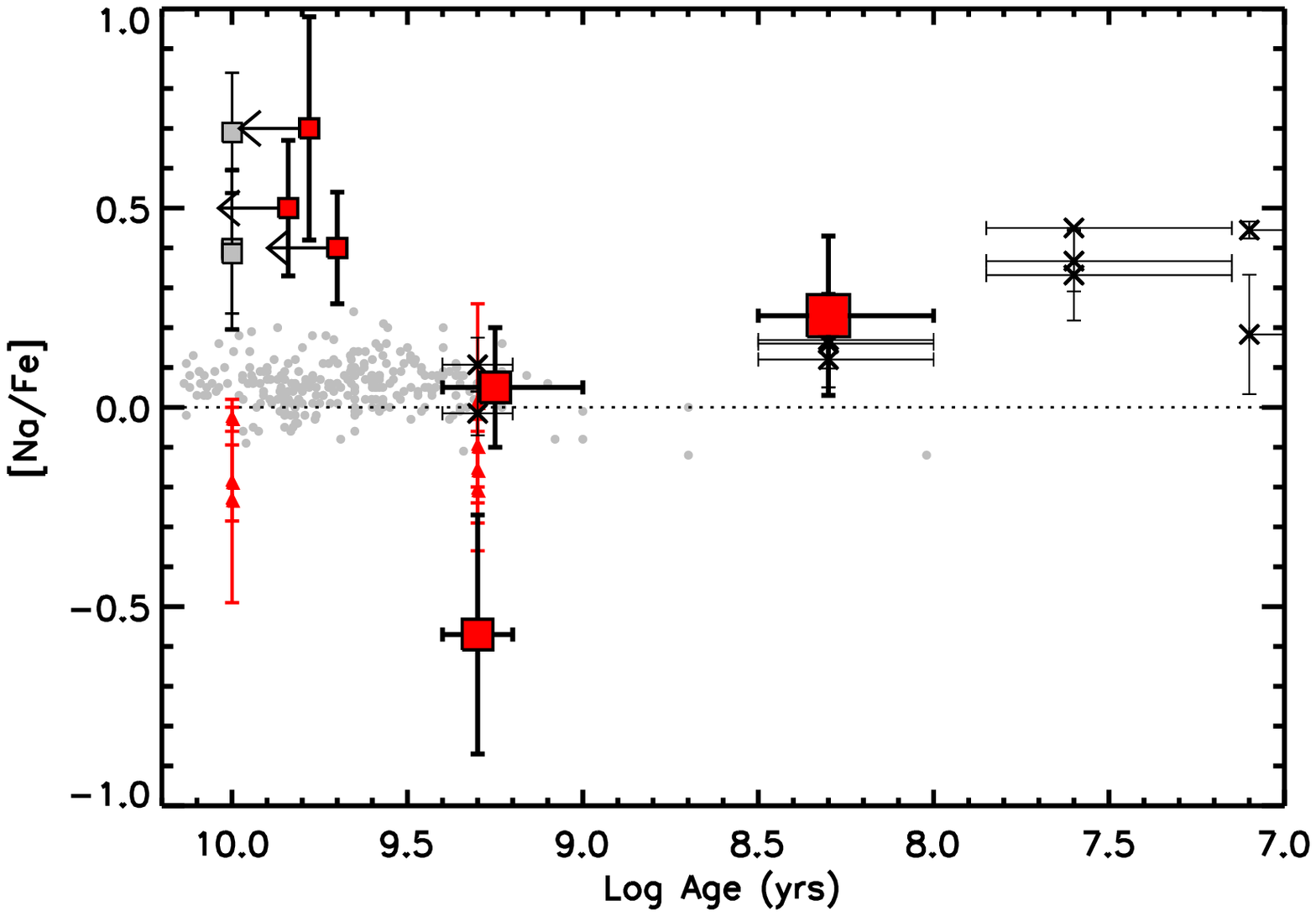}
\includegraphics[scale=0.4]{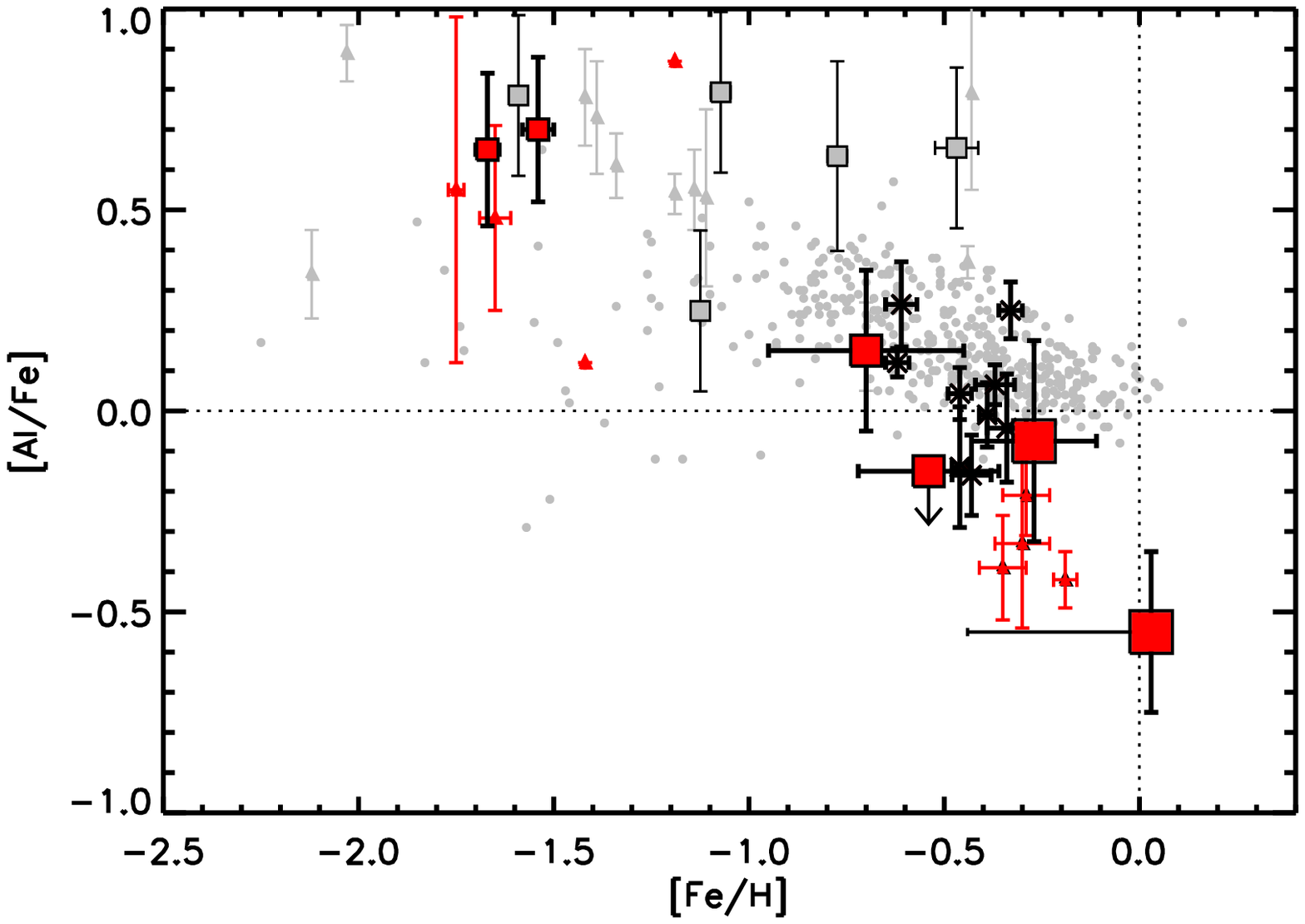}
\includegraphics[scale=0.4]{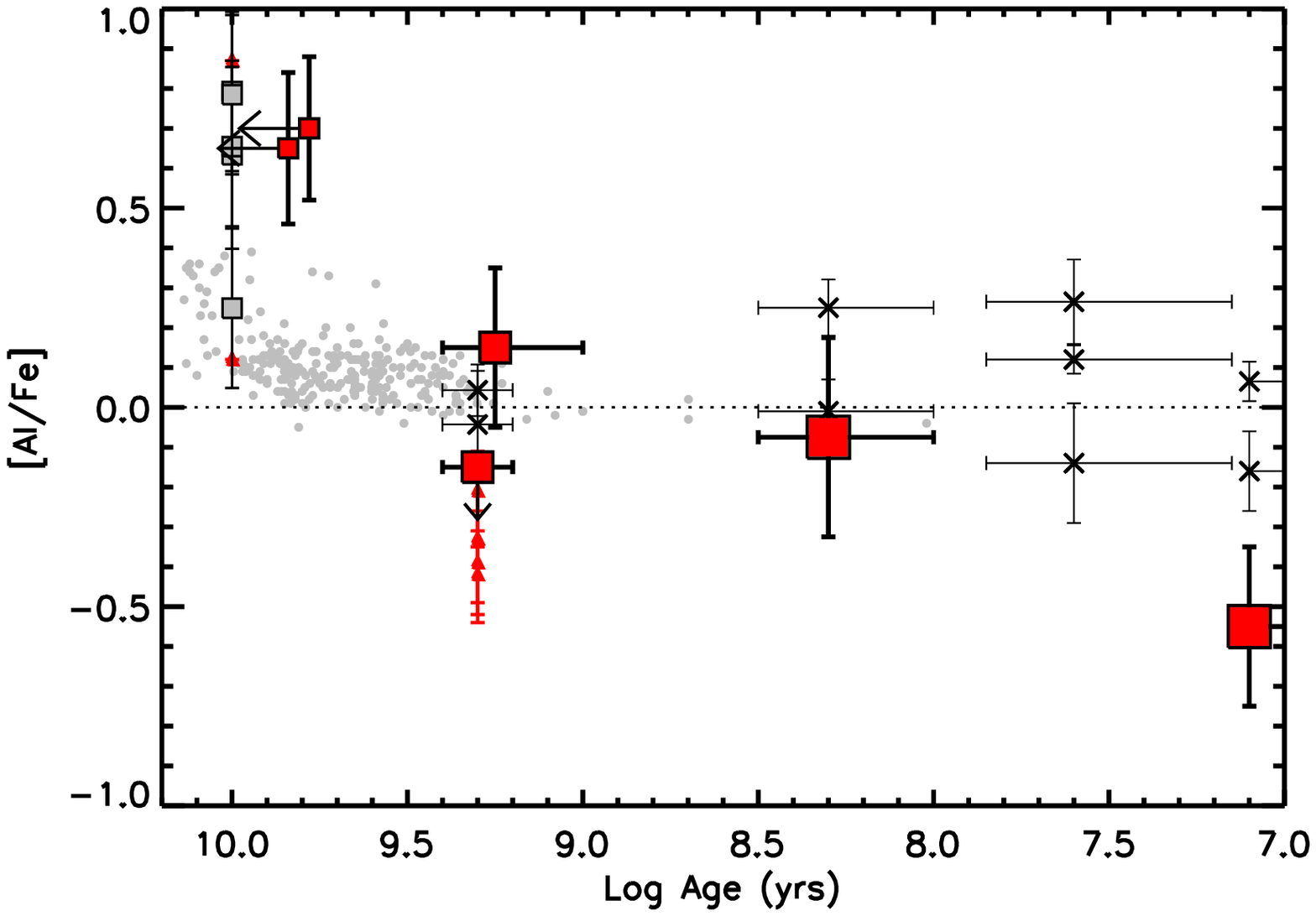}
\caption{Same as Figure \ref{fig:alpha} for the light elements Al and Na.  Symbols and data for single stars are the same as in Figure~\ref{fig:alpha}. Additional LMC individual star data (also triangles) are from  \cite{2009ApJ...695L.134M}.  Additional MW GC individual star data are from references compiled in \cite{2006AJ....131.1766C}.  }
\label{fig:light1}
\end{figure*}

\begin{figure*}
\centering
\includegraphics[scale=0.4]{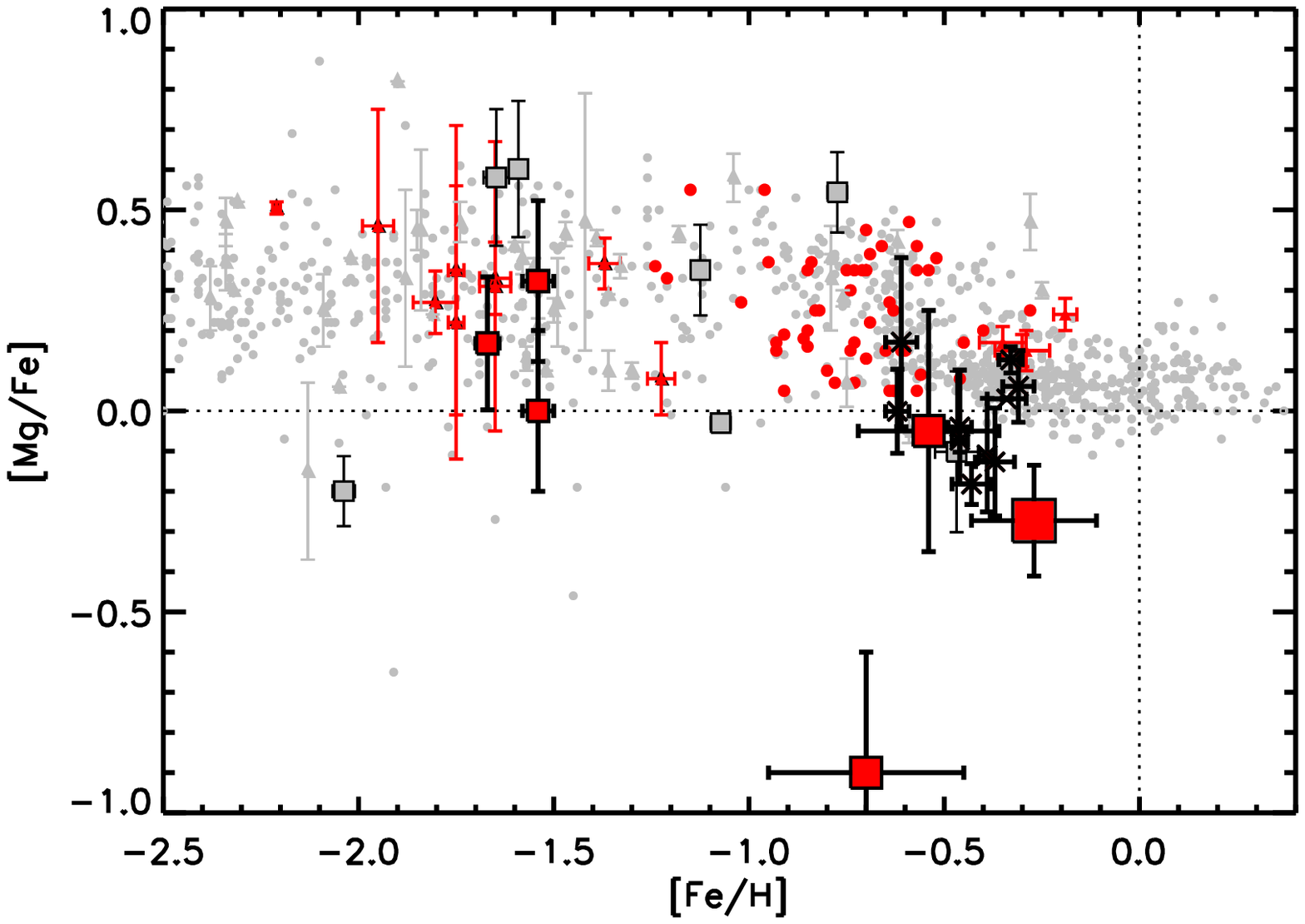}
\includegraphics[scale=0.4]{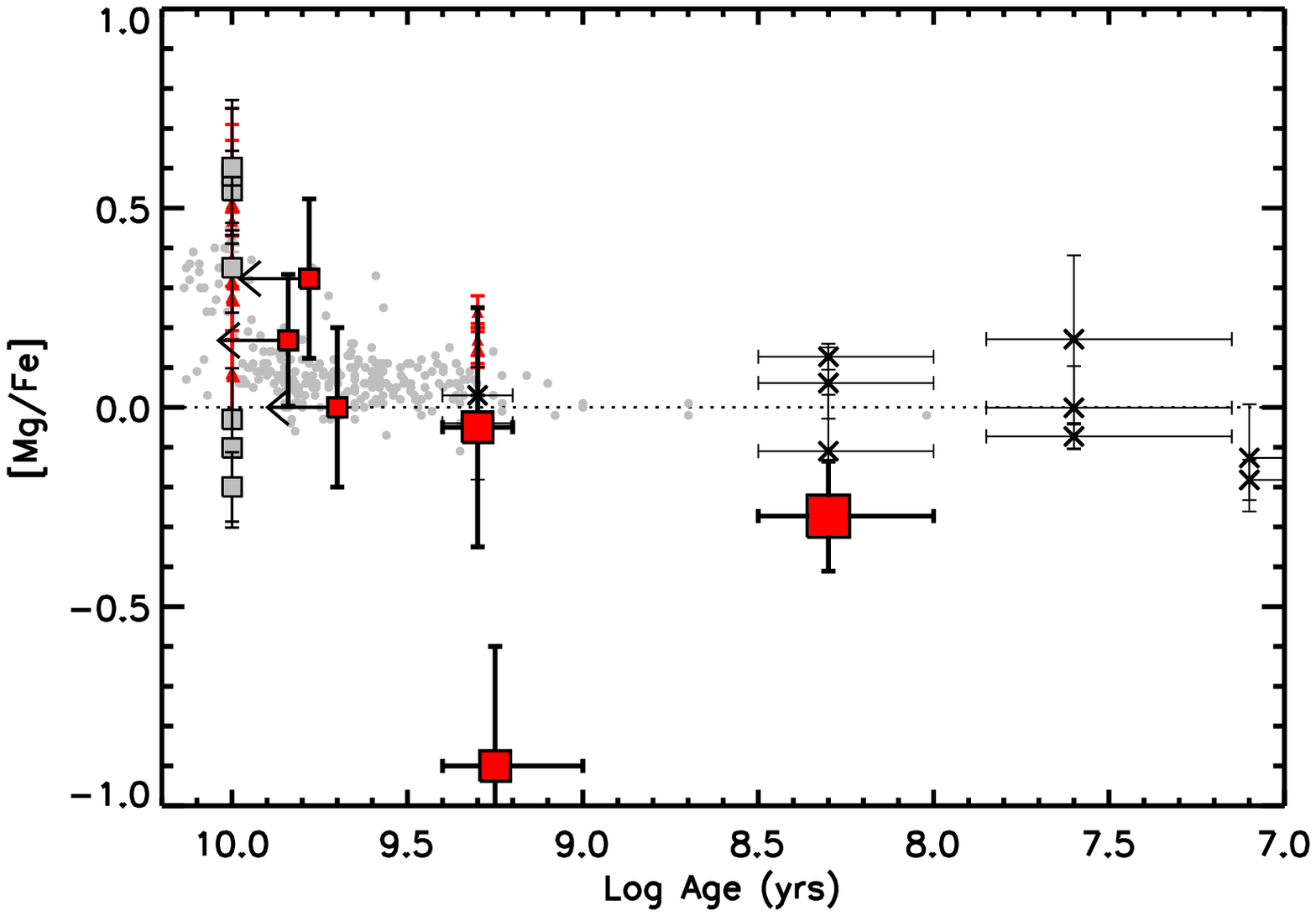}
\includegraphics[scale=0.4]{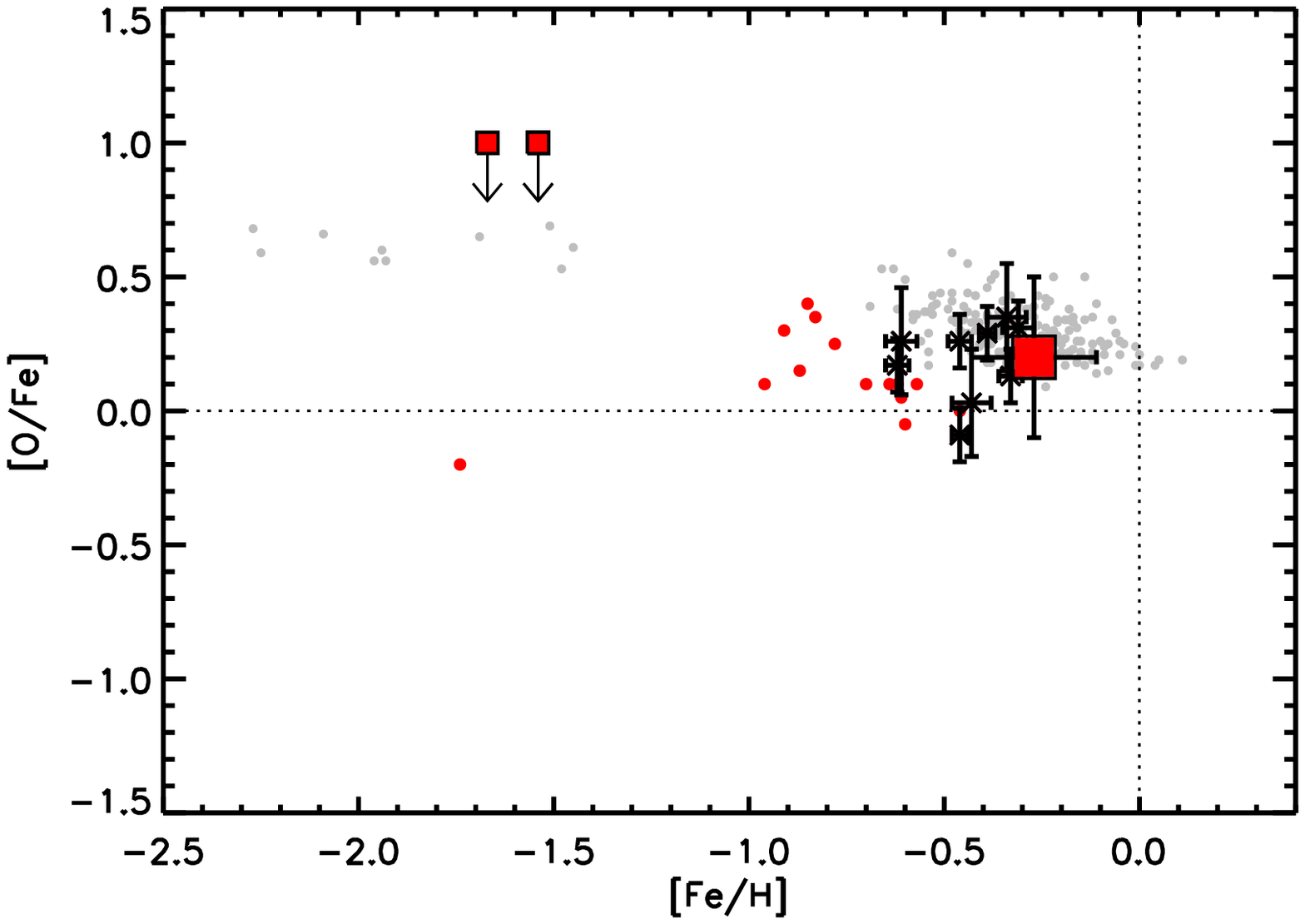}
\includegraphics[scale=0.4]{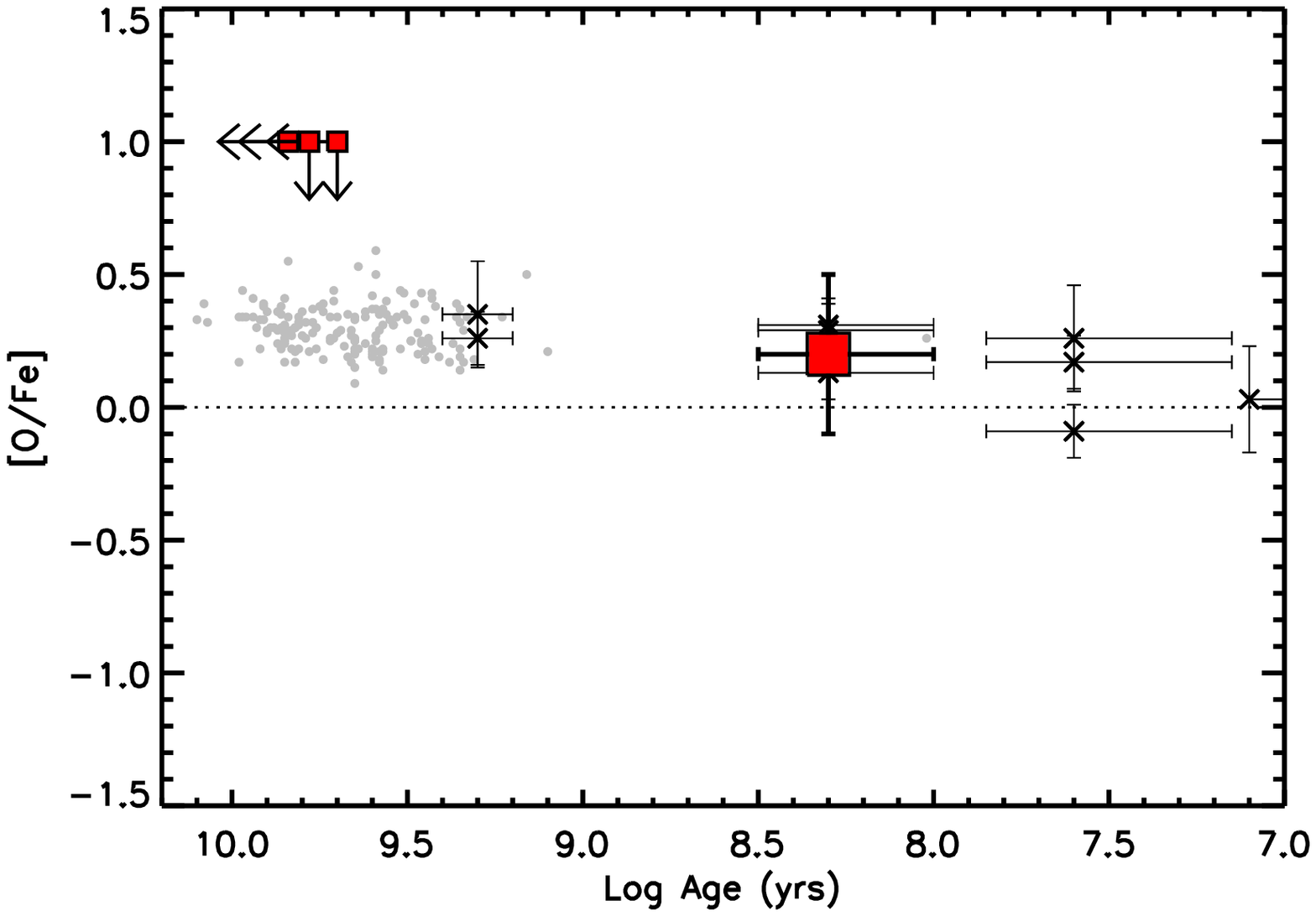}
\caption{Same as Figure \ref{fig:alpha} for the light elements O and Mg.   Symbols and data for single stars are the same as in Figure~\ref{fig:alpha}.   Additional LMC individual star data (also triangles) are from  \cite{2009ApJ...695L.134M}.  Note that the cluster IL [O/Fe] values  are measured from the 7771 \rAA triplet, and that no non-LTE correction can currently be applied. }
\label{fig:light2}
\end{figure*}

For all of our LMC clusters, the Na and Al abundances are shown in
Figure \ref{fig:light1}, and the Mg and O abundances are shown in
Figure \ref{fig:light2}.  The old LMC clusters in our sample clearly
have elevated [Na/Fe] ($\sim+0.5$ dex) compared to field stars and
younger clusters.  The intermediate age clusters (NGC 1978 and NGC
1718) have significantly lower [Na/Fe] with a wide range of values
($\sim$$-0.6$ to $+0.1$).  In the younger clusters, [Na/Fe] is only
available in the IL of NGC 1866, where [Na/Fe]=$+0.23 \pm 0.2$.

Because Na I is difficult to measure, it is worth noting the features
from which our results are derived. 
The most accessible Na I features in cluster IL spectra are the
5682/5688 \rAA and 6154/6160 \rAA doublets, which are relatively weak
($\sim$30 m\rAA), and so are measured with the line synthesis
component of ILABUNDS.  Stellar Na I abundances are measured with the
SYNTH routine in MOOG.  We have not corrected our cluster IL or
stellar abundances for non-LTE effects, but note that for cool giants
(T$_{eff}<$5000 K) the corrections tabulated by
\cite{1999A&A...350..955G} are at most $+$15 \% at high metallicity.
 
The range in [Na/Fe] found by \cite{2009ApJ...695L.134M} is consistent
with our result, as can be seen in Figure \ref{fig:light1}.  Our
measurements of [Na/Fe] from 10 individual stars in the young clusters
generally agree with the IL measurements, as can be seen in Figure
\ref{fig:light1}.

Also in Figure \ref{fig:light1} are our [Al/Fe] results.  Similar to
[Na/Fe], the old clusters clearly also have high [Al/Fe], with a mean
above $+0.5$ dex, as also true of some old MW clusters.  The
intermediate age clusters  have [Al/Fe] abundances similar to
[Na/Fe], with values significantly lower than for the old clusters.
For young clusters, the IL and individual stars both show lower
abundances than the intermediate age clusters.  This results in a
trend of decreasing [Al/Fe] with decreasing cluster age, as can be
seen in Figure \ref{fig:light1}.  Like Na I, the Al I measurements
come from line synthesis in both cluster IL and individual stars.  All
of our cluster IL abundances are obtained from the Al I 6696/6698 \rAA
doublet, while the stellar Al I abundances are obtained both from the
6696/6698 \rAA doublet and the Al I 7835.31 \rAA feature.

\begin{figure}
\centering
\includegraphics[angle=90,scale=0.35]{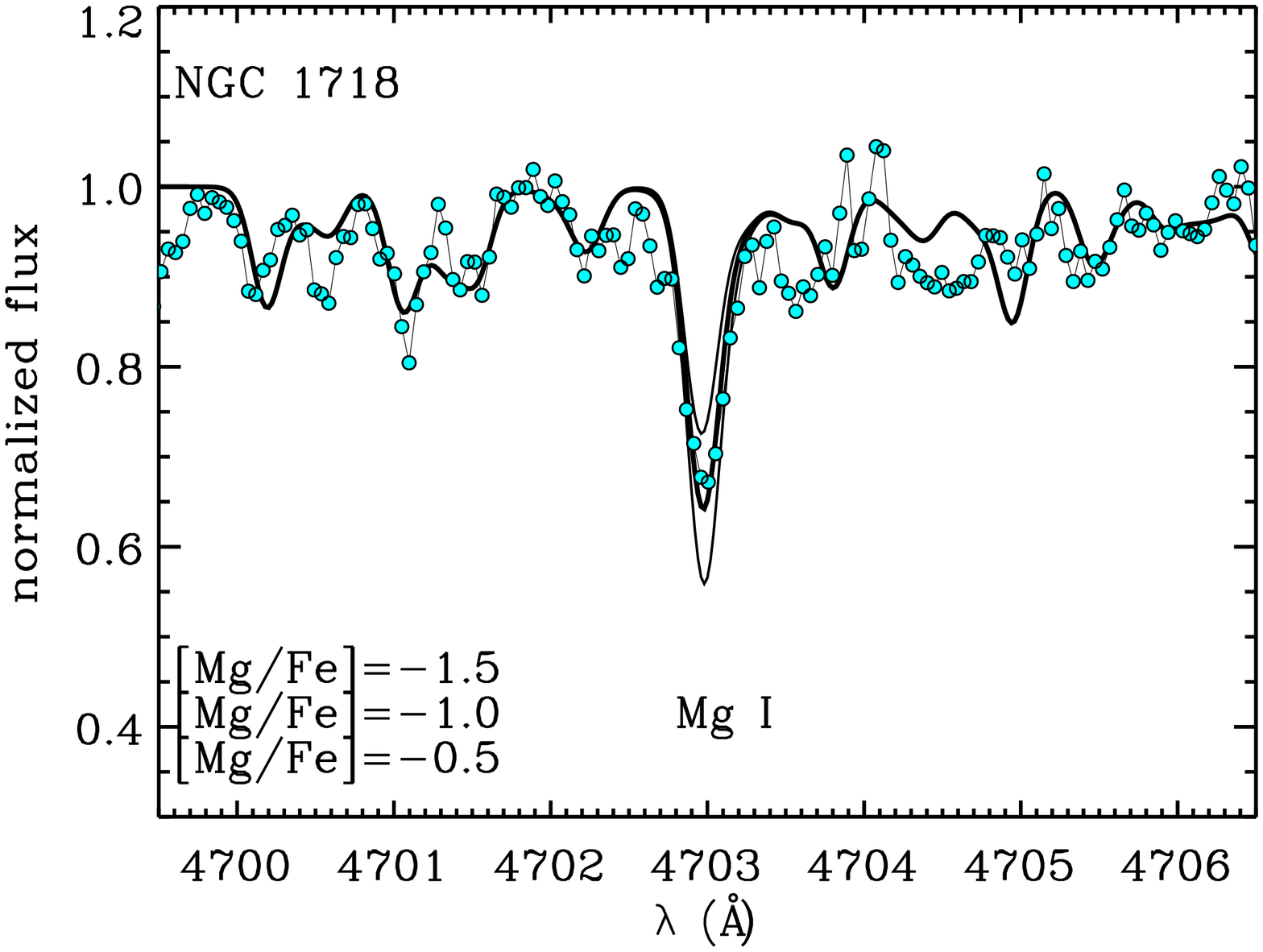}
\includegraphics[angle=90,scale=0.35]{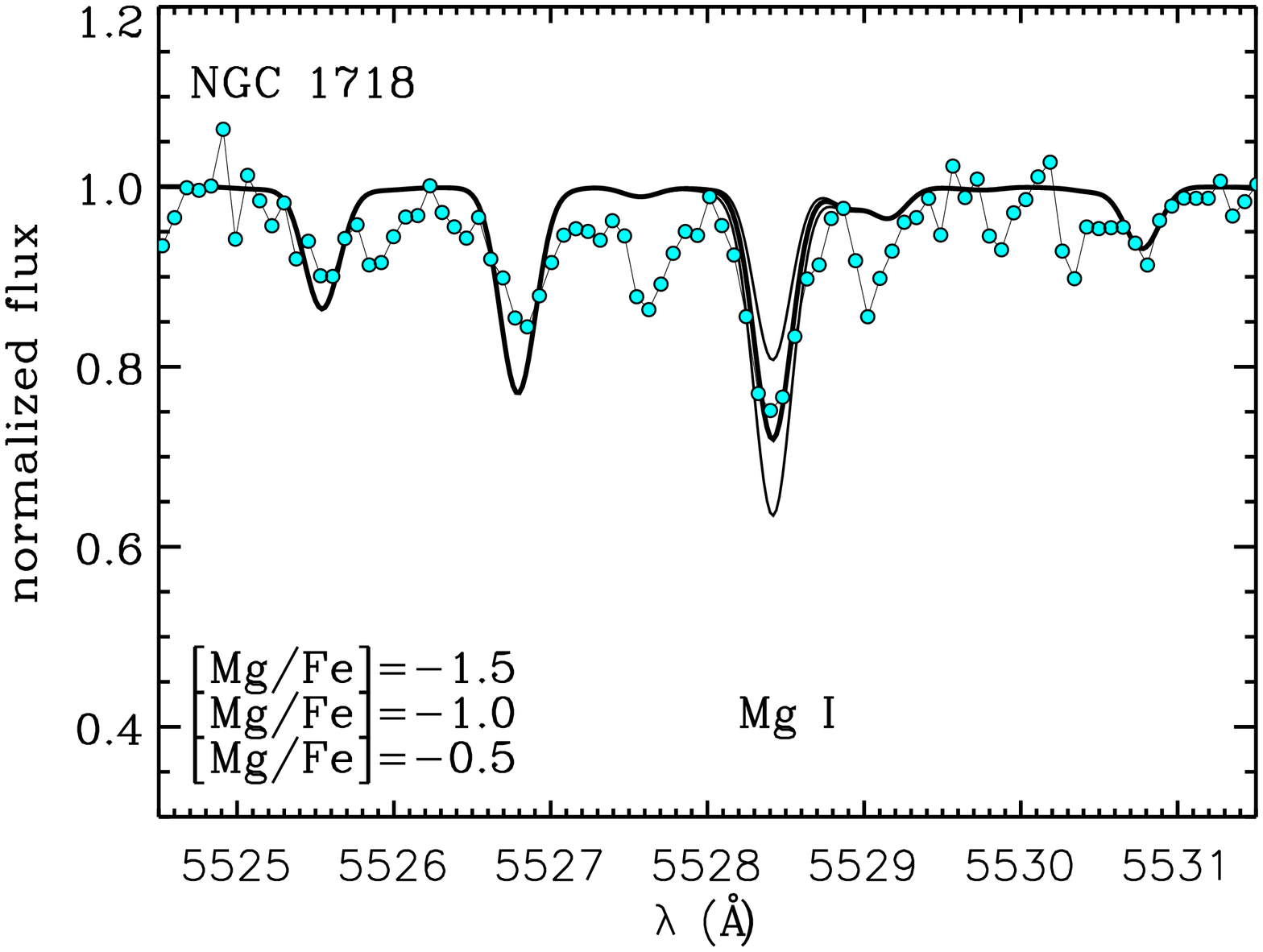}
\caption{ IL line syntheses for two Mg I features in the IL spectra of NGC 1718. Cyan circles show the data and thick black lines show syntheses of [Mg/Fe] = $-1.5, -1.0,$ and $-0.5$.  All synthesized spectra were created using a CMD with an age of 3 Gyr and a metallicity of [Fe/H]$=-0.73$, corresponding to our oldest acceptable solution for NGC 1718. Both the 4703 \rAA and 5528 \rAA features are best fit by the [Mg/Fe]=$-1.0$ synthesis.}
\label{fig:mg}
\end{figure}

Figure \ref{fig:light2} shows the results for [Mg/Fe].  [Mg/Fe] is 
generally consistent with  other $\alpha$-element abundances.
Mean values for [Mg/Fe] for the old clusters are in the range $+0.0$
to $+0.32$.  We do not see evidence for significantly depleted [Mg/Fe]
with respect to [$\alpha$/Fe] in the old clusters.

Over the full age range of the sample, [Mg/Fe] generally decreases
with decreasing cluster age.  The only exception to this trend is NGC
1718, for which we obtain a significantly lower value of [Mg/Fe]$=-0.9
\pm 0.30$.  The low Mg I in this cluster is interesting, 
although difficult to interpret.  We note that this cluster 
has one of lowest masses  in our sample. 
This may suggest that NGC 1718
formed in a poorly mixed environment in which 
high mass SNe II (M$>35 \msol$) had not contributed metals,
as the highest mass SNe are thought to produce Mg and O
most efficiently \citep{1995ApJS..101..181W}.
Our measurement appears robust because we
obtain very similar abundances using spectral synthesis for two Mg I
features which are separated in wavelength by $\sim$1000 \AA.  This is
shown in Figure \ref{fig:mg}, where we demonstrate that the best
fitting syntheses have [Mg/Fe]$\sim-1.0$.

For the young clusters in our sample, we are able to measure Mg I in
the cluster IL spectra of NGC 1866, and obtain [Mg/Fe]=$-0.27 \pm
0.2$.  We obtain [Mg/Fe]=$-0.03 \pm 0.12$ from the individual stars in
this cluster, which is consistent with the IL result, given the large
uncertainties.  From the individual stars in NGC 1711 and NGC 2100 we
obtain [Mg/Fe]=$+0.08 \pm 0.08$ and $-0.16 \pm 0.04$, respectively.
Previous measurements in individual stars (e.g. in NGC 1978)
are consistent with our results and 
are discussed further in \textsection \ref{sec:comparisons}.

O I is always difficult to measure in cluster stars, but is
particularly difficult in the IL because of the velocity dispersion of
the clusters (i.e. line broadening).  Our IL estimates for O I are
obtained by line synthesis of the 7771 \rAA triplet; the 6300 \rAA
forbidden line is consistently too weak.  For the old clusters, even
the 7771 \rAA triplet is quite weak and we only obtain limits on O of
[O/Fe]$<$+1.0 for NGC 2019, NGC 1916, and NGC 2005.  For NGC 1866
($\sim$100 Myrs) we measure [O/Fe]$=+0.2 \pm 0.3$. It is important to
note that the 7771 \rAA triplet can have significant non-LTE effects
in cool stars, which must be kept in mind when comparing to O
abundances from the 6300 \rAA forbidden line.  Unfortunately, there is
no obvious way to correct the 7771 \rAA IL abundances for non-LTE
effects, so we report the abundances with this caveat.

To these IL results, we can add several measurements from the stars in
young clusters.  The O I abundances for the individual stars were
measured from the 6300 \rAA forbidden line.  The stars in both the
intermediate age and young clusters have a constant value of
[O/Fe]$\sim+0.25$, similar to previous results for other LMC stars
\citepalias{pompeia08,mucc1866}.

In summary, 
we find evidence for star-to-star abundance variations in the old
clusters in our sample in the form of highly enriched [Na/Fe] and
[Al/Fe].  The [Mg/Fe] and [O/Fe] abundances are not indicative of
star-to-star abundance variations on their own, but are still
consistent with this picture.  We find that [Na/Fe], [Al/Fe] and
[Mg/Fe] generally decrease with decreasing cluster age.  In
\textsection \ref{sec:light-mass} we discuss the implications of our
measurements further, including the dependence of the IL abundance
with cluster mass.

\subsection{Fe-peak Elements}
\label{sec:fepeak}

Fe peak elements are well-studied in individual stars in the MW and
LMC.  In general, the abundances of Ni, Cr, Sc, V, and Co tend to
scale with Fe, so that the [X/Fe] ratios for these elements are
approximately solar for [Fe/H]$>-2.0$, or occasionally sub-solar in
the LMC \citepalias[e.g.][]{pompeia08}.  The abundance of Mn, however,
has a plateau value of [Mn/Fe]$\sim-0.4$ for [Fe/H]$< -1.0$, and then
increases to solar ratios between [Fe/H]=$-1.0$ and [Fe/H]=$+0$.

\begin{figure}
\centering
\includegraphics[scale=0.4]{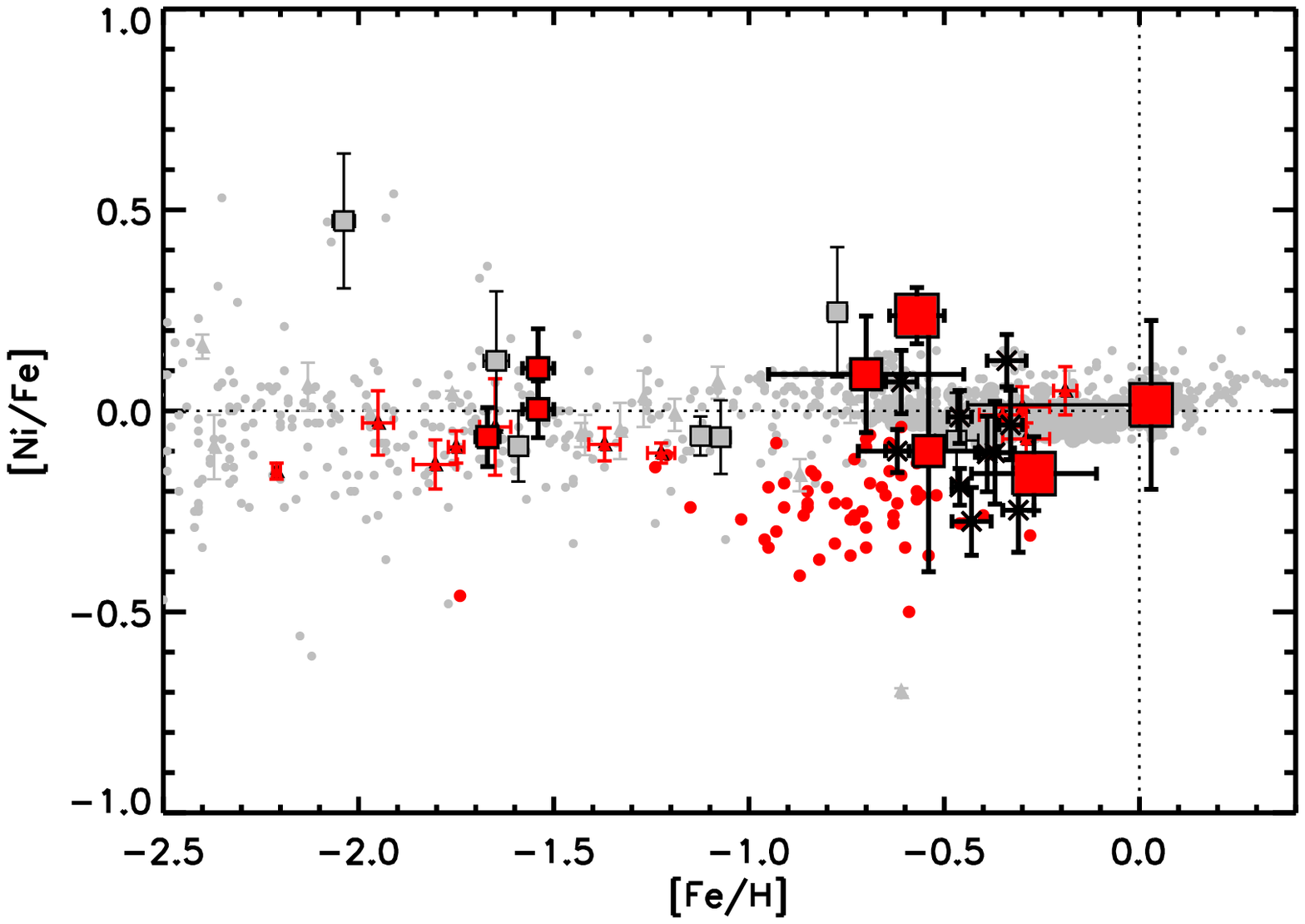}
\includegraphics[scale=0.4]{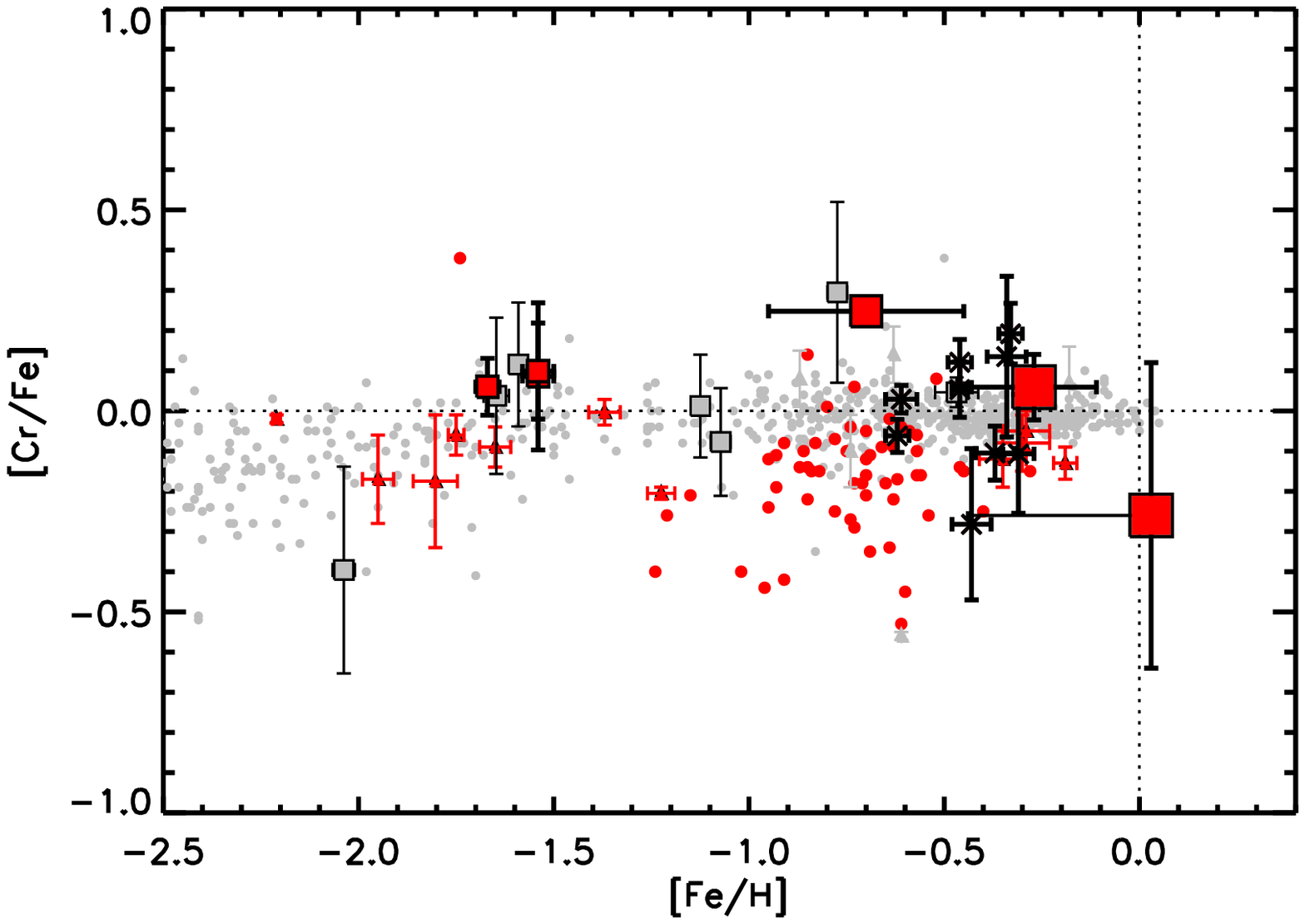}
\includegraphics[scale=0.4]{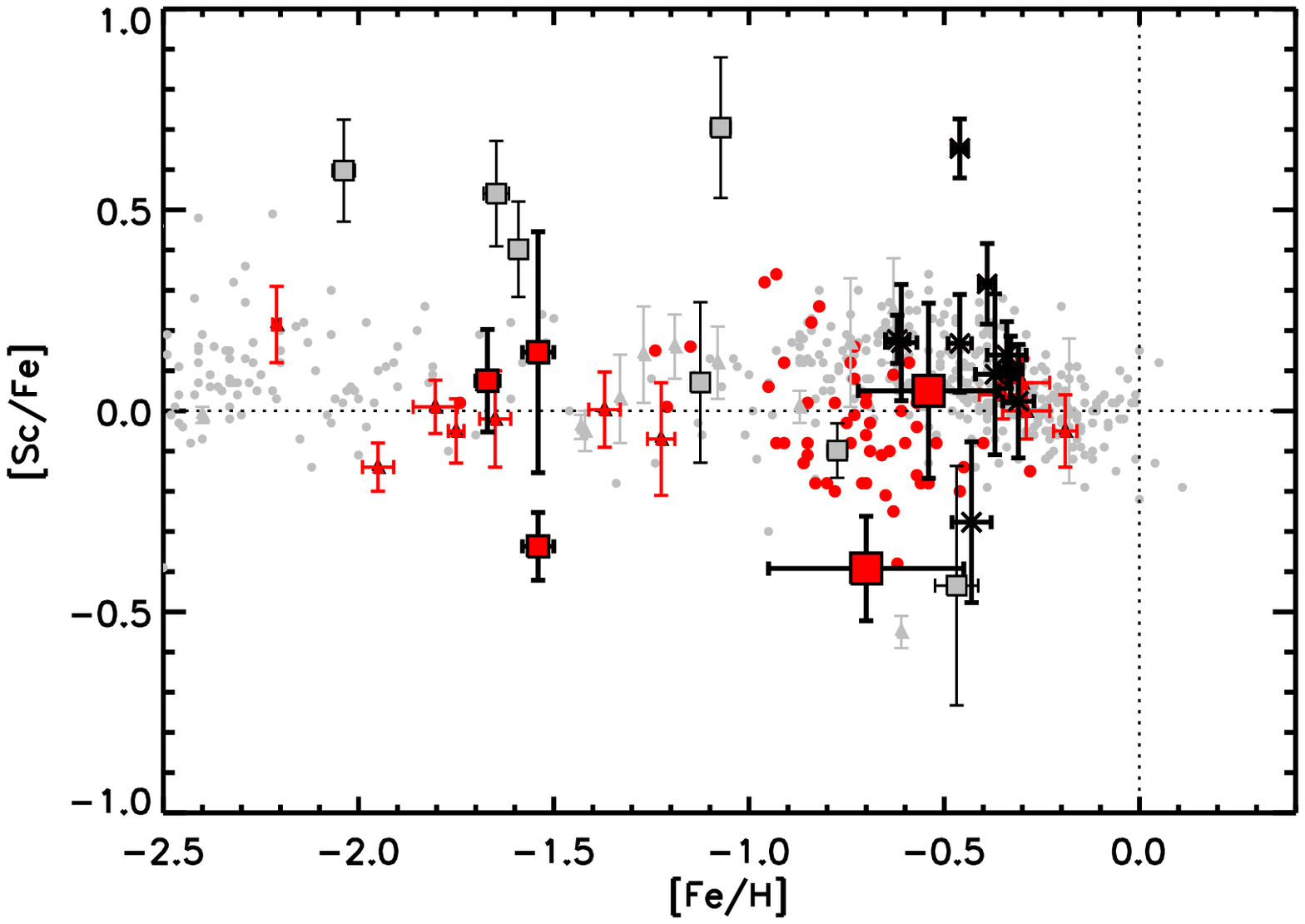}
\includegraphics[scale=0.4]{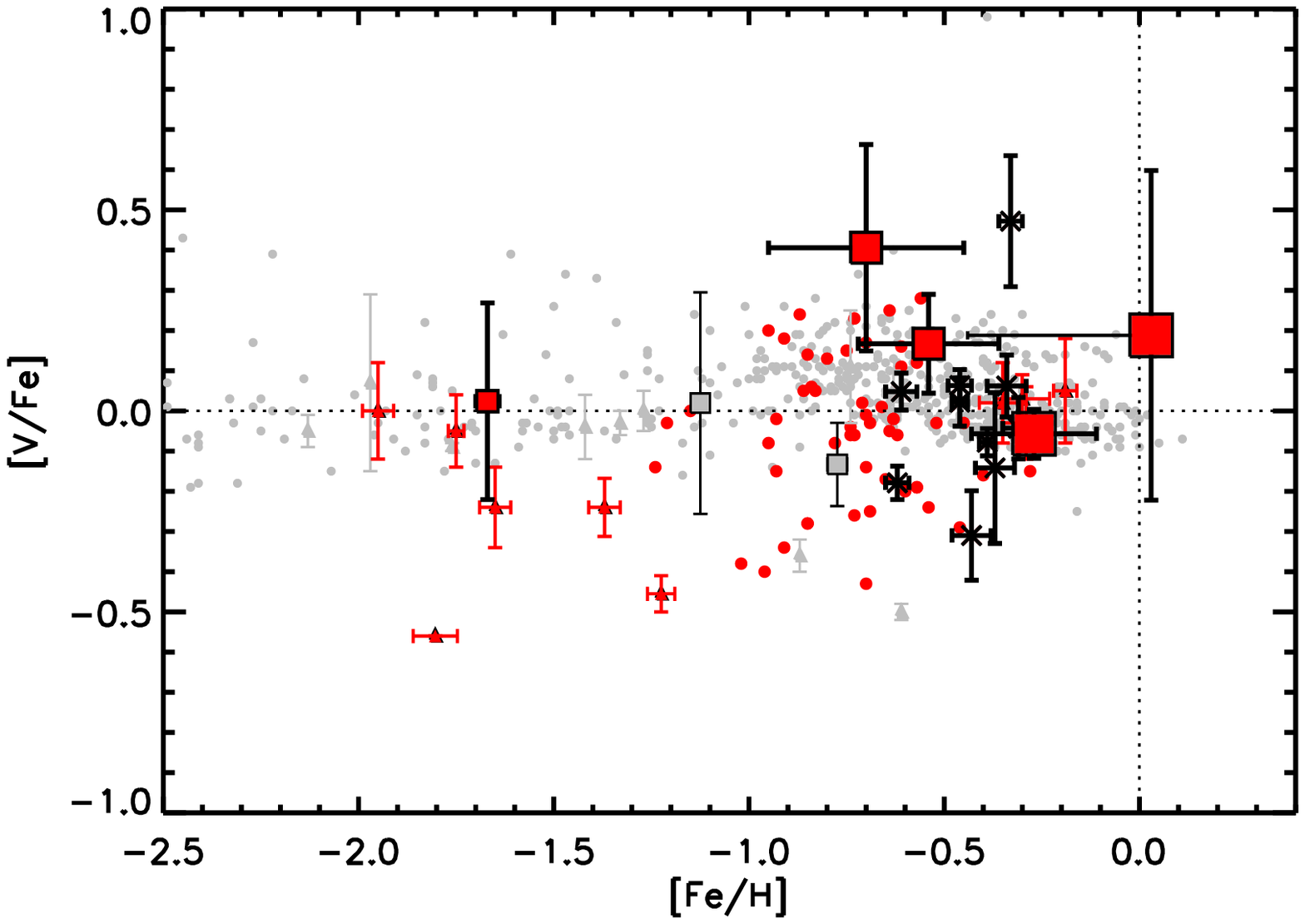}
\caption{Abundances for Fe-peak elements Ni, Cr, Sc and V.  Symbols are the same as in Figure \ref{fig:alpha}.  }
\label{fig:fepeak1}
\end{figure}

We measure Ni I, Cr I, Sc II and V I in most of our LMC sample.  We
find that the abundance ratios for these elements are consistent with
solar ratios across all of the ages and metallicities of the clusters,
as shown in Figure \ref{fig:fepeak1}.  The scatter between cluster IL
measurements is largest for Sc II, similar to what was found for IL
measurements in the MW and in clusters in M31 \citep{milkyway,
  m31paper}.

\begin{figure}
\centering
\includegraphics[scale=0.4]{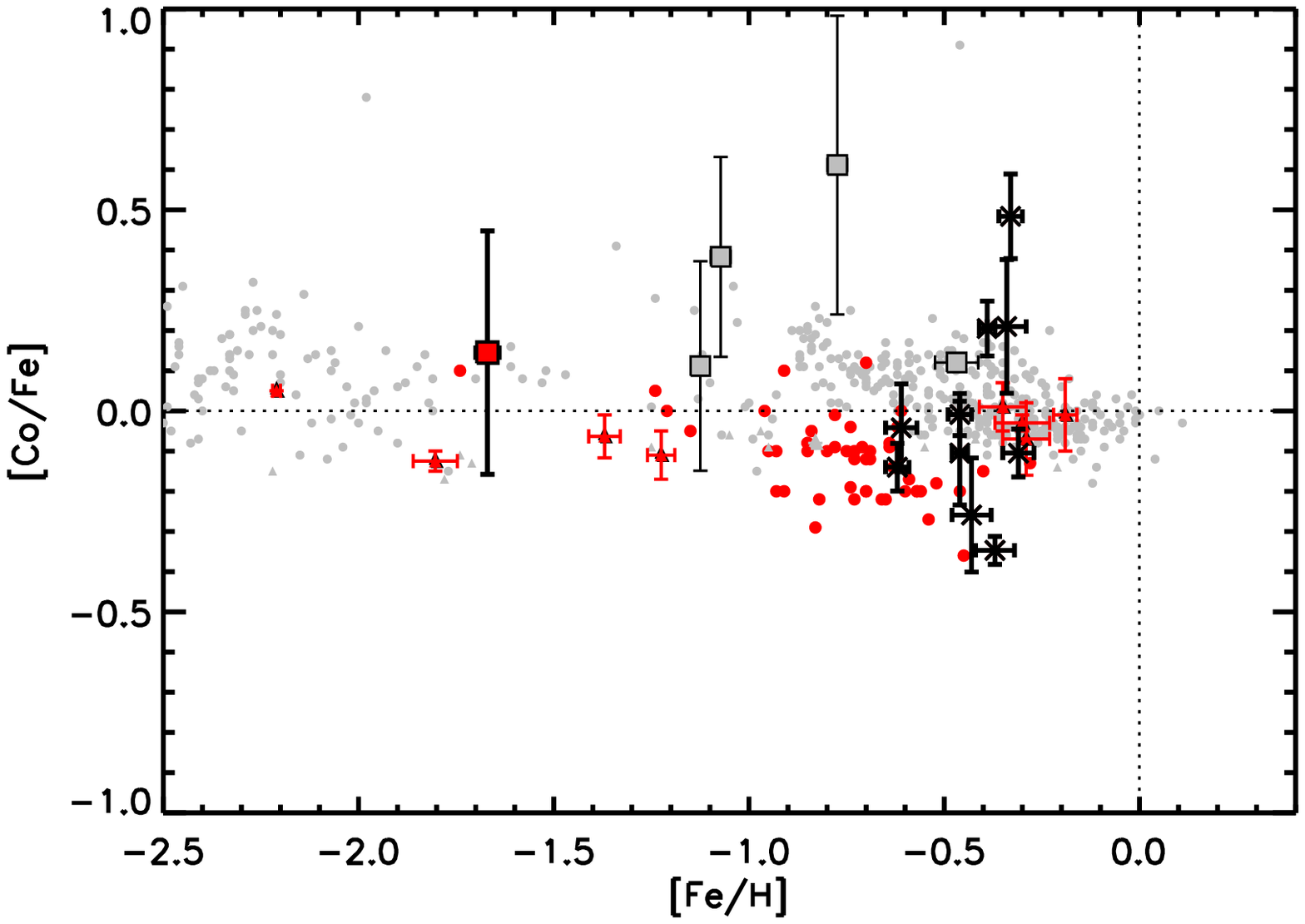}
\includegraphics[scale=0.4]{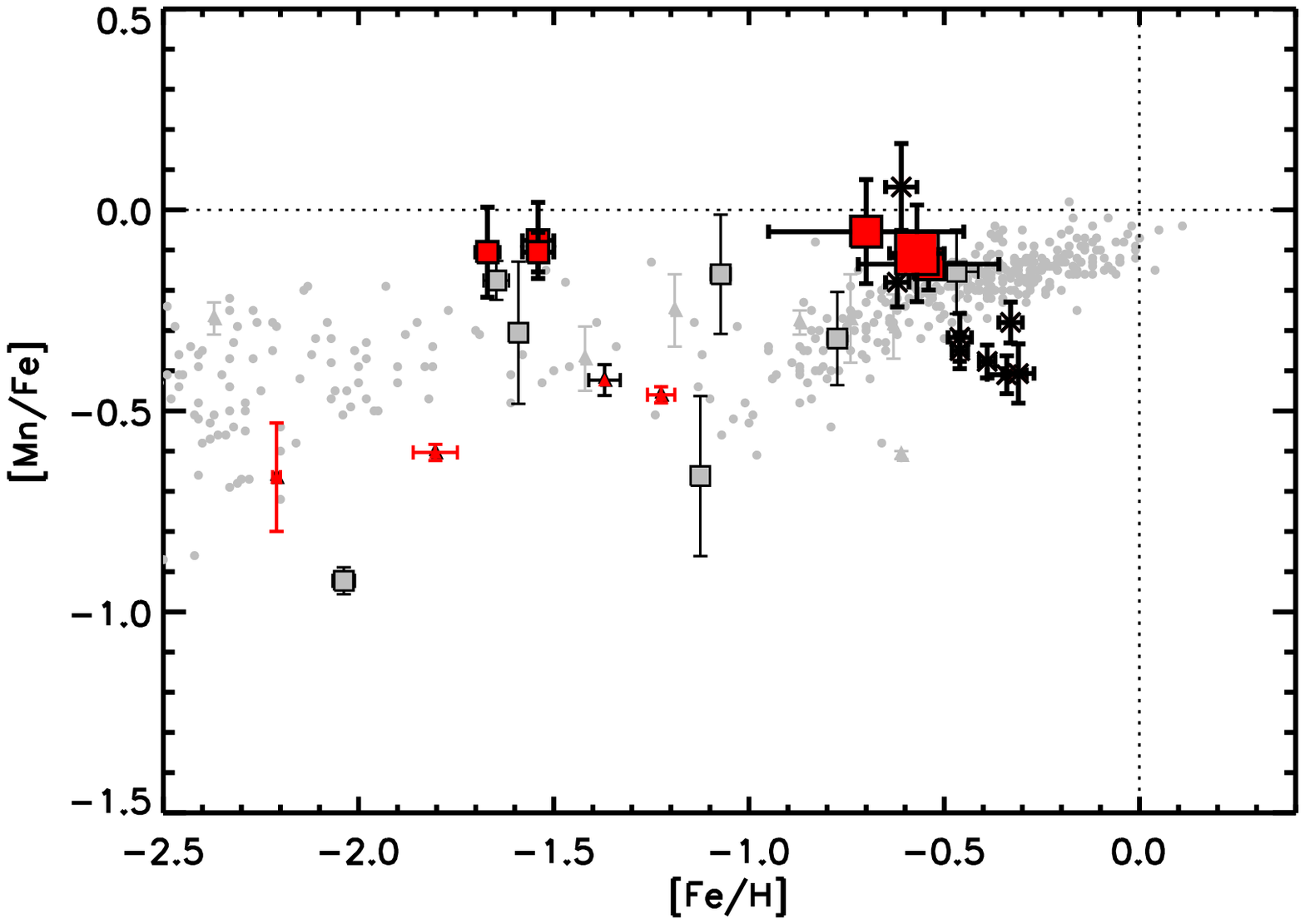}
\includegraphics[scale=0.4]{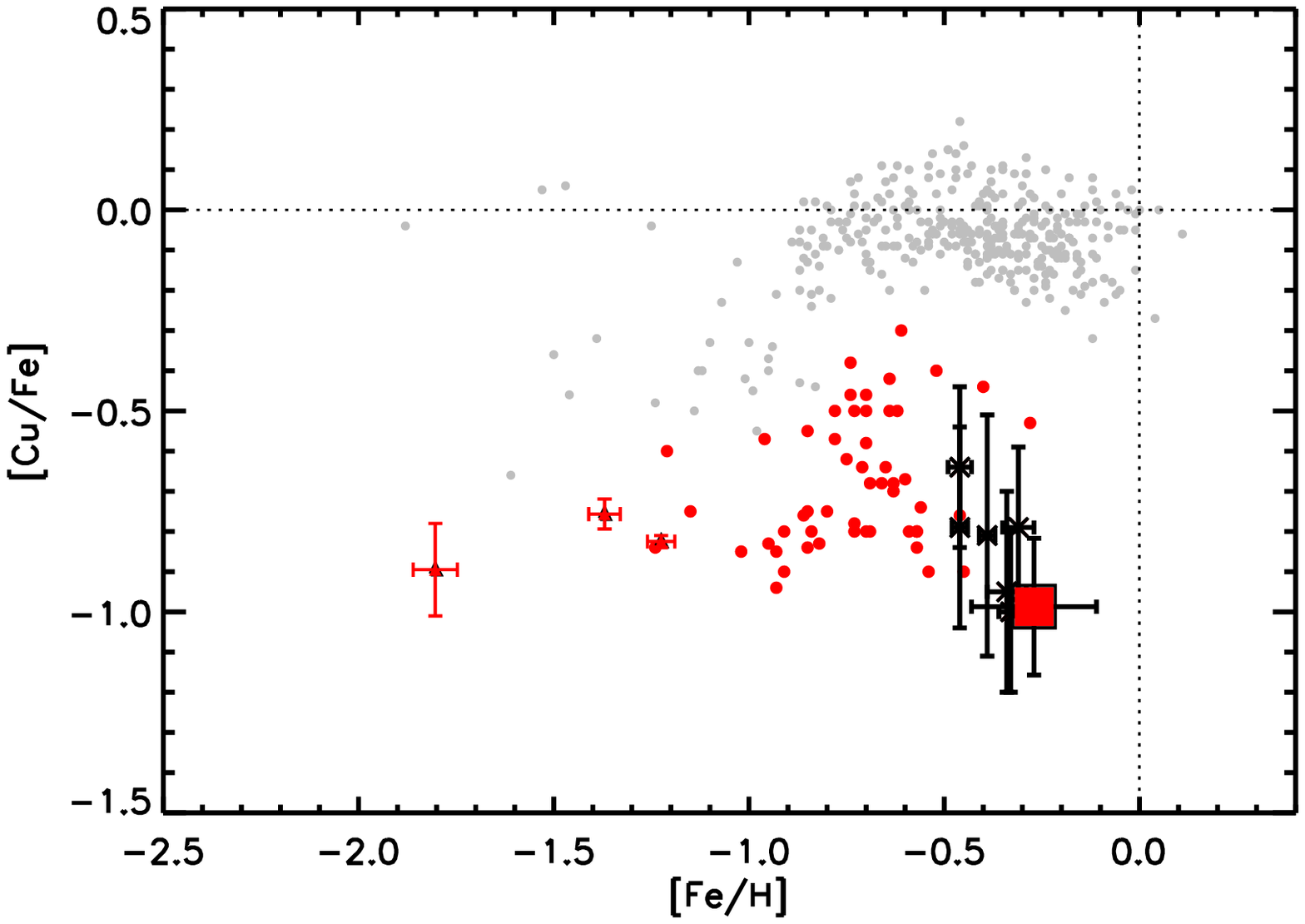}
\caption{Same as  Figure \ref{fig:fepeak1} for Co, Mn and Cu. Symbols are the same as in Figure \ref{fig:alpha}.  }
\label{fig:fepeak2}
\end{figure}

The Mn I abundances that we obtain for the LMC clusters are fairly
constant, with a mean value of [Mn/Fe]$\sim-0.1$ dex, as shown in
Figure \ref{fig:fepeak2}. The older, lower metallicity clusters do not
reach the MW low of [Mn/Fe]$\sim-0.4$.  However, they do overlap with
our MW IL measurements, so it is difficult to tell if the LMC ratios
differ systematically.
 
The [Mn/Fe] of the intermediate age and young clusters overlap with
the higher metallicity MW field stars of
\cite{2004AJ....128.1177V}. For the youngest clusters we have only
measured [Mn/Fe] in the individual stars, and find that it is offset
to lower ratios than MW field stars at similar metallicities (see
black crosses in Figure \ref{fig:fepeak2}).
We note that \cite{2003ApJ...592L..21M} and \cite{2007A&A...465..815S}
found a similar pattern of low [Mn/Fe] at high metallicity in the
Sagittarius dwarf galaxy. To explain the observations in Sagittarius,
 \cite{2003ApJ...592L..21M} and \cite{2008A&A...491..401C} proposed 
that both metallicity dependent Mn-production in  SNe II and SNe Ia 
as well as 
significant gas loss through galactic winds would be required.
The gas-loss required in this case could also 
 explain the reduced effective 
star formation rate between bursts that has been found for the
LMC.

In Figure \ref{fig:fepeak2} we also show the [Co/Fe] abundances. From
the cluster IL we can measure Co I in one cluster, NGC 2019.  The rest
of the Co I abundances that we measure are obtained from the
individual stars in the intermediate age and young clusters.  We find
[Co/Fe] to be approximately solar in our sample. The abundances of
\citetalias{pompeia08}, \citetalias{mucc08int}, and
\citetalias{johnson06} are consistent with our results.

The last element shown in Figure \ref{fig:fepeak2} is Cu. Our [Cu/Fe]
ratios are significantly sub-solar ($\sim -0.9$ dex) at all [Fe/H].
We are able to measure Cu I in the cluster IL spectra of NGC 1866.
All of the other measurements we obtain for Cu I are from the
individual stars in the intermediate age and young clusters. 
Previous measurements of [Cu/Fe] by \citetalias{johnson06} and
\citetalias{pompeia08} are also sub-solar at both high and low
metallicity, which is unlike the  [Cu/Fe]$\sim 0$ ratios
found in the MW for [Fe/H]$>-1$.
Cu production is thought to occur in SNe II, and is metallicity dependent
\citep[e.g.][]{1995ApJS..101..181W,
 2005NuPhA.758..284B}.   As the abundance
here is not seen to rise at any [Fe/H], it appears that SNe II with
higher [Fe/H] are not impacting  the overall metallicity of the LMC gas
over time.  This could be due to inflows of metal-poor gas, outflows
of metal-rich gas, or low star formation rates over time with few high
mass stars.
Low [Cu/Fe] was also measured for stars in the Sagittarius dwarf
galaxy by \cite{2005ApJ...622L..29M}, where it was argued that this
implied significant gas-loss after early generations of star formation.

In summary, most of the Fe-peak elements generally follow the abundance trends
that have been previously measured in the LMC.  We find depleted
[Cu/Fe] and [Mn/Fe] at high metallicity, which may reflect the lower
star formation efficiency of the LMC and metallicity-dependent SNe
yields.

\subsection{Neutron Capture Elements}
\label{sec:ncapture}

\begin{figure*}
\centering
\includegraphics[scale=0.4]{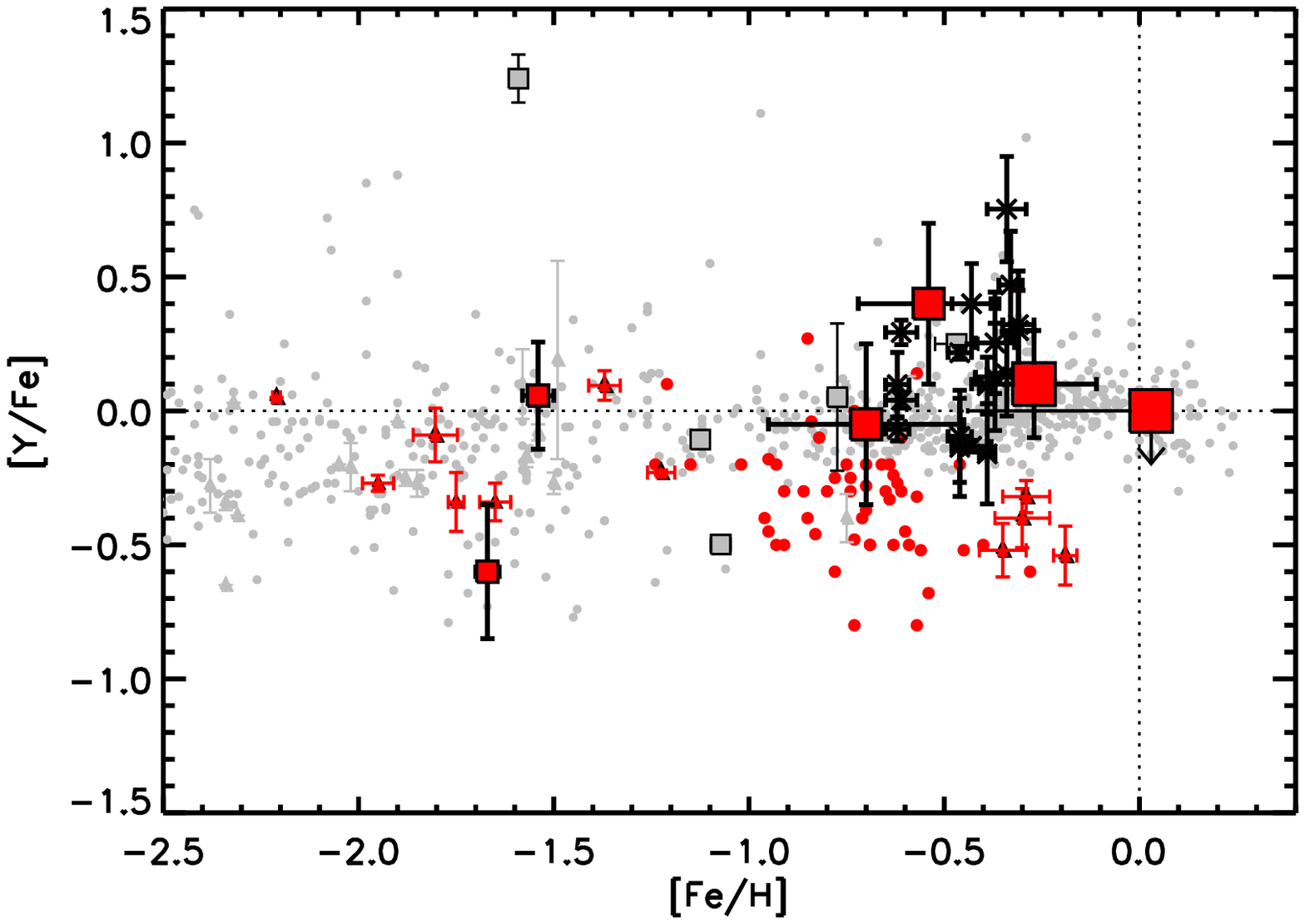}
\includegraphics[scale=0.4]{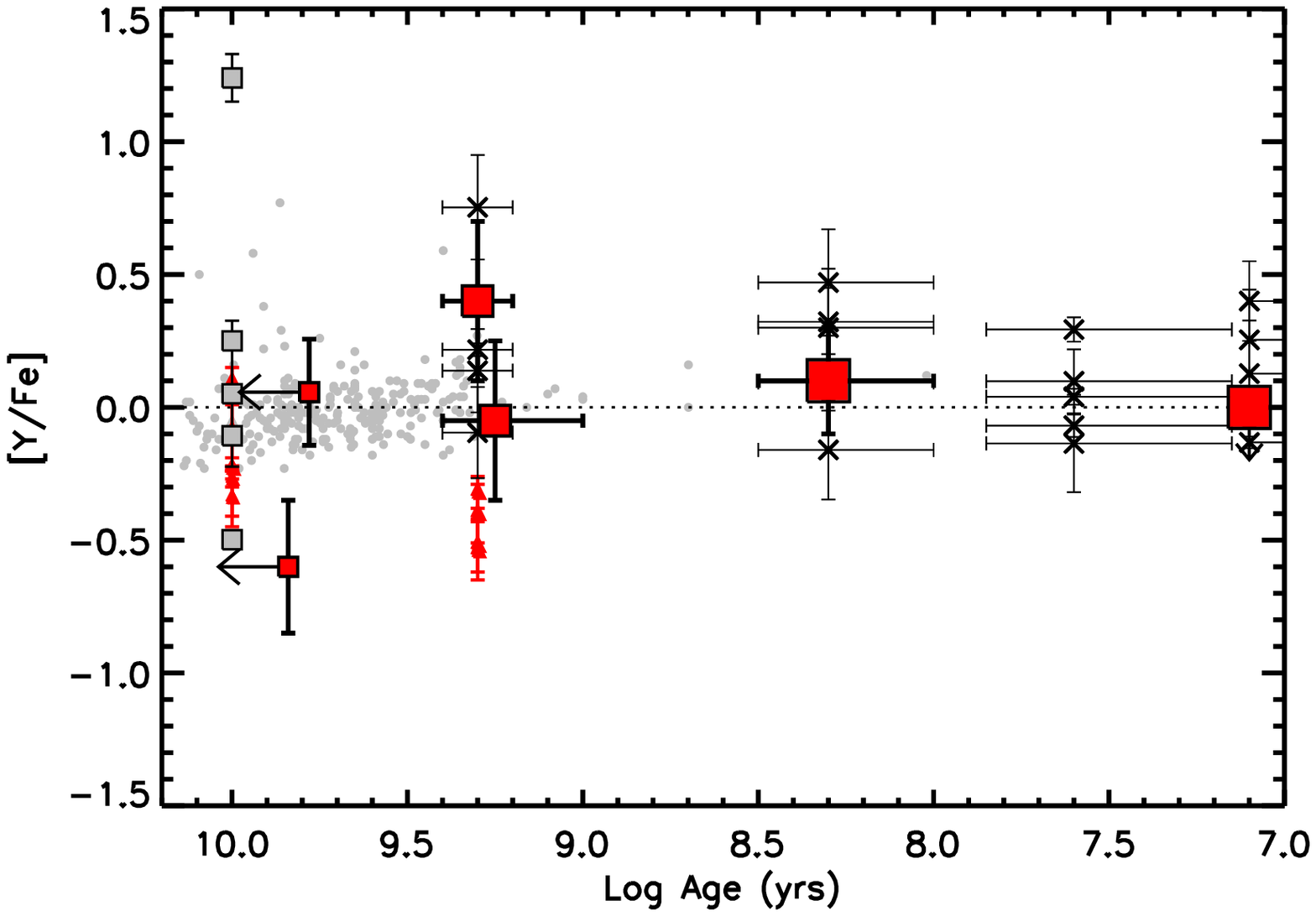}
\caption{Same as Figure \ref{fig:alpha} for the neutron-capture element Y.  For consistency with the Y II IL measurements, only the Y II stellar measurements are shown. Symbols are the same as in Figure \ref{fig:alpha}.   }
\label{fig:ncapture1}
\end{figure*}

\begin{figure}
\centering
\includegraphics[scale=0.4]{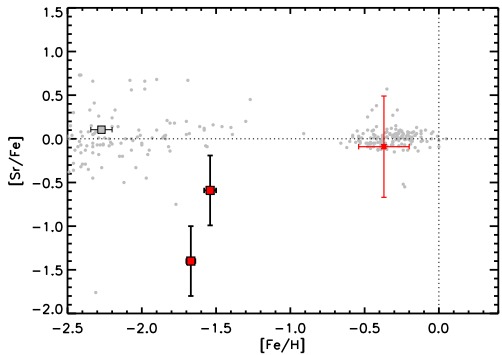}
\caption{Abundances for the neutron-capture element Sr. The small red cross shows the mean [Sr/Fe] of LMC F-type supergiants from \citetalias{1989ApJS...70..865R}.  Other symbols are the same as in Figure \ref{fig:alpha}.}
\label{fig:ncapture1b}
\end{figure}

\begin{figure*}
\centering
\includegraphics[scale=0.4]{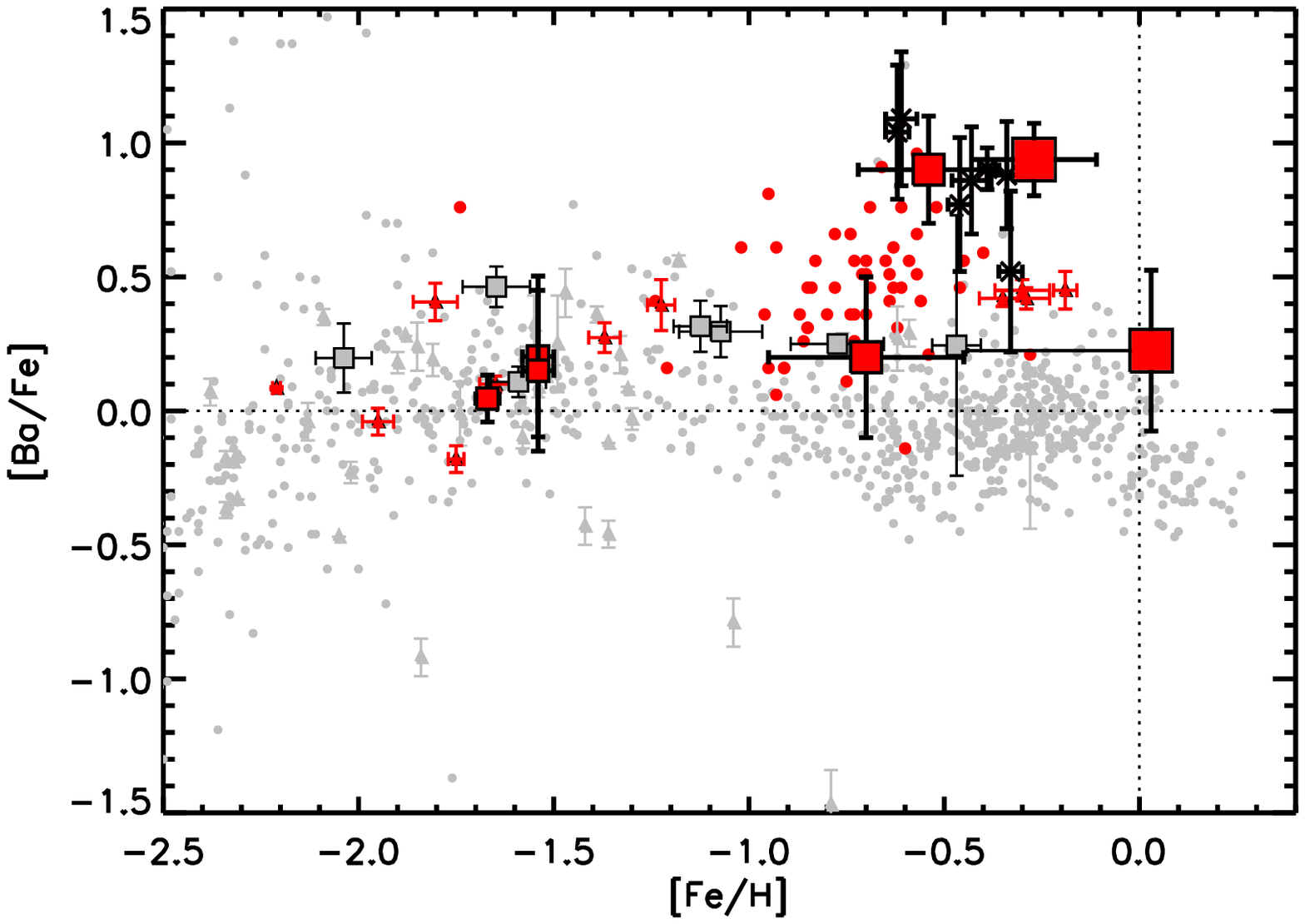}
\includegraphics[scale=0.4]{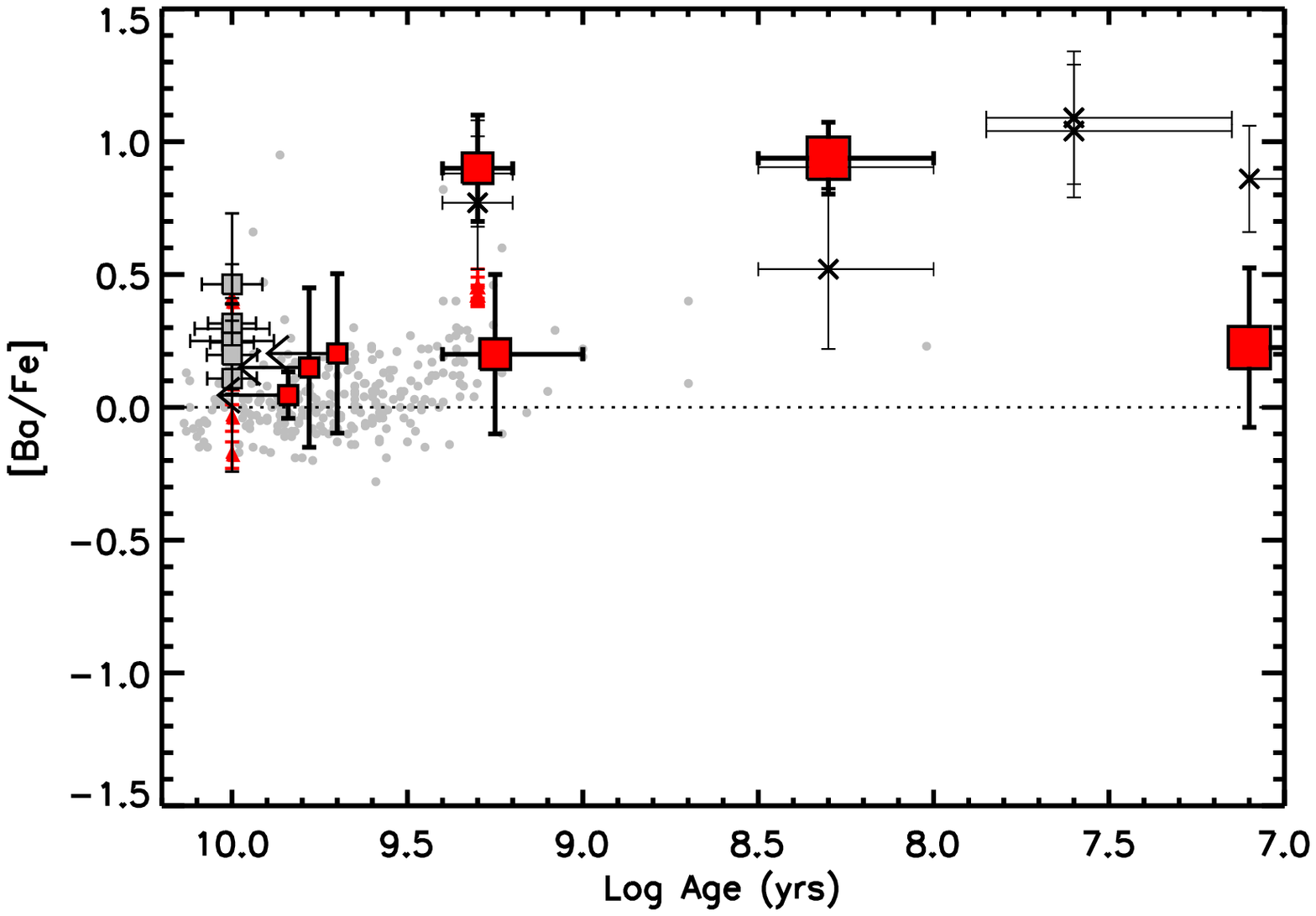}
\caption{Same as Figure \ref{fig:alpha} for the neutron-capture element Ba. Symbols are the same as in Figure \ref{fig:alpha}. }%
\label{fig:ncapture1c}
\end{figure*}

\begin{figure*}
\centering
\includegraphics[scale=0.4]{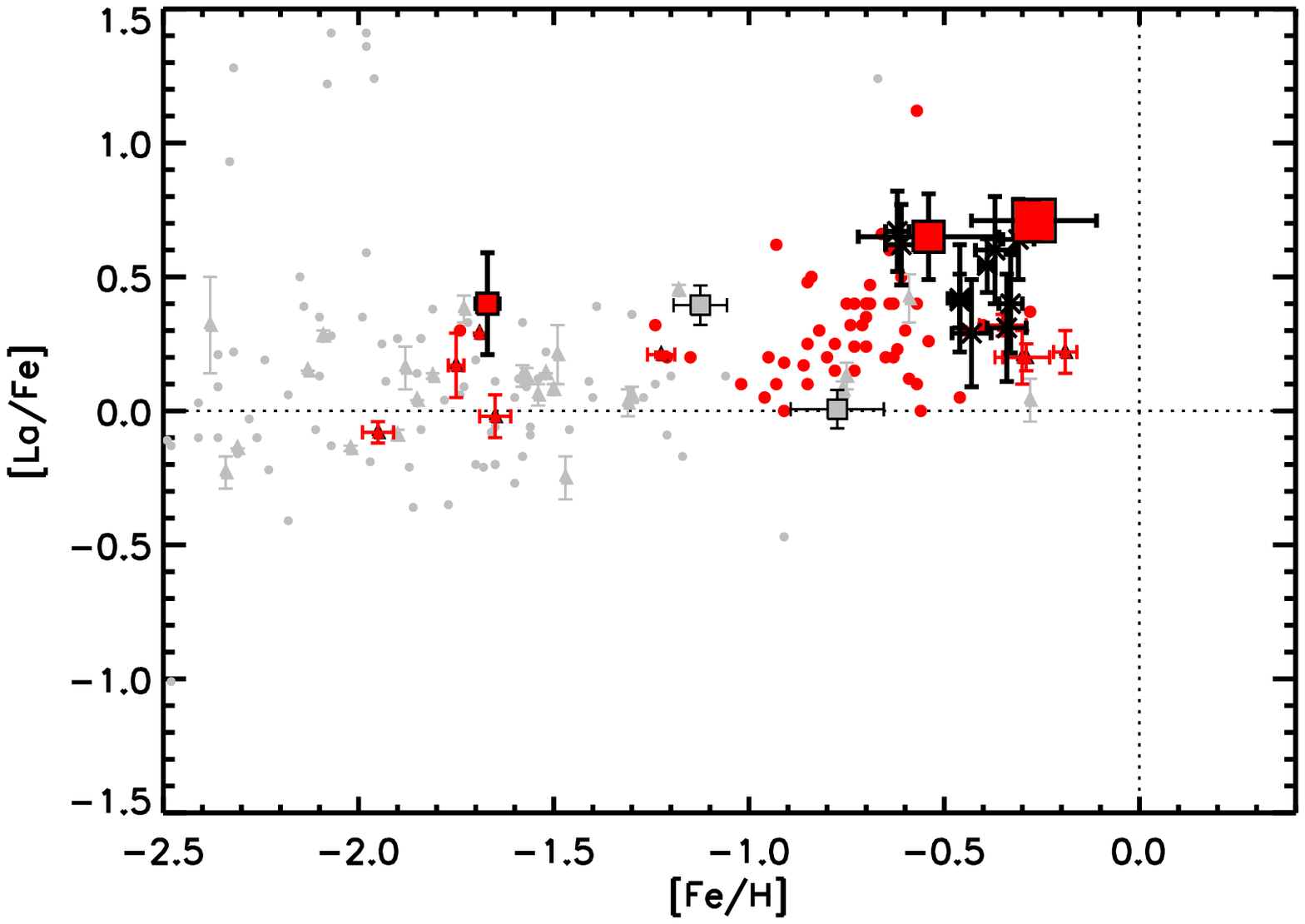}
\includegraphics[scale=0.4]{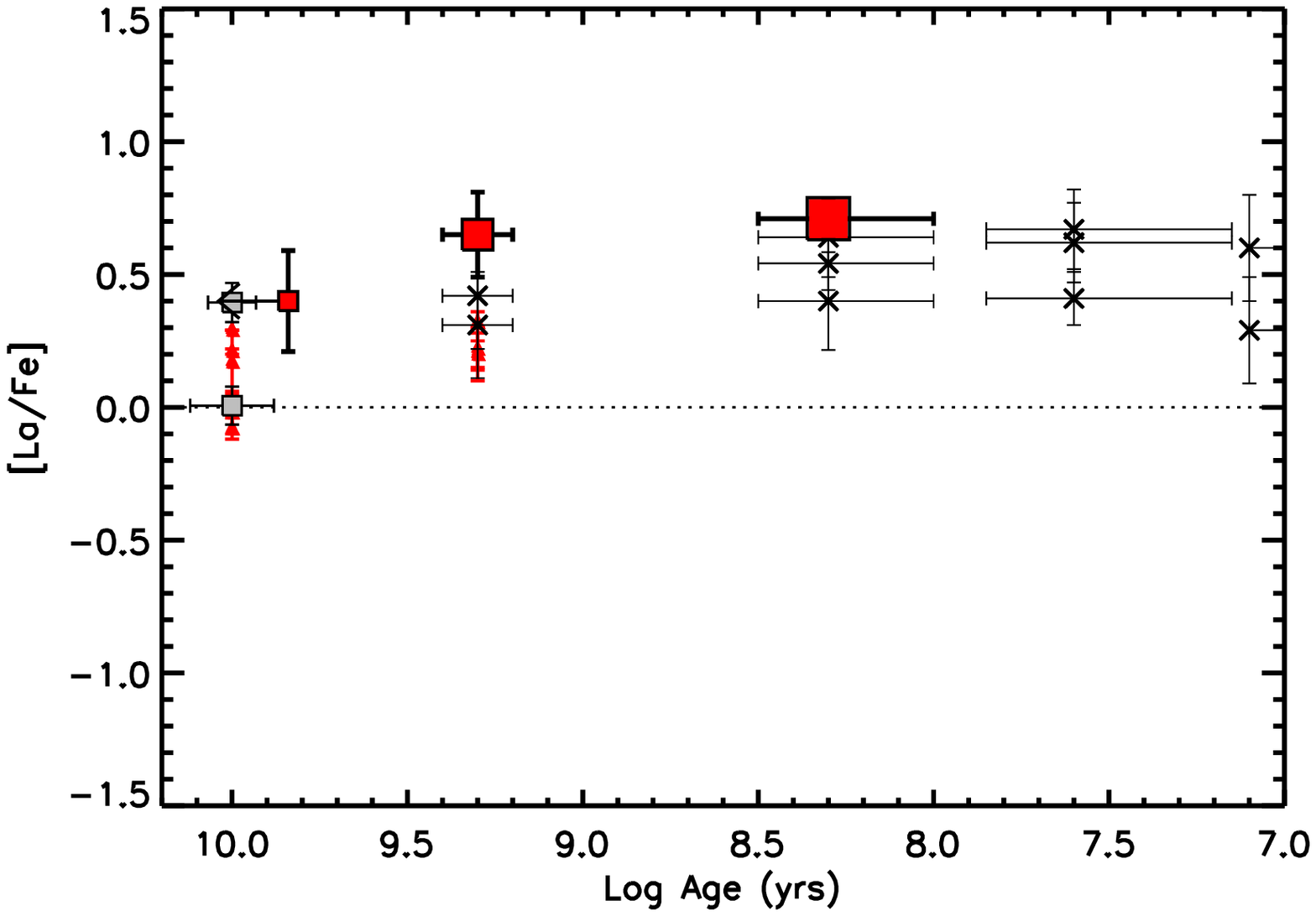}
\includegraphics[scale=0.4]{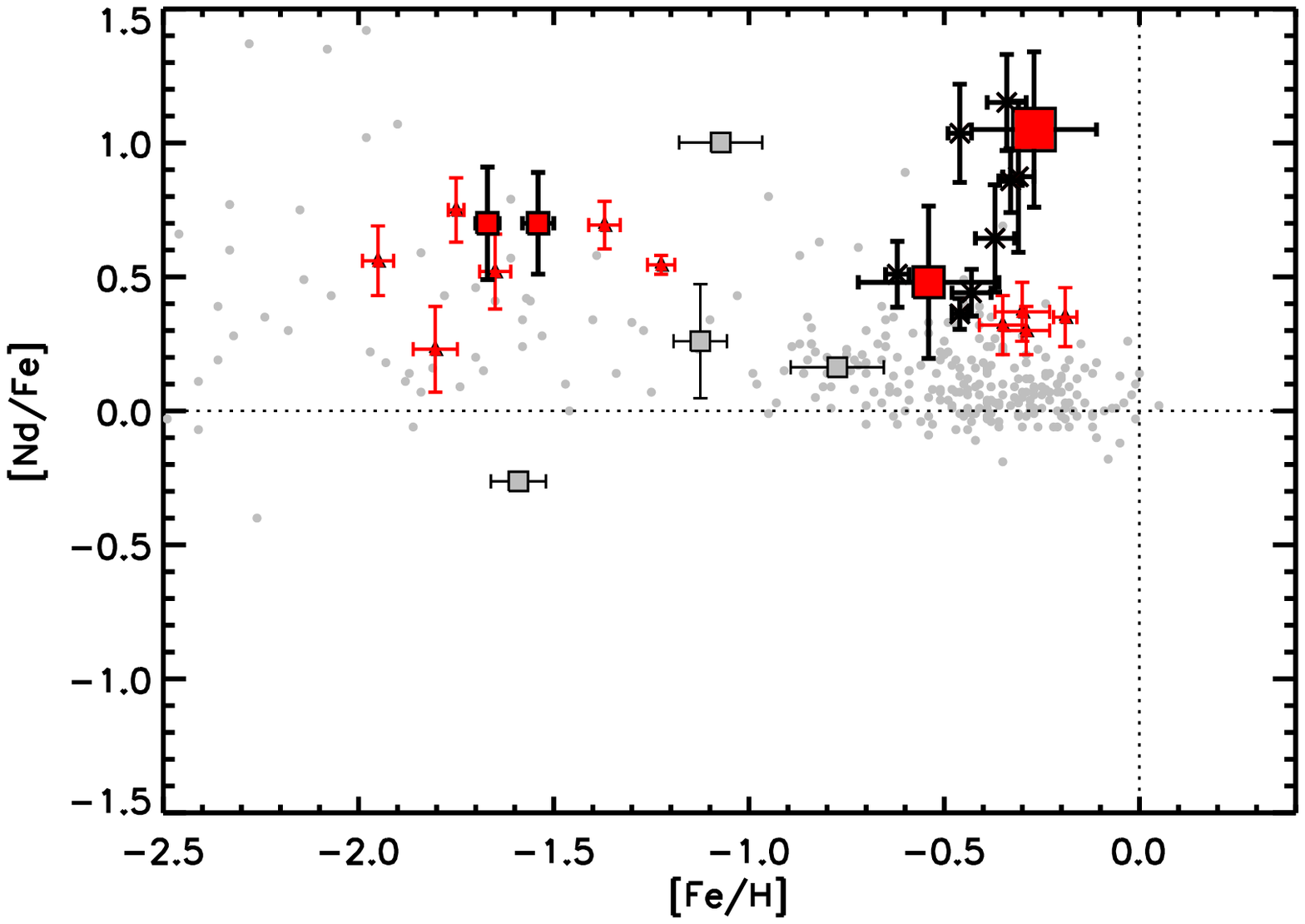}
 \includegraphics[scale=0.4]{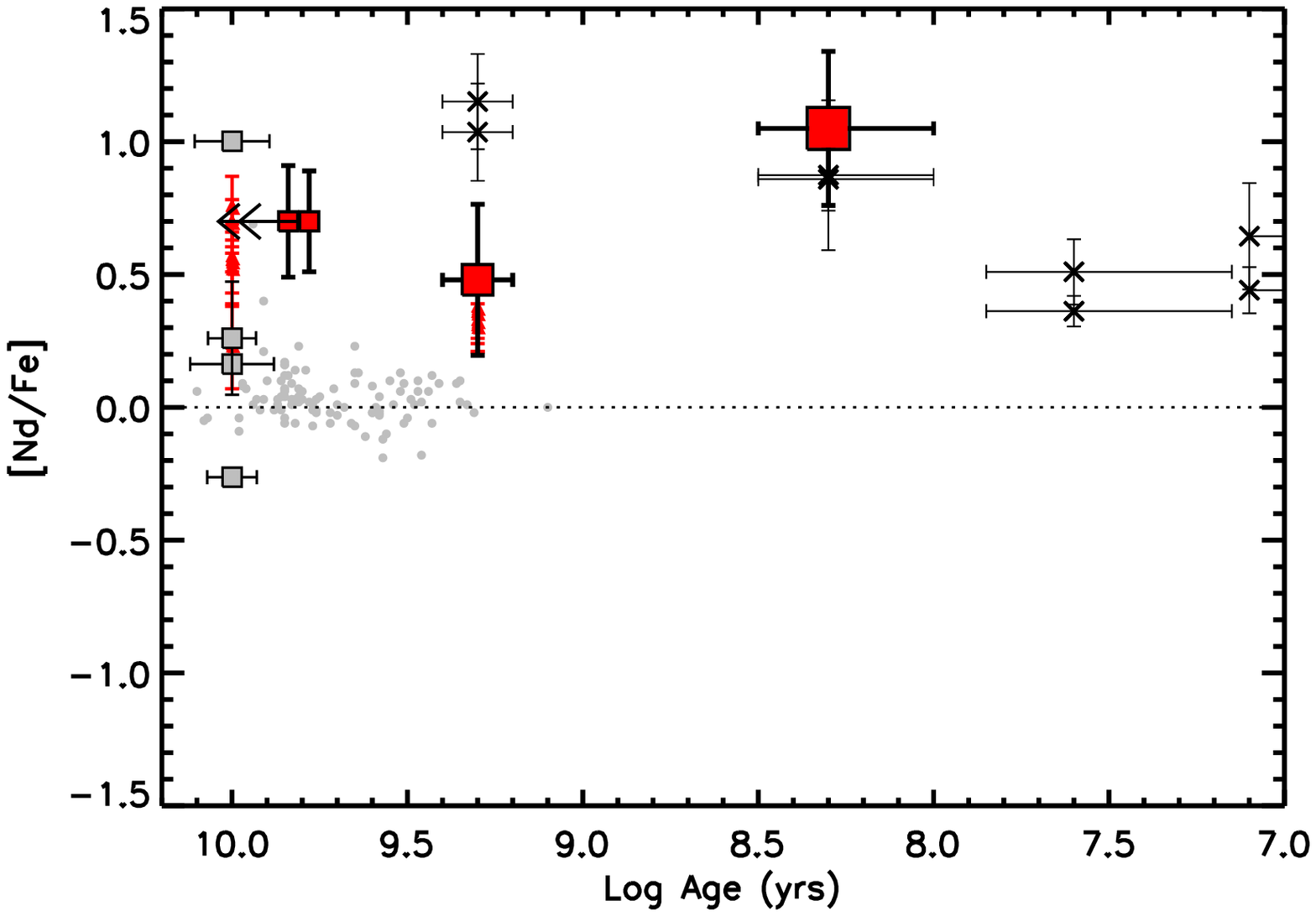}
\caption{Same as Figure \ref{fig:alpha} for the neutron-capture elements  La and Nd. Symbols are the same as in Figure \ref{fig:alpha}. }%
\label{fig:ncapture2}
\end{figure*}

\begin{figure*}
\centering
\includegraphics[scale=0.4]{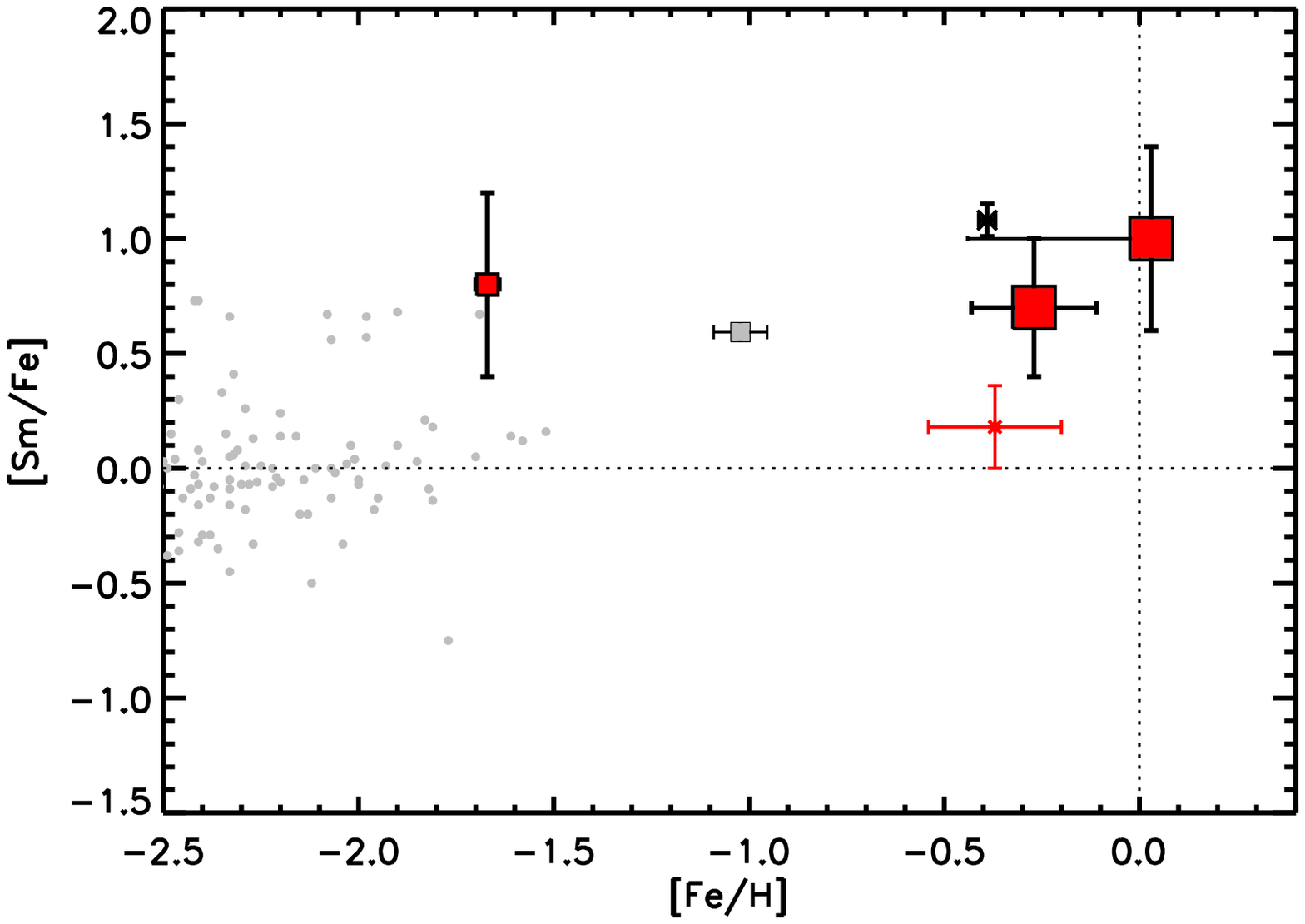}
\includegraphics[scale=0.4]{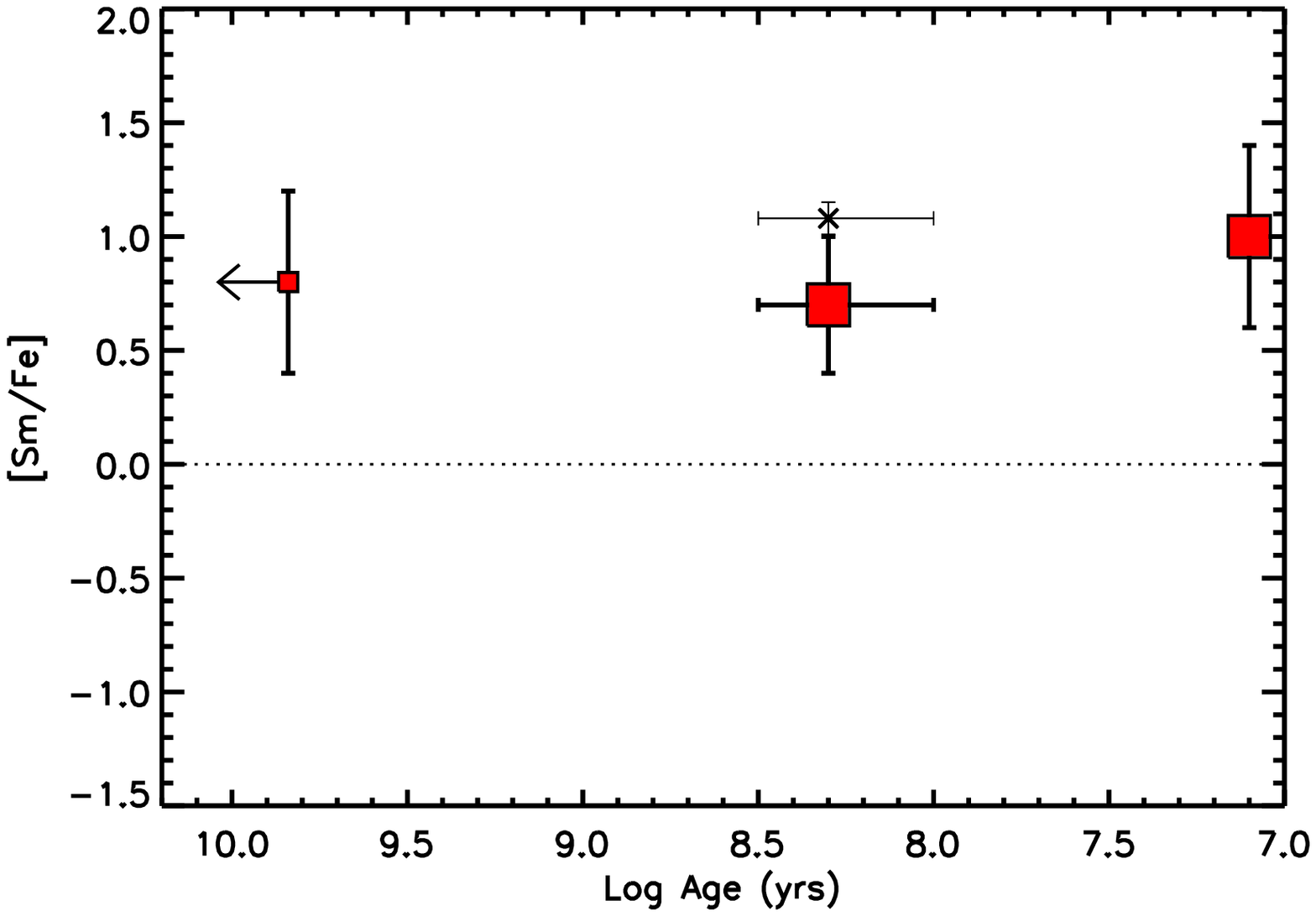}
\includegraphics[scale=0.4]{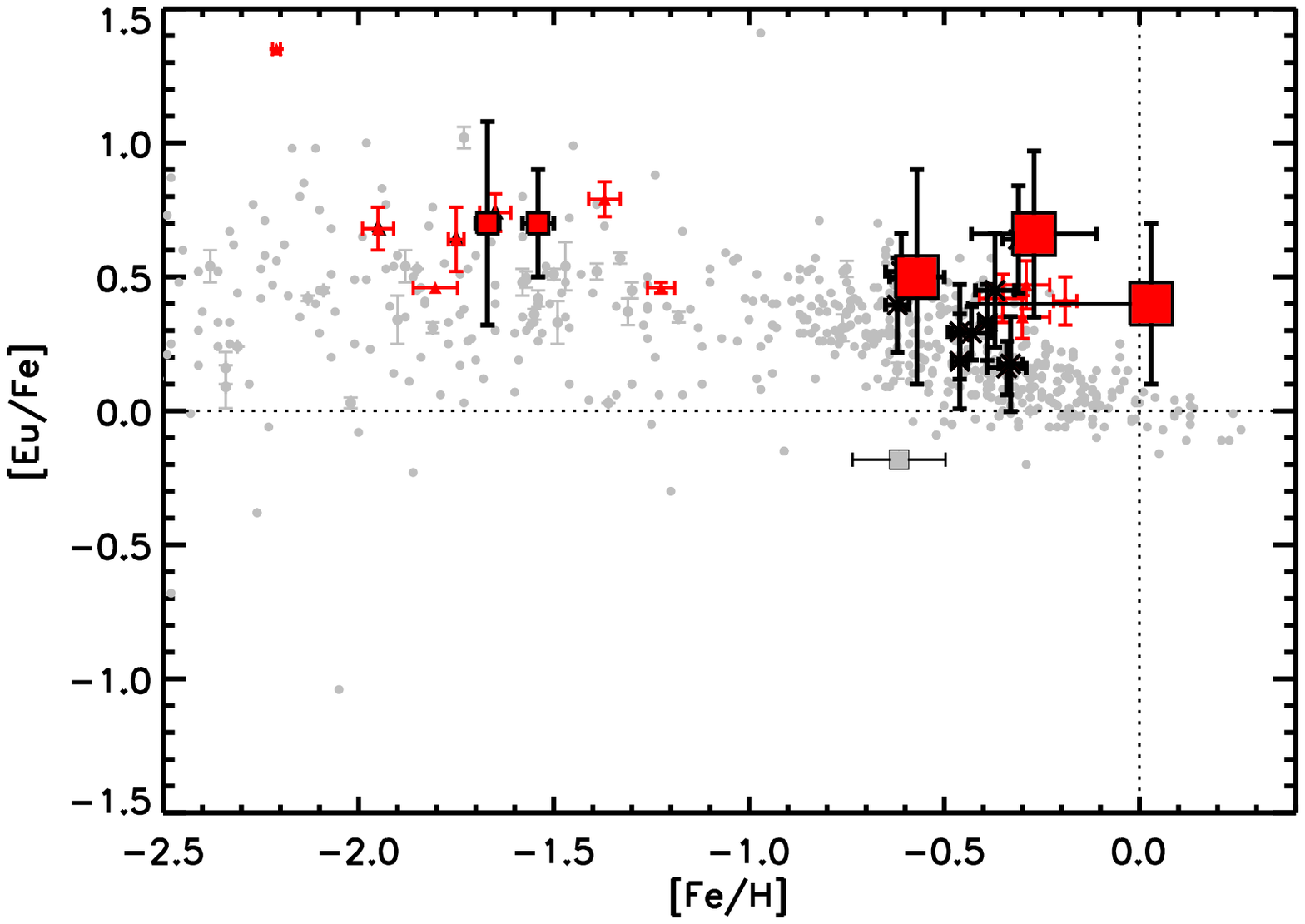}
\includegraphics[scale=0.4]{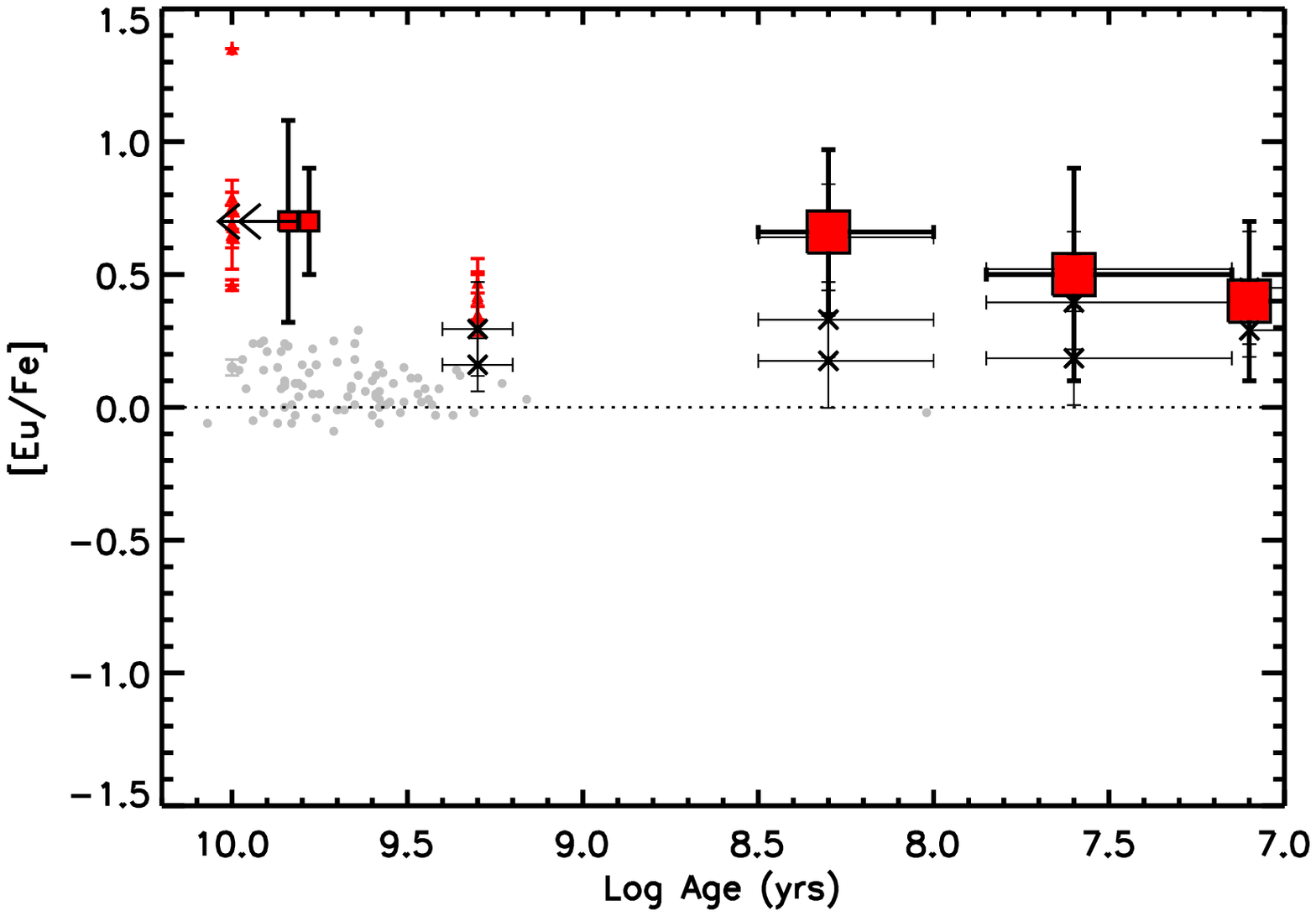}
\caption{Same as Figure \ref{fig:alpha} for the neutron-capture elements Sm and  Eu. Small red cross shows the mean [Sm/Fe] of LMC F-type supergiants from \citetalias{1989ApJS...70..865R}. Other symbols are the same as in Figure \ref{fig:alpha}.   }%
\label{fig:ncapture3}
\end{figure*}

The abundance ratios of neutron capture elements in different
environments are particularly useful for constraining chemical
evolution models, especially the contribution of AGB stars to the
interstellar medium \citep{2008ARA&A..46..241S}.  These elements have
been observed to be critically sensitive to the star formation history
of a galaxy and, like $\alpha$-elements, show different patterns in
dwarf galaxies than in the MW.

Many neutron capture elements have weak features in individual stellar
spectra, and so are particularly difficult to measure in cluster IL.
With EWs, we have been able to measure Ba II in most of the LMC
clusters but can only measure Y II, Eu II, La II, Nd II, Sm II, and Sr
II in 1$-$3 clusters each.  Therefore, the line synthesis component of
ILABUNDs was used to measure or put upper limits on these elements for
most of the sample.

There is some scatter in abundance ratios for the light {\it
  s}-process elements [Y/Fe] and [Sr/Fe] as shown in
Figure~\ref{fig:ncapture1} and \ref{fig:ncapture1b}.  Sr II was
measured in two old clusters; to our knowledge, these are the first Sr
II measurements in LMC clusters to date.  [Sr/Fe] is sub-solar in both
cases, and lower than both the IL MW measurement and the mean [Sr/Fe]
from LMC F-type supergiants \citepalias{1989ApJS...70..865R}.

From cluster IL, Y II measurements or upper limits were made for 6 of
the 8 clusters.  From the individual stars, we are able to measure
both Y I and Y II in the intermediate age and young clusters.  We find
some scatter in [Y/Fe] between clusters, and potentially some
star-to-star scatter within some of the younger clusters, although our
stellar samples in each cluster are small. In general, at high
metallicity and young ages the results are consistent with solar
ratios. We find similar [Y/Fe] in the LMC clusters as we did for the
MW clusters in \citetalias{milkyway}.
 
In Figure \ref{fig:ncapture1c}, we show our measurements of [Ba/Fe].
[Ba/Fe] is always super-solar in our sample, and the mean value is
higher for the young and intermediate age clusters than it is for the
old clusters.  The [Ba/Fe] obtained from individual stars in younger
clusters are also high: [Ba/Fe]$\sim+0.9$ dex.  At low metallicity and
older ages, the [Ba/Fe] we obtain is consistent with previous stellar
measurements in both the MW and the LMC.  At high metallicity and
younger ages, our measurements are similar to the high [Ba/Fe]
measured previously in the LMC
\citepalias{pompeia08,mucc08int,mucc1866}.  High values of [Ba/Fe]
suggest that intermediate mass AGB stars had a significant effect on
the chemical evolution of the LMC, which we discuss further in
\textsection~\ref{sec:gcheavy}.

We measure La II from cluster IL for NGC 2019, NGC 1978, and NGC 1866,
as shown in Figure~\ref{fig:ncapture2}. The rest of our La II
measurements are obtained from the individual stars in the younger
clusters.  We find [La/Fe] to be approximately constant for all of the
ages and metallicities in our sample, with a mean value of
[La/Fe]$\sim+0.5$.  The measurements of [La/Fe] of
\citetalias{pompeia08} and \citetalias{mucc08int} are consistent with
our results.

As shown in Figure \ref{fig:ncapture2}, we find more scatter in
[Nd/Fe] than we do for [La/Fe], but the pattern of high ratios for all
ages and metallicities is generally similar.  From the individual
stars in the youngest clusters, NGC 1711 and NGC 2100, there is some
indication that the [Nd/Fe] value decreased at late times in the
LMC. High values of [Nd/Fe] were also found in LMC clusters by
\citetalias{johnson06,mucc08int, mucc10old}, and in LMC field stars by
\citetalias{HILL95}.  Nd II is obtained by line synthesis of the IL of
four clusters, and by EW analysis for the individual stars.

Figure \ref{fig:ncapture3} shows some of the first Sm II measurements
in the LMC, particularly for old stars.  Although the sample is small,
like [La/Fe] and [Nd/Fe], [Sm/Fe] is relatively constant over the full
age and metallicity range. There is little [Sm/Fe] abundance
information available for comparison, but
\citetalias{1992ApJS...79..303L} and \citetalias{1989ApJS...70..865R}
also found Sm to be overabundant in LMC supergiants by $\sim+0.4$ dex
when compared to MW supergiants. Here we find evidence that this
overabundance is also present at low metallicities and old ages.

In Figure \ref{fig:ncapture3} we show our [Eu/Fe] abundances.  We find
[Eu/Fe] to be consistently enhanced ($\sim$0.5 dex) over solar ratios
for all of the clusters.  Eu II abundances are obtained from line
synthesis in both the cluster IL and the individual stars.  These
measurements are similar to what has been found for stars in other LMC
clusters \citepalias{johnson06,mucc08int,mucc10old}.
\citetalias{1992ApJS...79..303L} also find enhanced [Eu/Fe] with
respect to solar for very young LMC field stars, which is in contrast
to the declining or solar [Eu/Fe] found in MW stars of similar ages
and metallicity.

In \textsection~\ref{sec:gcheavy} we discuss further implications of
the LMC cluster neutron-capture abundances.  We have demonstrated with
the LMC cluster sample that many neutron-capture elements can be
analyzed in high resolution IL spectra, and that this analysis method
holds promise for strong constraints on these elements for distant,
unresolved clusters.

\section{Accuracy of LMC Analysis: Comparison to Stellar Results from the Literature}
\label{sec:comparisons}

\begin{figure}
\centering
\includegraphics[trim = 0mm 100mm 0mm 0mm, clip,angle=90,scale=0.50]{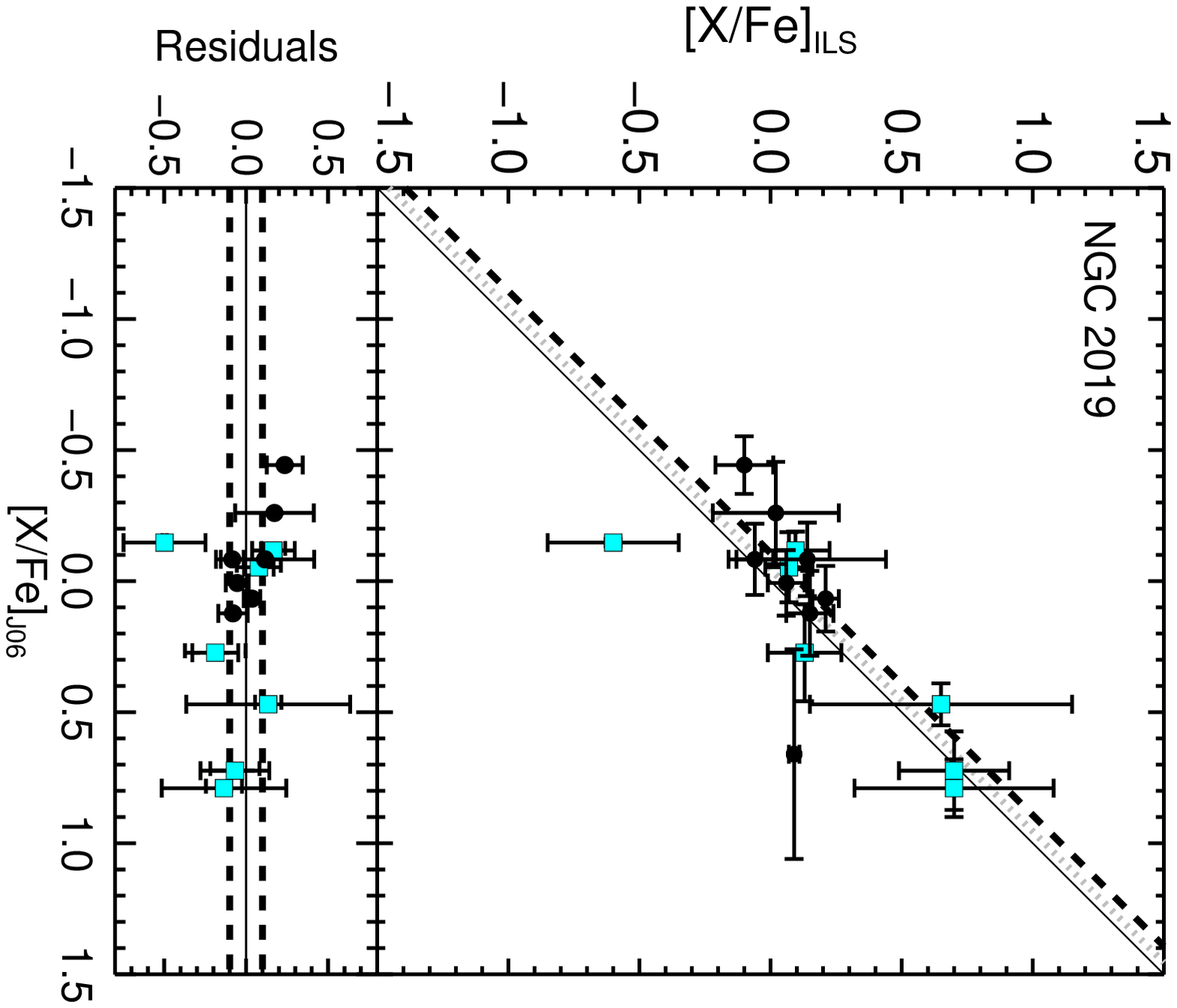}
\caption{Comparison of  abundance ratios from IL and  stellar analysis for NGC 2019.  Stellar abundances are from \citetalias{johnson06}.   Black and cyan points show abundances for neutral and ionized species, respectively.  Solid line shows the 1:1 line where points would lie if there were perfect agreement between IL and stellar results.  Dashed  and dotted lines  show  linear fits to the neutral and ionized species, respectively, with the slopes constrained to unity.  Bottom panel shows the residuals from the linear fit. The dashed lines correspond to residuals of $\pm 0.1$ dex to guide the eye.}
\label{fig:2019comparison}
\end{figure}

\begin{figure}
\centering
\includegraphics[trim = 0mm 100mm 0mm 0mm, clip,angle=90,scale=0.50]{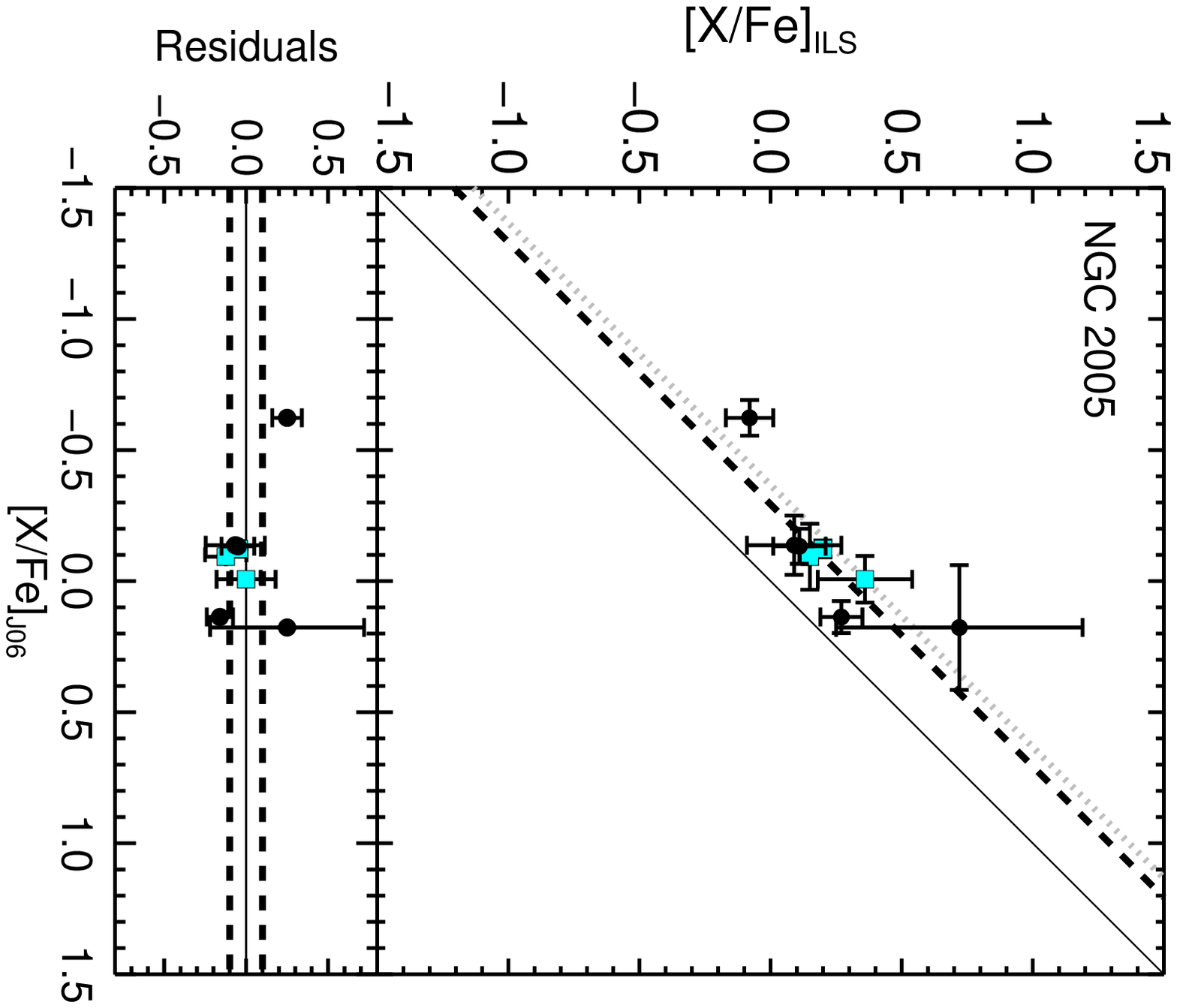}
\caption{Same as Figure \ref{fig:2019comparison} for  NGC 2005.  Stellar abundances are from \citetalias{johnson06}.   }
\label{fig:2005comparison}
\end{figure}

\begin{figure}
\centering
\includegraphics[trim = 0mm 100mm 0mm 0mm, clip,angle=90,scale=0.50]{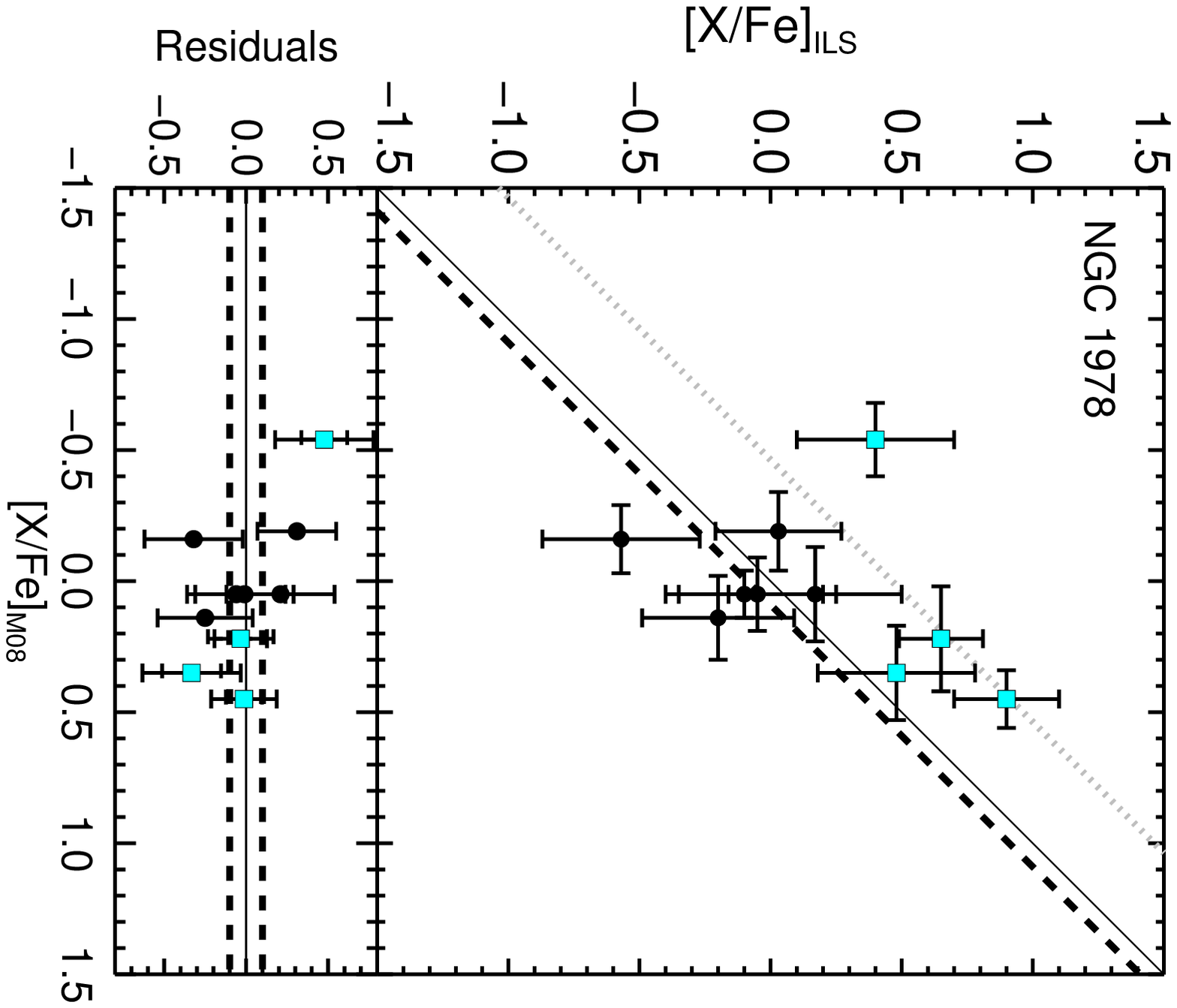}
\caption{Same as Figure \ref{fig:2019comparison} for NGC 1978.  Stellar abundances are from \citetalias{mucc08int}.  }
\label{fig:1978mucc}
\end{figure}

\begin{figure}
\centering
\includegraphics[trim = 0mm 100mm 0mm 0mm, clip,angle=90,scale=0.50]{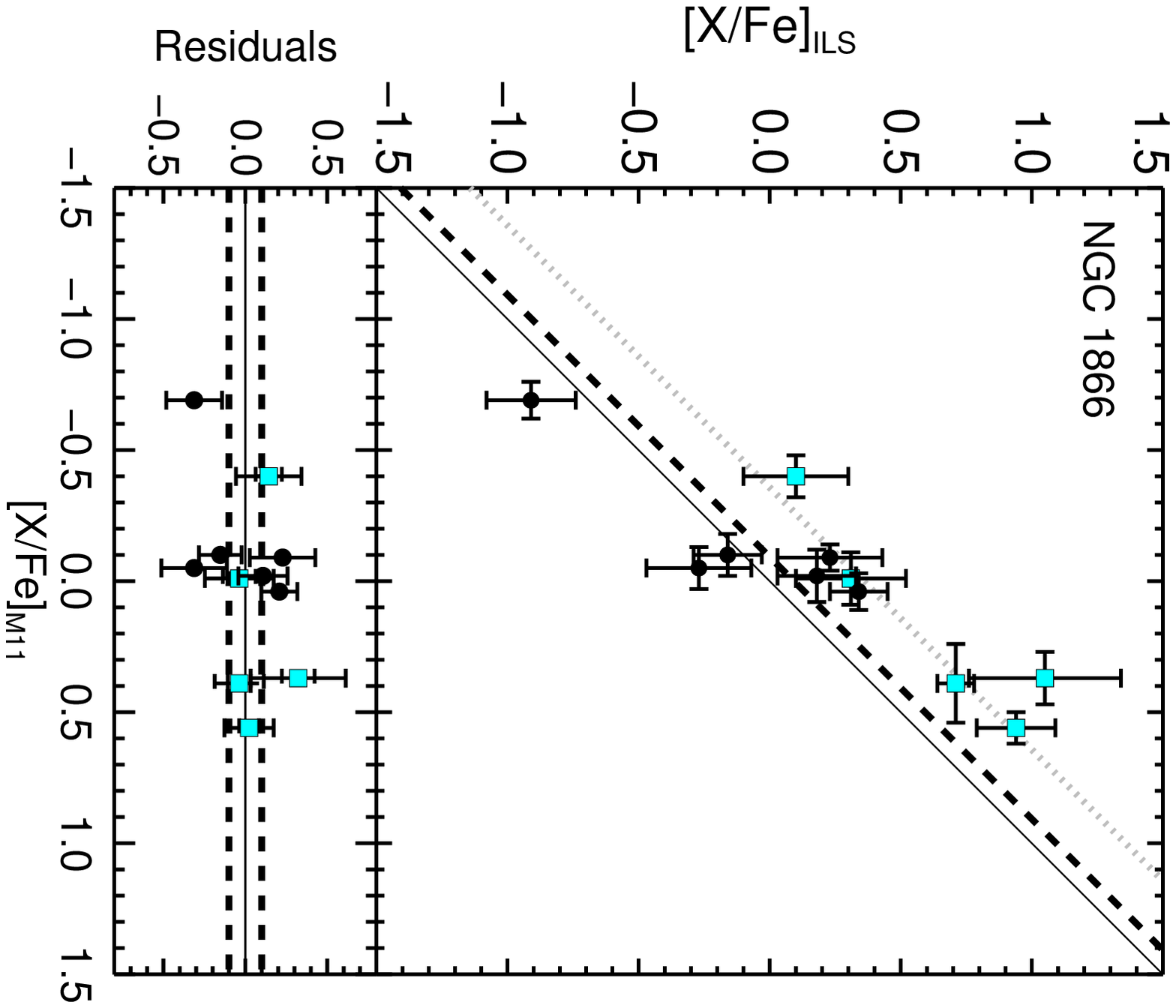}
\caption{Same as Figure \ref{fig:2019comparison} for NGC 1866.  Stellar abundances are from \citetalias{mucc1866}.  }
\label{fig:1866mucc}
\end{figure}

In \textsection \ref{sec:young} we demonstrated the accuracy of the IL analysis method with our own sample of individual stars in the youngest LMC clusters.  In this section we further demonstrate the accuracy with abundances from individual stars in the literature.

{\it NGC 2019 and NGC 2005. } 
The first comprehensive study of the detailed chemical composition in old LMC cluster stars was presented in \citetalias{johnson06}.  This work included  three RGB stars in NGC 2019 and  three in NGC 2005.  As discussed in \citetalias{paper3}, we find the [Fe/H] for  NGC 2019 is lower by $\sim$0.3 dex than in \citetalias{johnson06}, and higher for NGC 2005 by $\sim$0.25 dex.

As for the rest of the elements,   we illustrate a comparison of the [X/Fe] values for the two analyses of NGC 2019  in Figure~\ref{fig:2019comparison}.  This comparison is similar to that described in \textsection \ref{sec:young};  perfect agreement corresponds to the solid black 1:1 line, and  we identify a small systematic offset with a linear least squares fit.     The uncertainties in the measurements are shown by the  vertical error bars, which correspond to the error in the mean of the abundance of each element ($\sigma$/$\sqrt{N_{lines}-1}$). We note that  the abundances from \citetalias{johnson06}  come from a small sample of 3 stars, which have uncertainties comparable to the IL uncertainties   due to the difficulty in obtaining high S/N stellar spectra in the LMC.  The points in  Figure~\ref{fig:2019comparison}  therefore have horizontal error bars corresponding to the error in the mean  abundance for each species from  \citetalias{johnson06}.  We have removed the light elements Na, Al, and Mg from the fit because we do not expect them to track the measurements of \citetalias{johnson06} if star-to-star abundance variations are present (see   \textsection~\ref{sec:light}).

From  the dashed line in the top panel of Figure~\ref{fig:2019comparison},  we see that for NGC 2019, the abundance ratios are very consistent, and the systematic offsets in neutral and ionized species are small ($<$ 0.1 dex).  
Therefore, the overall abundance distribution pattern derived for NGC 2019 is very similar for the two analyses.  The outliers in the comparison are  [Si/Fe] and [Y/Fe].  This is probably because the dispersion in Si between the 2 stars of \citetalias{johnson06} is large, and that Y II is very difficult to measure.

Due to lower S/N of our IL spectra of NGC 2005, we were only able to measure abundances for  10 elements. As for NGC 2019, we remove the light elements Mg and Na from the comparison for NGC 2005.   When we compare the remaining elements to those of \citetalias{johnson06}, we find a  positive offset for the neutral and ionized species of $\sim+0.3$ dex, just as we found for Fe in \citetalias{paper3}.  When this offset is accounted for, the abundance ratios agree very well, as can be seen in the residual plot of Figure \ref{fig:2005comparison}. The exception is [Mn/Fe], which may be systematically high for all three old clusters (see \textsection~\ref{sec:fepeak}).

{\it NGC 1978. }
While we have already discussed a detailed comparison of stellar and IL analyses of our own analyses of NGC 1978 in \textsection \ref{sec:young}, it is also interesting to compare our results to the recent  work of \citetalias{mucc08int} and \cite{2006ApJ...645L..33F}, who presented  the abundances of $\sim$20 different elements for a sample of 11 stars.

    We  compare the [X/Fe] of our IL analysis to the stellar sample of \citetalias{mucc08int} in Figure~ \ref{fig:1978mucc}.    We find a small, $\sim-0.1$ dex offset in neutral species, and a large positive offset in ionized species,  although the dispersion is large $\sim$0.4.  The scatter is consistent with what we find from the comparison to our own stellar analysis in \textsection \ref{sec:young}, and is likely driven by the unavoidably poorer data quality of the IL spectra of NGC 1978.

{\it NGC 1866. } Like NGC 1978, for NGC 1866 we have already demonstrated good agreement between IL analysis and stellar analysis for our own sample. Here,  we  also compare our IL analysis to the stellar sample of \citetalias{mucc1866}, as shown in Figure \ref{fig:1866mucc}.
We find a small $\sim+0.1$ dex offset in the neutral species, and a larger offset of $\sim+0.4$ dex in the ionized species.  As for the other clusters in our sample, when these offsets are removed the abundance pattern is very similar in both analyses.  The remaining outlier is [Cu/Fe], which is one of the more uncertain measurements because it is obtained from a single line.

{\it NGC 2100. }  Using medium resolution spectra, 
\cite{1994A&A...282..717J} estimated  Ca, Ti, Cr and Fe for NGC 2100.  Their estimates for [Ca/Fe], [Ti/Fe] and [Cr/Fe]  are consistent with our measurements obtained with high resolution spectra.

In conclusion, as expected, we find  minor systematic offsets in our IL detailed abundance measurements when compared with previous  stellar measurements by other authors. However, when these offsets are accounted for, we find that the abundance patterns of the clusters,  as traced by the [X/Fe] ratios, are very consistent using both stellar and IL techniques.

\section{Cluster Formation and Light Elements}
\label{sec:light-mass}

\begin{figure}
\centering
\includegraphics[scale=0.45]{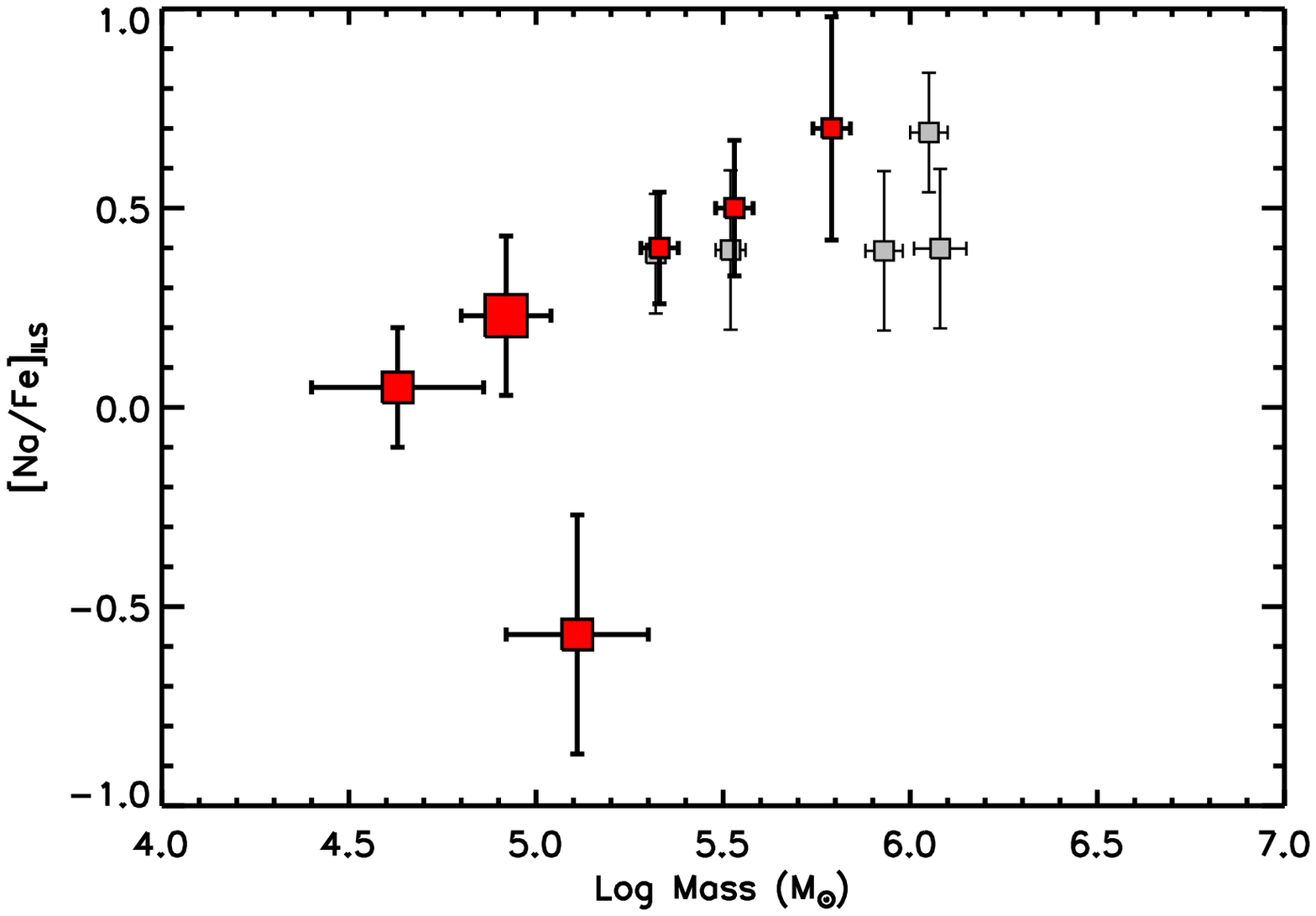}
\includegraphics[scale=0.45]{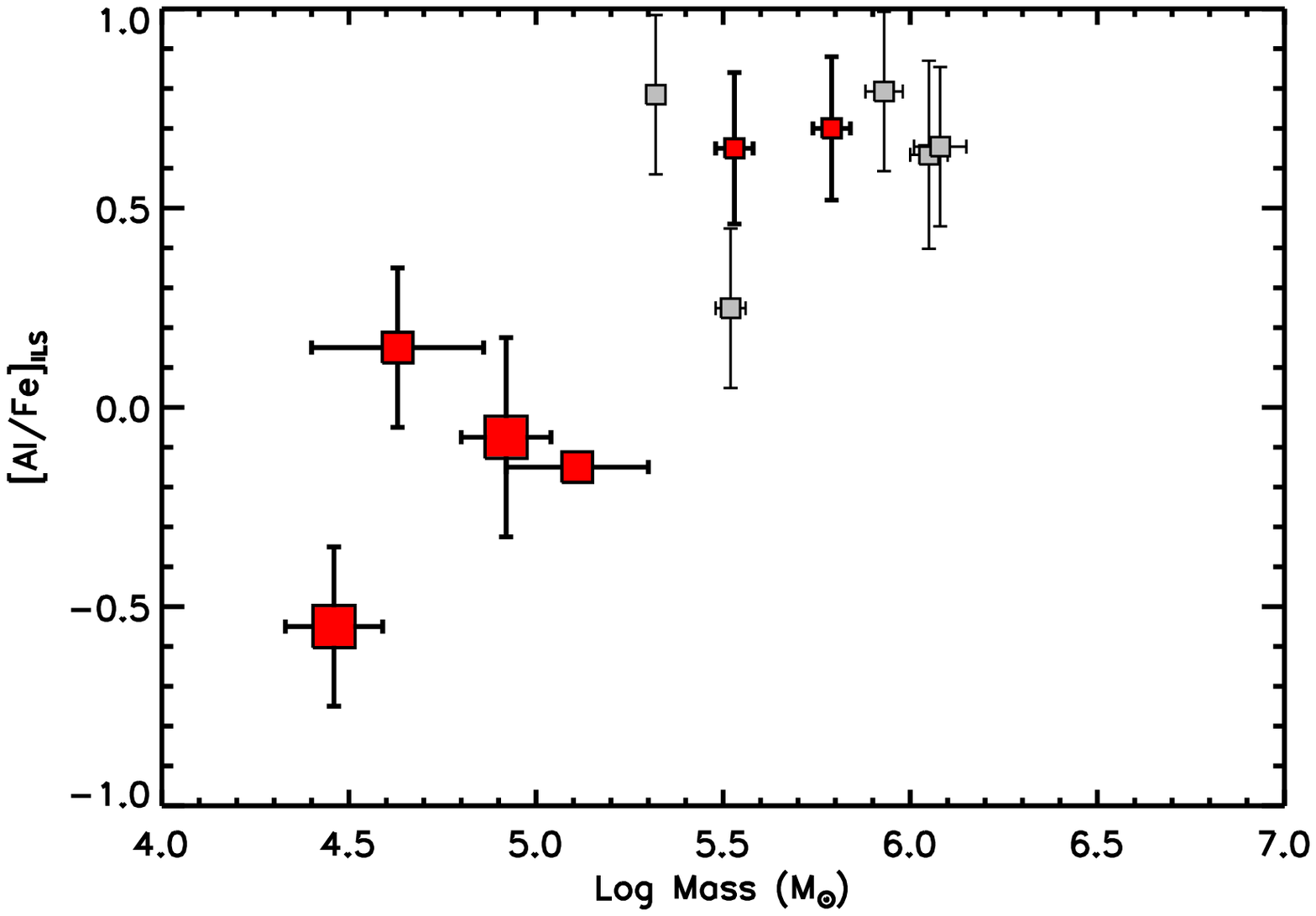}
\caption{The dependence of integrated light element abundances on total cluster dynamical mass. The symbols are the same as in Figure \ref{fig:alpha}.  All cluster masses are taken from \cite{2005ApJS..161..304M}, with the exceptions of NGC 6397, NGC 6752, and NGC 1978, which are taken from  \cite{1995ApJS..100..347D},  \cite{2003MNRAS.340..227B}, and \cite{1991ASPC...13..158M}, respectively.}%
\label{fig:mass}
\end{figure}

\begin{figure}
\centering
\includegraphics[scale=0.45]{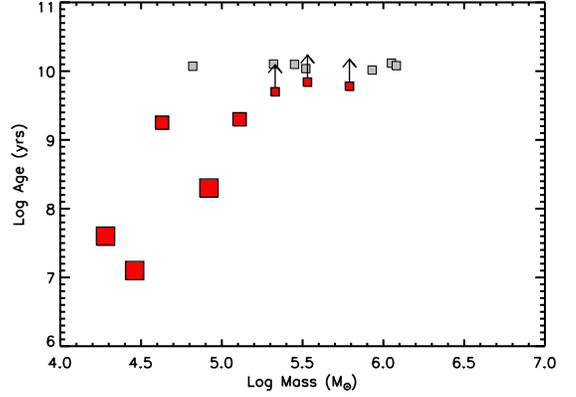}
\caption{The relationship between cluster age and cluster mass. The symbols are the same as in Figure \ref{fig:alpha}.  LMC cluster ages are taken from \citetalias{paper3}, and all MW cluster ages are from 
\cite{2009ApJ...694.1498M}.}%
\label{fig:mass-age}
\end{figure}

As presented in \textsection \ref{sec:light}, we have measured
enhanced Al and enhanced Na in the IL spectra for the old clusters the
LMC.  These results imply that these clusters have star-to-star
abundance variations in the light elements that are sensitive to
proton-capture nucleosynthesis.  No previous work has obtained enough
stars in these clusters to address this issue.

It is interesting to discuss our results in the context of recent
theoretical works that aim to connect cluster star-to-star abundance
variations with the multiple stellar populations that have been
observed in old and intermediate age clusters
\citep[e.g.][]{2007ApJ...661L..53P,2008ApJ...681L..17M,2009A&A...497..755M,2011ApJ...737....3G}.
One scenario is that a second generation of cluster stars forms out of
material that has been polluted with hot bottom burning products from
the first generation intermediate mass ($4-8 \msol$) AGB stars.  Using
this framework, \cite{2011arXiv1101.2208C} find a correlation between
total cluster mass and the fraction of current cluster mass that was
formed out of pure AGB star ejecta.  \cite{2011arXiv1101.2208C} obtain
this result by parametrizing the observed Na-O correlations in MW
clusters, but also find consistency with the observations of LMC
intermediate age clusters of \citetalias{mucc08int}.

We find a correlation, with some scatter, of Na and Al abundances
with cluster mass, as shown in Figure \ref{fig:mass}. This is
consistent with a scenario where more massive clusters have a higher
fraction of stars that are composed of pure AGB ejecta, as discussed
in \cite{2011arXiv1101.2208C}.  The trends are related to the trends
of decreasing Na and Al with cluster age that were discussed in
\textsection \ref{sec:light}, because the younger LMC clusters have
smaller masses, as shown in Figure \ref{fig:mass-age}.  However, we
note that there would not be high [Na/Fe] and [Al/Fe], and thus no
trend with mass, if star-to-star abundance variations were not present
in the old clusters in the sample.
The trend for the intermediate age clusters is more ambiguous because
of the mass-age relationship of our current sample.  
We note that \cite{2011arXiv1101.2208C} find that the stars in
NGC 1978 (age $\sim$2 Gyr) that were observed by
\citetalias{mucc08int} show a weak but detectable Na-O correlation
(and thus star-to-star abundance variations), which is consistent with
our result for clusters in this age range.

\begin{figure*}
\centering
\includegraphics[trim = 2mm 50mm 2mm 5mm, clip,scale=0.92]{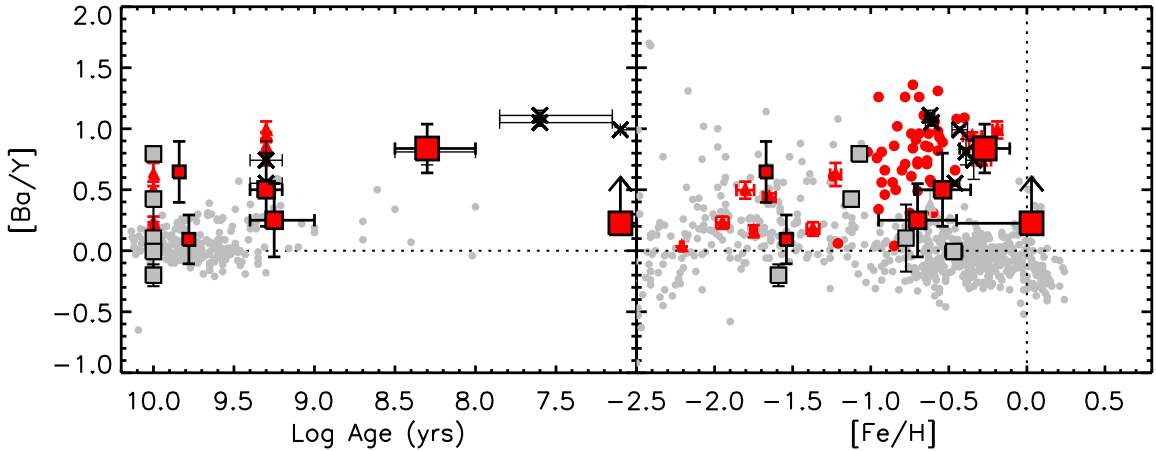}
\caption{The ratio of heavy to light {\it s}-process elements ([Ba/Y]) as a function of age and [Fe/H].    Symbols are the same as in Figure \ref{fig:alpha}.}
\label{fig:ncapture4}
\end{figure*}

It is interesting to look for star-to-star abundance variations in the
young clusters because the pollution timescale ($\sim10^8$ yrs) for
the first generation of stars is thought to be comparable to the ages
of some of these clusters
\citep[e.g.][]{2011ApJ...726...36C,2009A&A...507L...1D,2007A&A...464.1029D,2007MNRAS.379.1431D}.
Strong constraints on this timescale come from observations of age
spreads of $\sim10^8$ years at the main sequence turnoffs in some
$\sim2$ Gyr age LMC clusters \citep{2009A&A...497..755M,
  2011ApJ...737....3G}.  Like the intermediate age clusters, the
youngest clusters in our sample have Na and Al ratios that are similar
to what a pristine first generation of stars should have
\citep{2011arXiv1101.2208C}.  These ratios and the mass-age
relationship make it difficult, if not impossible, to determine if
star-to-star abundance variations are present in the cluster by using
the integrated light.  We note that \citetalias{mucc1866}, with a
large sample of individual stars in NGC 1866, claim that there is no
evidence for star-to-star abundance variations in this cluster (age
$\sim$150 Myr). Unfortunately, the numbers of individual stars that we
have observed in each cluster in our sample are too small to determine
if star-to-star variations are present. Therefore we conclude that our
present data do not allow us to definitively answer the question of
whether star-to-star abundance variations are present in massive
clusters with ages of $\lesssim$100 Myrs.

\section{Galaxy Formation and Heavy Elements}
\label{sec:gcheavy}

As discussed in \S\ref{sec:ncapture}, we have measured many of the first
abundances of 7 neutron-capture elements (Sr, Y, Ba, La, Nd, Sm, Eu)
in these LMC clusters.  Relative to Fe, the most notable results in
our sample are that Ba, La, Nd, Sm, and Eu are significantly enhanced
in the young clusters.  While enhanced heavy element abundances have
been found in young LMC stars for many years
\citep{1989ApJS...70..865R,1992ApJS...79..303L,1993A&A...272..116S},
this is the first indication that these elements are also enhanced in
clusters with ages $<$300 Myrs.

Neutron capture elements are generally divided into {\it r}- and {\it s}-process
categories based on the neutron density of the site that dominates their
production.  However both high density, rapid-neutron capture
({\it r}-process) sites and lower density, slow-neutron capture ({\it s}-process)
sites can both contribute to the overall abundance of heavy elements
over time.  For example, while Eu is often referred to as an {\it r}-process
element because 97\% of the Eu in the Sun was produced in the {\it
  r}-process, a larger fraction of Eu could come from {\it s}-process
sites in an environment with a different (e.g. slower) star formation
history.  To better constrain the star formation history of a
population, it is therefore useful to look at the ratios of different
heavy elements to each other, rather than just to Fe in order to
disentangle the relative dominance of {\it s}- and {\it r}-process sites.  
In addition, the relative abundances of {\it s}-process light (Sr, Y) and heavy
(Ba, La) peaks can provide interesting constraints on the relative
contributions of metal-poor and metal-rich ABG stars,
because light {\it s}-process elements are produced less in low
metallicity AGB stars \citep{1999ARA&A..37..239B,2008ARA&A..46..241S}.
This is because at low metallicity the lighter {\it s}-process element production is
bypassed because the neutron flux per seed Fe nuclei is higher than at
high metallicity when more Fe nuclei are present.

\begin{figure*}
\centering
\includegraphics[trim = 2mm 50mm 2mm 5mm, clip,scale=0.92]{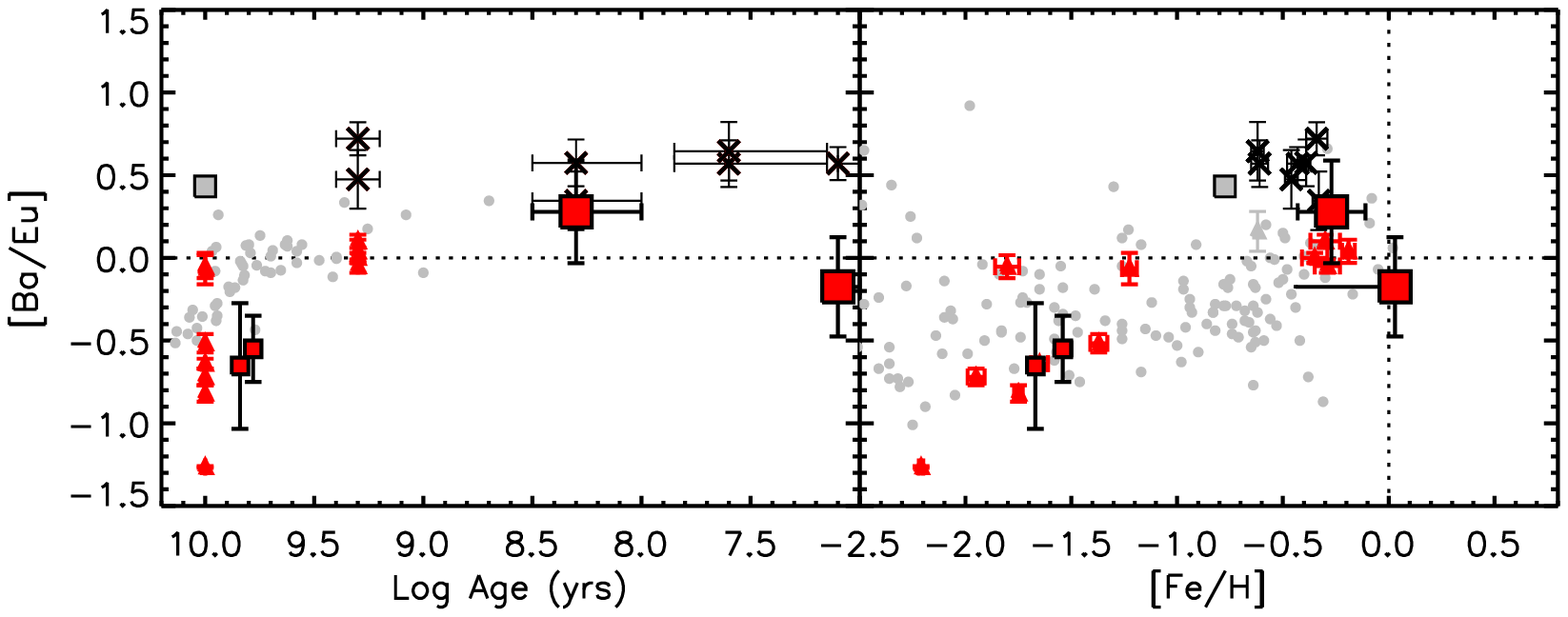}
\includegraphics[trim = 2mm 50mm 2mm 5mm, clip,scale=0.92]{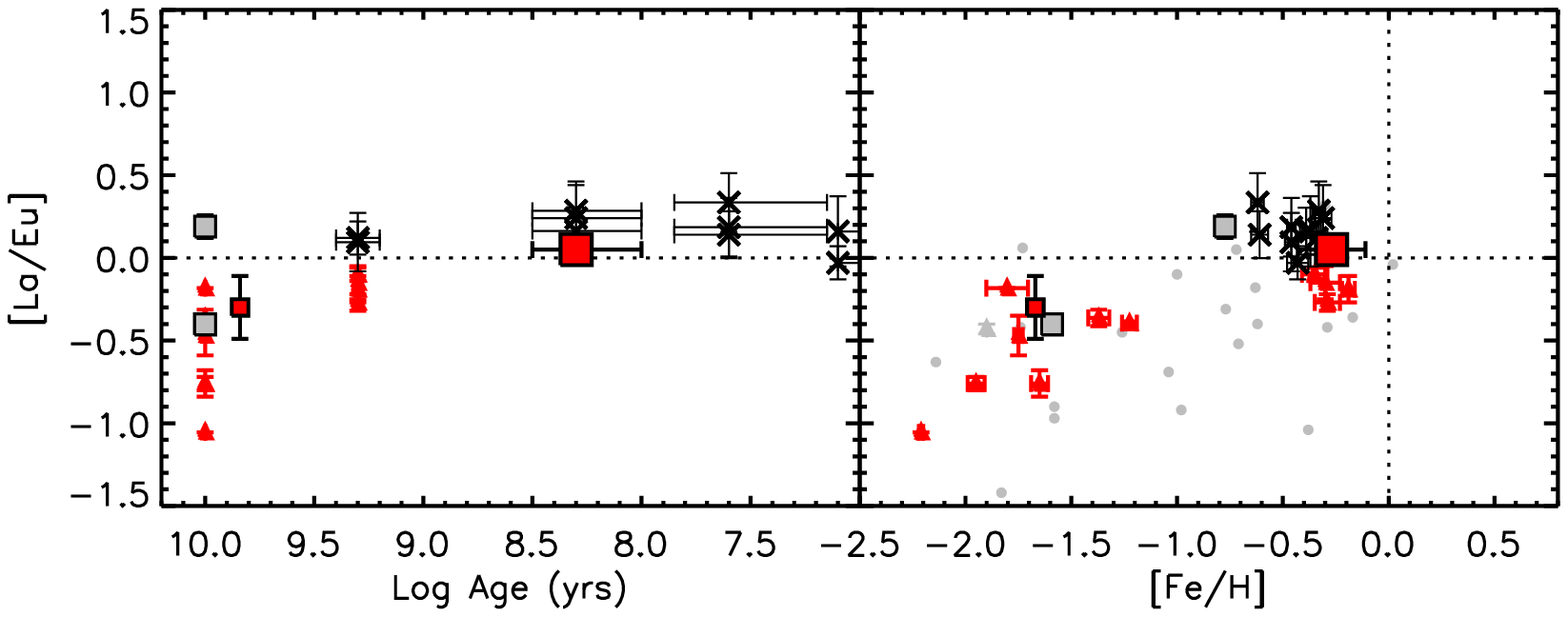}
\caption{Top panels show the ratios of heavy {\it s}-process to {\it r}-process, [Ba/Eu], as a function of age and [Fe/H]. Bottom panels show the ratio of  heavy {\it s}-process to  {\it r}-process elements  [La/Eu].   Symbols are the same as in Figure \ref{fig:alpha}. }
\label{fig:ncapture5}
\end{figure*}

\begin{figure*}
\centering
\includegraphics[trim = 2mm 50mm 2mm 5mm, clip,scale=0.92]{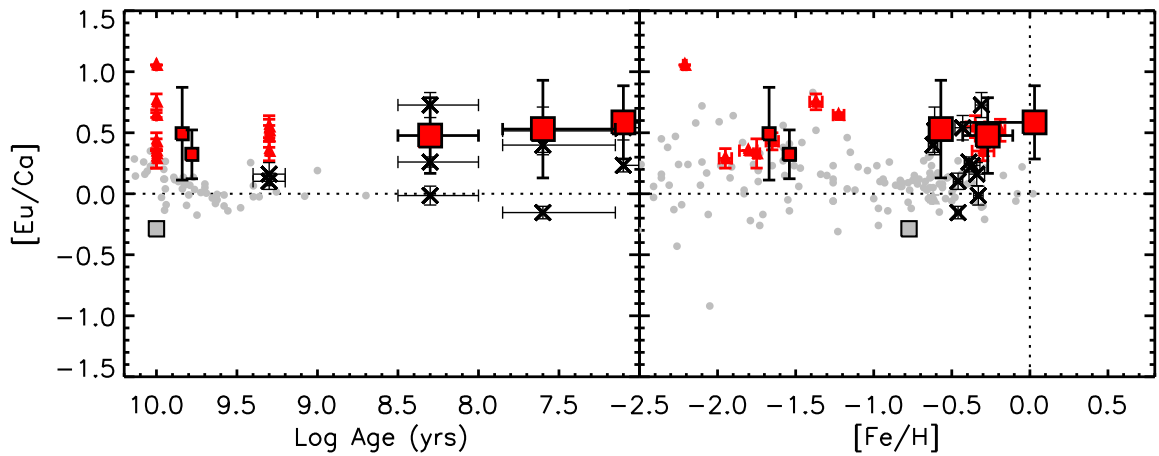}
\caption{The ratios of  
 {\it r}-process elements to $\alpha$-elements ([Eu/Ca]) as a function of age and [Fe/H].    Symbols are the same as in Figure \ref{fig:alpha}.}
\label{fig:ncapture6}
\end{figure*}

The ratio [Ba/Y] is a convenient diagnostic of the {\it s}-process in this
way.  In the MW, [Ba/Y] is fairly constant
over all ages and [Fe/H] values, implying a high star formation rate
at early times, and that heavy and light {\it s}-process elements were
produced in similar amounts at all times (i.e. by high metallicity AGB
stars). In the LMC, we find that the young clusters in our
sample have [Ba/Y]$>$ +0.5, as shown in Figure~\ref{fig:ncapture4}.
This high  [Ba/Y] result  implies that low
metallicity AGB stars dominated the {\it s}-process yields over a long
range in [Fe/H]  --- the star formation rate was low at early times and
metallicity increased slowly.  

Because high-metallicity AGB stars produce heavy and light {\it
  s}-process elements in similar amounts, the imprint of low star
formation rate at early times (high [Ba/Y]) will remain in younger,
metal rich stars.  In the LMC, high [Ba/Y] at {\it any} [Fe/H]
therefore demonstrates that it had a low star formation rate at early
times.  The fact that [Ba/Y] seems to increase in the LMC with
increasing [Fe/H] and decreasing age further implies that low
metallicity AGB stars have contributed significantly to the
cluster-forming gas at late times as well.  This proves that the LMC
has undergone much slower star formation than the MW, even though we
see evidence of recent star-bursting episodes.
In general, this is consistent with the evidence from Cu and Mn (see
\S\ref{sec:fepeak})  that the LMC may have experienced a significant
reduction in overall metallicity due to outflows of metal-rich gas, or
inflows of metal-poor gas.

The old clusters in our sample have low [Ba/Eu] and [La/Eu], shown in
Figure~\ref{fig:ncapture5}. This is indicative of neutron-capture
element enrichment dominated by the {\it r}-process
\citep{1999ApJ...525..886A,2004ApJ...617.1091S} at early times,
similar to the MW halo.  This is compatible with our results for
$\alpha$-element enrichment in the old clusters, because the {\it
  r}-process and $\alpha$-element production are both thought to occur
in SNe II.
The intermediate-age and young clusters, particularly the individual
member stars, have higher [Ba/Eu] and [La/Eu] than the old clusters,
which reflects the contribution of {\it s}-process products at higher
metallicity and later times.

In Figure~\ref{fig:ncapture6}, we show the ratio of {\it r}-process
elements to $\alpha$-elements, represented by [Eu/Ca].
The ratios of these two groups of elements are interesting because
they both are in large quantities in SNe II.
We observe [Eu/Ca]$\sim+0.5$ for all clusters, which is consistent
with previous work on LMC clusters \citepalias{mucc08int,mucc10old}.
The enhanced [Eu/Ca] are interesting for two reasons.  The first  is that all of the older clusters
at the lowest metallicities observed in the LMC studied thus far show
enhanced [Eu/Ca], while in the MW there is a lot of scatter in this
ratio at lower metallicities.
This scatter in the MW [Eu/$\alpha$] abundances is believed to be
evidence that not all SNe II produce the same amount of {\it
  r}-process products, while they do seem to produce very consistent
amounts of $\alpha$-elements \citep{2008ARA&A..46..241S}.
\citetalias{mucc10old} concluded that consistently high [Eu/$\alpha$]
at low metallicity implies an unusually efficient {\it r}-process in
the LMC relative to the MW.
  The second reason is
that Eu abundances are  enhanced relative to Ca and Fe at high
metallicity and young age.  This may be a result of the second burst
of star formation and corresponding increase in SNe II and {\it r}-process products that kept
[Eu/Fe] elevated.  However, it is also possible that a significant contribution to  the Eu
abundance comes from {\it s}-process sites, especially at young age and high metallicity.  This would be consistent with
the overall dominance of {\it s}-process production of Ba, La, and Nd that
is clear from the ratios of these elements to Eu, as discussed above.
Our data do not conclusively distinguish between these alternatives.

\begin{figure*}
\centering
\includegraphics[scale=0.8]{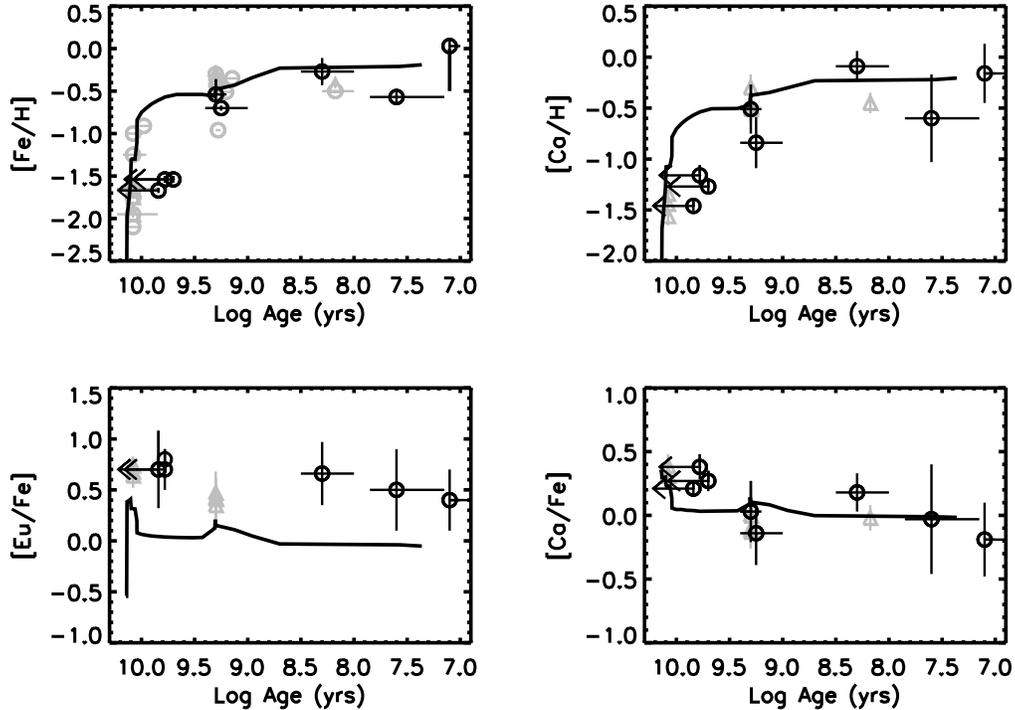}
\caption{ LMC age and abundance relationships for [Fe/H], [Ca/H], [Eu/Fe], and [Ca/Fe].  Our IL LMC cluster data    are plotted as black  circles.  Gray circles are cluster abundances from \citetalias{johnson06}, \citetalias{mucc08int}, \citetalias{mucc10old}, and \citetalias{mucc1866}.  Black lines show the analytical chemical evolution models  of \cite{pt98} for the LMC.    }%
\label{fig:chemev}
\end{figure*}

\section{Comparison with a Basic Dual-Burst Model }
\label{sec:model}

With our sample of clusters we can probe the chemical evolution of the
LMC in a self-consistent way over a larger range in age and [Fe/H]
than has been possible before.  In Figure \ref{fig:chemev} we compare
the LMC training set abundances to the analytical chemical evolution
model for the LMC by \cite{pt98}.  The model assumes supernova yields
and time delays identical to models for the solar neighborhood, and
employs bursting star formation rates, infall, and galactic winds to
match the age and [Fe/H] data for the LMC that was available at the
time. As seen in the age-[Fe/H] plot in Figure \ref{fig:chemev}, the
model is a reasonably good fit to the age-[Fe/H] relationship for the
LMC training set clusters, as well as the cluster samples of
\citetalias{johnson06}, \citetalias{mucc08int},
\citetalias{mucc10old}, and \citetalias{mucc1866}.

Our IL analysis now provides information on [X/Fe] ratios covering the
same dynamic range in age as the well-established LMC age-[Fe/H]
relationship.  At the time, \cite{pt98} could only compare their
models with abundances from young LMC supergiants, and so they only
had power to constrain their models in [X/Fe] and [Fe/H] space. In
general, \cite{pt98} found reasonable agreement, but noted that there
was a large scatter in supergiant abundances, and it was unclear if the scatter was real or due to 
measurement uncertainties.  We can now test the model of \cite{pt98}
with elements other than Fe, with finer age resolution than is
possible for individual field stars.  In Figure \ref{fig:chemev}, we
show examples for [Ca/H], [Eu/Fe] and [Ca/Fe] as a function of
age. The agreement for age and [Ca/H] or [Ca/Fe] is generally good and
similar to the good agreement for age and [Fe/H].  However, agreement
is poorer for [Eu/Fe].  In general, the model is offset to lower
abundances, which could mean that the overall supernova yields assumed
are too low. However, even if we applied a positive offset to the
model, it would still fail to produce the high [Eu/Fe] found at the
youngest ages in the LMC.

Our abundance data provides unprecedented constraints that are needed
for tests of new chemical evolution models for the LMC. In future work
we will investigate models for the chemical evolution of the LMC,
focussing on the parameter space that is consistent with the
age-[X/Fe] relationships for all $\sim$20 elements that we have
obtained in this work.

\section{Summary}
\label{sec:summary}

We have presented chemical abundances of $\sim$20 elements in 8 clusters in the LMC. These clusters have ages of 0.05 to 12 Gyrs and $-1.7 <$ [Fe/H]$< +0.0$.  Abundances were primarily measured by fitting individual  absorption line EWs.   For absorption lines that were too weak or too low S/N to measure with  EWs, we developed an additional technique using a  $\chi^2$-minimization scheme for  IL synthesized spectra.  In addition to the cluster IL abundances, we have also obtained abundances for 10 individual stars in the youngest clusters in our sample. 
 The new results found in this work can be summarized as follows:

\begin{itemize}

\item We verify that the IL method provides  [X/Fe] accurate to $\lesssim$0.1 dex in  clusters with ages $<$ 2 Gyr.

\item  We provide the first detailed chemical abundances for 18 elements in the old cluster NGC 1916,  for 12 elements in the intermediate age cluster NGC 1718, for 22 elements in the young cluster NGC 1711, and for 19 elements in the young cluster NGC 2100. 

\item We obtain the first measurements of Sr II in old LMC stellar populations, with the result that  [Sr/Fe] is sub-solar in the old LMC clusters of our sample.

\item We obtain the first measurements of Sm II in old LMC stellar populations, with the result that [Sm/Fe] is super-solar for the three clusters in our sample in which it was measured.

\item We find evidence of a  spread  in $\alpha$-enhancement at old age and similar [Fe/H] (NGC 2019 and NGC 1916),  which implies that  ISM in the LMC was not well-mixed at the time these old clusters formed.

\item We find evolution in [$\alpha$/Fe] with age and [Fe/H] in the LMC. This reflects the growing contribution of  SNe Ia enrichment products in the LMC between 10 and 2 Gyr ago.

\item We measure elevated [Na/Fe] and [Al/Fe] in our sample of old clusters, including the first evidence that NGC 1916, NGC 2019, and NGC 2005 harbor star-to-star abundance variations.
   
\item We find decreasing [Na/Fe], and [Al/Fe] with decreasing cluster age and decreasing cluster  mass, which could mean that the fraction of pure AGB ejecta incorporated into second generation stars is dependent on the total cluster mass.

\item We find low [Mn/Fe] and  [Cu/Fe] at high metallicity and young age in the cluster sample.  This could reflect the low  star formation efficiency of the LMC and possibly metal-rich outflows of gas.

\item We obtain elevated La, Nd, Sm, and Eu  ratios for all of the clusters  in our sample.

\item We obtain  increasing [Ba/Fe] and [Ba/Y] with decreasing cluster age, which is evidence that there has been significant contributions to the LMC ISM of ejecta from low-metallicity AGB stars throughout the history of the LMC.  This is consistent with prolonged or low efficiency star formation in the LMC.

\item We find low  [Ba/Eu] and [La/Eu] for the  old clusters in our sample, which indicates that {\it r}-process products dominated the neutron-capture chemical enrichment at early times in the LMC. 

\item We find high  [Ba/Eu] and [La/Eu] in the younger clusters in our sample, which indicates that there was significant enrichment with {\it s}-process products at later times in the LMC.

\item We obtain consistently high [Eu/Ca] for all of the clusters in our sample.

\end{itemize}

As a whole, our results are consistent with low efficiency or
prolonged star formation between the bursts of star formation in which
these massive clusters formed
\citep[e.g.][]{hz2009,2008AJ....135..836C,1986A&A...156..261B,1991AJ....101..515O}.
The early, rapid burst of star formation in the LMC is reflected in
high [$\alpha$/Fe] and high {\it r}-process heavy element composition.
The period of prolonged, or inefficient star formation is reflected in
the steady increase in the {\it s}-process fraction toward high
metallicities and younger ages, and perhaps in our observations of low
[Mn/Fe] and [Cu/Fe] at high metallicity.

Several general scenarios could produce such a star formation and
chemical evolution history.  One possibility is a ``leaky-box''
scenario in which enriched gas from a first generation of star
formation and SNe II is lost in outflows.  However that is difficult to
quantitatively distinguish from a situation in which there is a
combination of both outflows and inflows of low-metallicity gas (see
modeling of the LMC age-[Fe/H] relationship by \cite{2008AJ....135..836C}).
Clearly there is current evidence for both ongoing  accretion and gas stripping in the LMC
(e.g. see \cite{2010ApJ...721L..97B,2010ApJ...723.1618N,2011arXiv1106.0044O}).  Finally, the analytical dual-burst chemical evolution
model of \cite{pt98} describes the well known age-metallicity
relationship for the LMC very well.  However, this new data with fine
age and [Fe/H] resolution reveals that the simple model does not
describe the chemical evolution of other elements as well.  Our data
set provides new constraints for future, more sophisticated, models of
the star formation history of the LMC.

\acknowledgements
This research was supported by NSF grant AST-0507350. J.E.C. thanks N. Calvet, P. GuhaThakurta, M. Mateo, M. S. Oey, and G. Evrard for careful reading of an early version of the manuscript. The authors thank D. Osip for help with the scanning algorithm for MIKE on the Magellan Clay Telescope.



\begin{deluxetable}{rrcrr}
\small
\tablecolumns{5}
\tablewidth{0pc}
\tablecaption{LMC Cluster Properties \label{tab:previous_results}}
\tablehead{
\colhead{Cluster} & \colhead{Age\tablenotemark{a}} &    \colhead{[Fe/H]} & \colhead{[$\alpha$/Fe]\tablenotemark{b}} &\colhead{Ref.\tablenotemark{c}} \\ &\colhead{ILS} &\colhead{ILS} &\colhead{Lit} }

\startdata
\sidehead{Old ($>$ 5 Gyrs)}

\hline
\\
NGC 1916	&	$>$5 &      $-$1.54  $\pm$ 0.04	 &\nodata		 		\\
NGC 2019	&$>$7 &     $-$1.67 $\pm$ 0.03	     &$+$0.20	&1\\
NGC 2005		 &$>$5      &$-$1.54 $\pm$  0.04			&$+$0.05	&1 \\
\\
\sidehead{Intermediate Age (1-3 Gyrs)}

\hline
\\
NGC 1978		&1.0-3.0 &    $-$0.54  $\pm$ 0.18
&$+$0.02, $+$0.38  	& 2, 3 \\
		                                                        
NGC 1718		&1.0-2.5 &   $-$0.70   $\pm$ 0.03			&\nodata	 \\
\\
\sidehead{Young ($<$ 1 Gyr)}

\hline
\\
NGC 1866    &0.10-0.50 &   $-$0.27    $\pm$ 0.16		&$+$0.08,$+$0.00   & 3,4	\\
NGC 1711				&0.06-0.10& $-$0.57 $\pm$    0.07		&\nodata & 	\\
NGC 2100				&$<$0.04&          $-0.40<$[Fe/H]$<$$+$0.03 		&$-$0.06 &5	\\

\enddata
\scriptsize

\tablenotetext{a}{Ages listed in Gyrs}
\tablenotetext{b}{Mean of [Si/Fe], [Ca/Fe], and [Ti/Fe] abundances
  from 1, 2, 4 and 5, and mean of [O/Fe] from  3.}
\tablenotetext{c}{References for column 5:
  1. \citetalias{johnson06}, 2. \citetalias{mucc08int},  3. \citetalias{hill00}, 4. \citetalias{mucc1866}, 5.\cite{1994A&A...282..717J}  }

\end{deluxetable}

\clearpage

\LongTables


\begin{deluxetable}{rrrrccccccccc}

\tablecolumns{13}
\tablewidth{0pc}
\tablecaption{Line Parameters and Integrated Light Equivalent Widths for LMC GCs \label{tab:linetable_nf}}
\tablehead{\colhead{Species} & \colhead{$\lambda$}   & \colhead{E.P.} & \colhead{log {\it gf}}   & \colhead{EW(m\AA)} & \colhead{EW(m\AA)} & \colhead{EW(m\AA)} & \colhead{EW(m\AA)} & \colhead{EW(m\AA)}& \colhead{EW(m\AA)}& \colhead{EW(m\AA)}& \colhead{EW(m\AA)}& \colhead{type}\\ 
\colhead{} & \colhead{(\AA)} & \colhead{(eV)} &\colhead{} & \colhead{1916}&\colhead{2005}&\colhead{2019}&\colhead{1978}&\colhead{1718}&\colhead{1866}&\colhead{1711}&\colhead{2100}&\colhead{}}

\startdata

  O  I &7771.941 &  9.146 &  0.369 &  0.00 &  0.00 &  0.00 &  \nodata & \nodata& 113.7 & \nodata & \nodata  & SYN\\
  O  I &7775.391 &  9.146 &  0.001 &  \nodata &  \nodata &  \nodata &  \nodata & \nodata& 92.0 & \nodata & \nodata & SYN\\
\\

Na  I &5682.650 &  2.100 & -0.700 & 27.7 & 56.7 & 39.9 & 121.2 & 66.5 & 73.6 & \nodata & \nodata & SYN\\
 Na  I &5688.220 &  2.100 & -0.460 & 78.2 & 61.0 & 55.4 &114.6 & 92.6 & 87.1 & \nodata & \nodata & SYN\\
 Na  I &6154.230 &  2.100 & -1.570 & \nodata & \nodata & \nodata & \nodata & \nodata & 47.6 & \nodata & \nodata & SYN \\
 Na  I &6160.753 &  2.100 & -1.270 & \nodata & \nodata & 25.8 & 74.8 & 45.9 & 35.6 & \nodata & \nodata & SYN \\
\\
\enddata

\tablecomments{~ Lines listed twice correspond to those measured in adjacent orders with overlapping wavelength coverage. Abundances that are calculated using EWs are labeled ``EW'' in Column 13, and abundances that are calculated using line synthesis are labeled ``SYN.''  Abundances  that were calculated with hyperfine splitting of energy levels are noted as ``HFS'' in Column 13. Log {\it gf} values are taken from \citetalias{mb08}, \citetalias{milkyway},  \citet{m31paper}, \cite{2009A&A...497..611M} and references therein,  as well as the Kurucz atomic and molecular line database \citep{1997IAUS..189..217K}. Hfs line lists and {\it A} and {\it B} constants were taken from \cite{1994ApJS...91..749M}, \cite{1995AJ....109.2757M}, \cite{1995AJ....109.2736M}, \cite{1998AJ....115.1640M}, \citetalias{johnson06} and \citep{1997IAUS..189..217K}.  }
 
\end{deluxetable}

\begin{deluxetable}{llllrrccccccccc}
\scriptsize
\tablecolumns{15}
\tablewidth{0pc}
\tablecaption{Line Parameters and Stellar Equivalent Widths \label{tab:linetable_stars}}
\tablehead{\colhead{Species} & \colhead{$\lambda$}   & \colhead{E.P.} & \colhead{log {\it gf}}   & \colhead{EW} & \colhead{EW} & \colhead{EW} & \colhead{EW} & \colhead{EW}& \colhead{EW}& \colhead{EW}& \colhead{EW}& \colhead{EW}&  \colhead{EW}& \colhead{Note}\\ \colhead{} & \colhead{(\AA)} & \colhead{(eV)} &\colhead{}  & \colhead{(m\AA)}& \colhead{(m\AA)} & \colhead{(m\AA)} & \colhead{(m\AA)} & \colhead{(m\AA)} & \colhead{(m\AA)} & \colhead{(m\AA)} & \colhead{(m\AA)} & \colhead{(m\AA)} & \colhead{(m\AA)}  \\ \colhead{} & \colhead{} & \colhead{} &\colhead{} & \colhead{1978}&\colhead{1978}&\colhead{1866}&\colhead{1866}&\colhead{1866}&\colhead{1711} &\colhead{1711}&\colhead{1711}&\colhead{2100} &\colhead{2100}  \\ \colhead{} & \colhead{} & \colhead{} &\colhead{} & \colhead{737}&\colhead{730}&\colhead{954}&\colhead{1653}&\colhead{1667}&\colhead{831} &\colhead{988}&\colhead{1194}&\colhead{c12} &\colhead{b22} }
\startdata
  O  I &6300.311 &  0.000 & -9.750 &  SYN   &  SYN   & SYN   & SYN & SYN & SYN & SYN& SYN& \nodata & SYN \\

\\ 
Na  I &5682.650 &  2.100 & -0.700 &SYN &SYN &SYN &SYN&SYN&SYN &SYN &  \nodata &\nodata&  \nodata\\
 Na  I &5688.220 &  2.100 & -0.460 & \nodata  &SYN&SYN&SYN &SYN &SYN& SYN & \nodata&  \nodata & \nodata\\
 Na  I &6154.230 &  2.100 & -1.570 &SYN&SYN& SYN& SYN&SYN& SYN&  SYN&  SYN&  SYN& SYN \\
 Na  I &6160.753 &  2.100 & -1.270 &SYN &  SYN& SYN&SYN&SYN&  SYN&  SYN&  SYN&SYN& SYN \\
\\ 
Mg  I &5711.090 &  4.340 & -1.630 &  \nodata &136.6 &  \nodata &163.3 &129.5 &163.8 & 68.4 &  \nodata & 89.6& 198.5 \\
\enddata
\tablecomments{~ Lines listed twice correspond to those measured in adjacent orders with overlapping wavelength coverage.  Unless noted as ``SYN'', all abundances are calculated with the listed EWs.  Abundances  that were calculated with hyperfine splitting of energy levels are noted as ``HFS'' in Column 15. References for line parameters are listed in Table \ref{tab:linetable_nf}.}
\end{deluxetable}

\clearpage


\begin{deluxetable}{l|rr|rr}
\centering
\tablecolumns{5}
\tablewidth{0pc}
\tablecaption{Comparison of Cluster IL to Stellar Results \label{tab:starsvsil}}
\tablehead{
\colhead{Cluster}&\multicolumn{2}{c}{Neutral}&\multicolumn{2}{c}{Ionized}\\
&
\colhead{Offset} &\colhead{Scatter} &\colhead{Offset} &\colhead{Scatter}}

\startdata

NGC 1978	 & -0.08	&   0.40   &  0.05 & 0.33   \\
NGC 1866	&-0.05 	&  0.17   &     0.16    & 0.22  \\
NGC 1711	& 0.17	&    0.18  & \nodata & \nodata       \\
NGC 2100 	& -0.07 	&  0.29  &   0.03  & 0.26    \\

\enddata

\end{deluxetable}

\begin{deluxetable}{rrrrr}
\tablecolumns{5}
\tablewidth{0pc}
\tablecaption{NGC 2019 Abundances \label{tab:abundance_table_2019}}
\tablehead{
\colhead{Species}  &\colhead{log$_{10}\epsilon($X$)$} & \colhead{$\sigma_{\rm{lines}}$}  & \colhead{[X/Fe]\tablenotemark{1}}& \colhead{$N_{\rm{lines}}$} }
\startdata
 O  I &    \nodata &   Lim &   $<$1.00&      1  \\
 Na  I &    5.00&   0.17&    0.50&      3  \\
 Mg  I &    6.03&   0.17&    0.17&      4  \\
 Al  I &    5.35&   0.19&    0.65&      1  \\
 Si  I &    5.93&   0.02&    0.09&      2  \\
 Ca  I &    4.85&   0.05&    0.21&     16  \\
 Sc II &    1.48&   0.13&    0.09&      5  \\
 Ti  I &    3.38&   0.09&    0.15&     12  \\
 Ti II &    3.36&   0.14&    0.13&      8  \\
 V  I &    2.35&   0.24&    0.02&      3  \\
 Cr  I &    4.03&   0.07&    0.06&      8  \\
 Mn  I &    3.61&   0.11&   -0.10&      6  \\
 Fe  I &    5.83&   0.03&   -1.67&     49  \\
 Fe II &    5.85&   0.04&   -1.65&      5  \\
 Co  I &    3.39&   0.30&    0.14&      2  \\
 Ni  I &    4.50&   0.07&   -0.06&     16  \\
 Sr II &   -0.15&   0.40&   -1.40&      1  \\
  Y II &   -0.06&   0.25&   -0.60&      1  \\
 Ba II &    0.57&   0.09&    0.07&      5  \\
 La II &    -0.14&   0.19&    0.40&      1  \\
 Nd II &    0.48&   0.21&    0.70&      2  \\
 Sm II &    0.14&   0.40&   0.8&         1  \\
 Eu II &   -0.45&   0.38&    0.70&      1  \\
\enddata
\tablenotetext{1}{For Fe this quantity is [Fe/H].}
\tablecomments{The error in the mean for $N_{\rm{lines}}$ is defined as $\sigma_{\rm{lines}}=\sigma_{\rm{X}}/\sqrt{N_{\rm{lines}}-1}$.   }
\end{deluxetable}

\begin{deluxetable}{rrrrr}
\tablecolumns{5}
\tablewidth{0pc}
\tablecaption{NGC 2005 Abundances \label{tab:abundance_table_2005}}
\tablehead{
\colhead{Species}  &\colhead{log$_{10}\epsilon($X$)$} & \colhead{$\sigma_{\rm{lines}}$}  & \colhead{[X/Fe]\tablenotemark{1}}& \colhead{$N_{\rm{lines}}$} }
\startdata
 O  I &    \nodata&   Lim&    $<$1.00&      1  \\
 Na  I &    5.03&   0.14&    0.40&      2  \\
 Mg  I &    5.99&   0.20&   -0.00&      1  \\
 Ca  I &    5.04&   0.08&    0.27&      6  \\
 Sc II &    1.93&   0.30&    0.42&      1  \\
 Ti  I &    4.08&   0.47&    0.72&      2  \\
 Ti II &    3.99&   0.18&    0.63&      8  \\
 Cr  I &    4.19&   0.18&    0.09&      3  \\
 Mn  I &    3.77&   0.09&   -0.08&      2  \\
 Fe  I &    5.96&   0.04&   -1.54&     34  \\
 Fe II &    6.23&   0.03&   -1.27&      4  \\
 Ni  I &    4.80&   0.10&    0.11&      3  \\
 Ba II &    1.10&   0.30&    0.47&      1  \\
\enddata

\end{deluxetable}

\clearpage
\begin{deluxetable}{rrrrr}
\tablecolumns{5}
\tablewidth{0pc}
\tablecaption{NGC 1916 Abundances \label{tab:abundance_table_1916}}
\tablehead{
\colhead{Species}  &\colhead{log$_{10}\epsilon($X$)$} & \colhead{$\sigma_{\rm{lines}}$}  & \colhead{[X/Fe]\tablenotemark{1}}& \colhead{$N_{\rm{lines}}$} }
\startdata
 O  I &    \nodata &   Lim&    $<$1.00&      1  \\
 Na  I &    5.33&   0.28&    0.70&      2  \\
 Mg  I &    5.67&   0.20&    0.32&      1  \\
 Al  I &    5.53&   0.18&    0.70&      1  \\
 Si  I &    6.62&   0.18&    0.65&      2  \\
 Ca  I &    5.15&   0.10&    0.38&      7  \\
 Sc II &    1.70&   0.08&    0.19&      2  \\
 Ti  I &    3.91&   0.16&    0.55&      2  \\
 Ti II &    4.34&   0.19&    0.98&      6  \\
 Cr  I &    4.20&   0.12&    0.10&      5  \\
 Mn  I &    3.74&   0.05&   -0.10&      6  \\
 Fe  I &    5.96&   0.04&   -1.54&     50  \\
 Fe II &    6.49&   0.11&   -1.01&      3  \\
 Ni  I &    4.69&   0.07&    0.00&      7  \\
 Sr II &    0.84&   0.40&   -0.59&      1  \\
  Y II &    0.73&   0.20&    0.06&      1  \\
 Ba II &    0.78&   0.30&    0.15&      2  \\
Nd II &    0.61&   0.19&    0.70&      1  \\
 Eu II &   0.21&   0.20&    0.70&      1  \\
\enddata
\end{deluxetable}

\begin{deluxetable}{rrrrrrr}
\tablecolumns{7}
\tablewidth{0pc}
\tablecaption{NGC 1718 Abundances \label{tab:abundance_table_1718}}
\tablehead{
\colhead{Species}  &\colhead{log$_{10}\epsilon($X$)$} & \colhead{$\sigma_{\rm{lines}}$}& \colhead{$\sigma_{\rm{age}}$}  &\colhead{$\sigma_{\rm{Tot}}$}& \colhead{[X/Fe]\tablenotemark{1}}& \colhead{$N_{\rm{lines}}$}\\ \colhead{} & \colhead{} }
\startdata
 Na  I &    5.52&   0.16&   0.15&   0.22&    0.05&      3  \\
 Mg  I &    5.93&   0.30&      \nodata& 0.30 &  -0.90&      2  \\
 Al  I &    5.82&   0.20&      \nodata&  0.30 &  0.15&      1  \\
 Ca  I &    5.47&   0.25&   0.05&   0.25&   -0.14&      3  \\
 Sc II &    2.42&   0.13&   0.13&   0.18&   -0.39&      1  \\
 Ti  I &    4.90&   0.01&   0.06&   0.06&    0.70&      3  \\

 Cr  I &    5.19&   0.30 &   \nodata &  0.30 &    0.25&      1  \\
 Mn  I &    4.64&   0.08&   0.13&   0.15&   -0.05&      2  \\
 Fe  I &    6.80&   0.04&   0.03&   0.05&    -0.70  & 69\\
 Fe II &    7.24&   0.17&   0.06&   0.18&    -0.26  & 4 \\
 Ni  I &    5.62&   0.25&   0.14&   0.29&    0.09&      3  \\
  Y II &    1.46&   0.30&    \nodata& 0.30 &  -0.05&      1  \\
 Ba II &    1.67&   0.30&      \nodata&  0.30 &  0.20&      1  \\
\enddata

\end{deluxetable}

 \begin{deluxetable}{rrrrrrr}
\tablecolumns{7}
\tablewidth{0pc}
\tablecaption{NGC 1978 Abundances \label{tab:abundance_table_1978}}
\tablehead{
\colhead{Species}  &\colhead{log$_{10}\epsilon($X$)$} & \colhead{$\sigma_{\rm{lines}}$}& \colhead{$\sigma_{\rm{age}}$}  &\colhead{$\sigma_{\rm{Tot}}$}& \colhead{[X/Fe]\tablenotemark{1}}& \colhead{$N_{\rm{lines}}$}\\ \colhead{} & \colhead{} }
\startdata
 Na  I &    5.07&   0.30&   \nodata&   0.30&   -0.57&      3  \\
 Mg  I &    6.95&   0.30&  \nodata& 0.30&    -0.05&      1  \\
 Al  I &    \nodata&   Lim&   Lim&   Lim&   $<$-0.15&      1  \\
 Ca  I &    5.81&   0.23&   0.06&   0.24&    0.03&      5  \\
 Ti  I &    4.17&   0.27&   0.09&   0.29&   -0.20&      4  \\
 Ti II &    4.98&   0.22&   0.21&   0.31&    0.16&      2  \\
  V  I &    3.64&   0.31&   0.12&   0.33&    0.17&      3  \\
 Mn  I &    4.72&   0.14&   0.06&   0.15&   -0.14&      5  \\
 Fe  I &    6.96&   0.06&   0.18&   0.19 &-0.54&     36\\
 Fe II &    7.55&   0.09&   0.11&  0.14 &  0.05&    4\\
 Ni  I &    5.57&   0.30&      \nodata& 0.3&   -0.10&      1  \\
  Y II &    2.08&   0.30&  \nodata& 0.30   &    0.40&      1  \\
 Ba II &    2.54&   0.20&  \nodata& 0.20&      0.90&      1  \\
 La II &    1.25&   0.16&  \nodata& 0.16&       0.65&      1  \\
 Nd II &    1.40&   0.30&   \nodata&0.30&       0.48&      1  \\
\enddata

\end{deluxetable}

\begin{deluxetable}{rrrrrrr}
\tablecolumns{7}
\tablewidth{0pc}
\tablecaption{NGC 1866 Abundances \label{tab:abundance_table_1866}}
\tablehead{
\colhead{Species}  &\colhead{log$_{10}\epsilon($X$)$} & \colhead{$\sigma_{\rm{lines}}$}& \colhead{$\sigma_{\rm{age}}$}  &\colhead{$\sigma_{\rm{Tot}}$}& \colhead{[X/Fe]\tablenotemark{1}}& \colhead{$N_{\rm{lines}}$}\\ \colhead{} & \colhead{} }
\startdata
 O  I &    8.83&    0.30     &    \nodata  &   0.30&    0.20&      2  \\

 Na  I &    5.58&   0.20&   \nodata &   0.20&    0.23&      3  \\
 Mg  I &    6.91&   0.14&   0.14&   0.20&   -0.27&      3  \\
 Al  I &    6.93&   0.25&   \nodata&   0.25&   -0.08&      1  \\
 Si  I &    7.50&   0.11&   \nodata &   0.11&    0.34&      3  \\
 Ca  I &    6.14&   0.06&   0.14&   0.15&    0.18&      7  \\
 Ti  I &    5.00&   0.17&   0.13&   0.22&    0.45&      2  \\
 Ti II &    5.01&   0.10&   0.19&   0.21&    0.31&      2  \\
  V  I &    3.59&   0.22&   0.06&   0.23&   -0.06&      2  \\
Fe  I &    7.23&   0.04&   0.16&   0.16& -0.27  &    50\\
 Fe II &    7.27&   0.14&   0.11&  0.18&   -0.23&     8\\
 Ni  I &    5.72&   0.09&   0.09&   0.13&   -0.16&      7  \\

Cu I & 2.95 & 0.13 & 0.11 & 0.17 & -0.91 & 2 \\
  Y II &    1.95&   0.20&   \nodata&   0.20&    0.10&      1  \\

 Ba II &    2.90&   0.06&   0.14&   0.15&    0.94&      2  \\
 La II &    1.56&   0.04&   0.06&   0.07&    0.71&      2  \\
 Nd II &    2.14&   0.29&   \nodata&   0.29&    1.05&      2  \\
 Sm II &    1.18&   0.30&   \nodata&   0.30&    0.70&      1  \\
 Eu II &    0.82&   0.31&   \nodata &   0.31&    0.66&      1  \\
\enddata

\end{deluxetable}

\begin{deluxetable}{rrrrrrr}
\tablecolumns{7}
\tablewidth{0pc}
\tablecaption{NGC 1711 Abundances \label{tab:abundance_table_1711}}
\tablehead{
\colhead{Species}  &\colhead{log$_{10}\epsilon($X$)$} & \colhead{$\sigma_{\rm{lines}}$}& \colhead{$\sigma_{\rm{age}}$}  &\colhead{$\sigma_{\rm{Tot}}$}& \colhead{[X/Fe]\tablenotemark{1}}& \colhead{$N_{\rm{lines}}$}\\ \colhead{} & \colhead{} }
\startdata
 Ca  I &    5.71&   0.30&   0.30&   0.43&   -0.03&      5  \\
 Mn  I &    4.71&   0.12&   0.12&   0.17&   -0.11&      3  \\
 Fe  I &    6.93&   0.08&   0.07&   0.11&   -0.57&    25 \\
 Fe II &    6.62&   0.18&   0.05&   0.19&   -0.88&     4 \\
 Ni  I &    5.90&   0.05&   0.07&   0.09&    0.24&      5  \\
 Eu II &    0.47&   0.40&   \nodata&   0.40&    0.50&      1  \\
\enddata

\end{deluxetable}

\begin{deluxetable}{rrrrrrr}
\tablecolumns{7}
\tablewidth{0pc}
\tablecaption{NGC 2100 Abundances \label{tab:abundance_table_2100}}
\tablehead{
\colhead{Species}  &\colhead{log$_{10}\epsilon($X$)$} & \colhead{$\sigma_{\rm{lines}}$}& \colhead{$\sigma_{\rm{age}}$}  &\colhead{$\sigma_{\rm{Tot}}$}& \colhead{[X/Fe]\tablenotemark{1}}& \colhead{$N_{\rm{lines}}$}\\ \colhead{} & \colhead{} }
\startdata
Al  I &    5.67&   0.20&   \nodata&   0.20&   -0.55&      2  \\
 Si  I &    7.56&   0.30&   \nodata&   0.30&    0.20&      2  \\
 Ca  I &    6.16&   0.29&   \nodata&   0.29&   -0.19&     10  \\
 Ti  I &    5.09&   0.36&   \nodata&   0.36&    0.16&      5  \\
 Ti II &    4.42&   \nodata&   \nodata&  0.36&   -0.16&      1  \\
  V  I &    4.22&   0.42&   \nodata&   0.41&    0.19&      9  \\
 Cr  I &    5.41&   0.39&   \nodata&   0.39&   -0.26&      4  \\
 Mn  I &    4.65&   0.25&   \nodata&   0.25&   0.77&      5  \\
 Fe  I &    7.53&   0.08&   \nodata&   0.08&   0.03&    30\\
 Fe II &    7.18&   0.22&   \nodata&   0.22&   -0.32&     5\\
 Ni  I &    6.28&   0.22&   \nodata&   0.22&    0.01&      9  \\
  Y II &    \nodata &   Lim&   Lim&   Lim&   $<$-0.00&      1  \\
 Ba II &    2.08&   0.30&   \nodata&   0.30&    0.22&      1  \\
Sm II &    1.86&   0.40&   \nodata&   0.40&    1.00&      1  \\
 Eu II &    0.77&   0.30&   \nodata&   0.30&    0.40&      1  \\
\enddata

\end{deluxetable}

\clearpage

\begin{deluxetable}{ll|ll|lll|lll|ll}
\tablecolumns{12}
\tablewidth{0pc}
\tablecaption{Abundance Ratios Measured in Individual Cluster Stars\label{tab:abundance_table_stars}}
\tablehead{
\colhead{Species}  && \colhead{1978-737}  & \colhead{1978-730}& \colhead{1866-954} & \colhead{1866-1653} &\colhead{1866-1667} & \colhead{1711-831} & \colhead{1711-1194} & \colhead{1711-988} & \colhead{2100-c12} & \colhead{2100-b22}   }
\startdata
 O I  &               [O/Fe] &   0.26 &   0.35 &   0.29 &   0.31 &   0.13 &   0.17 &   0.26 &   $-$0.09 &   \nodata &   0.03   \\
      &             $\sigma$ &   0.10 &   0.20 &   0.10 &   0.10 &   0.10 &   0.10 &   0.20 &   0.10 &  \nodata &   0.20   \\
      &     N$_{\rm{Lines}}$ &   1 &   1 &   1 &   1 &  1 &   1 &   1 &   1 &   0 &   1   \\
                 \\  
Na I  &              [Na/Fe] &   0.11 &  -0.01 &   0.12 &   0.17 &   0.16 &   0.37 &   0.45 &   0.33 &   0.44 &   0.18   \\
      &             $\sigma$ &   0.07 &   0.06 &   0.07 &   0.12 &   0.06 &   0.08 &   0.00 &   0.11 &   0.02 &   0.15   \\
      &     N$_{\rm{Lines}}$ &   3 &   4 &   2 &   5 &   4 &   4 &   2 &   4 &   2 &   2   \\
                 \\
\enddata
\tablenotetext{1}{For Fe this quantity is [Fe/H].  }
\tablecomments{Uncertainties are defined as $\sigma=\sigma_{\rm{X}}/\sqrt{N_{\rm{lines}}-1}$.   }
\end{deluxetable}

\clearpage


\begin{deluxetable}{r|r|r}
\centering
\tablecolumns{5}
\tablewidth{0pc}
\tablecaption{Mean IL $\alpha$ Abundances for Milky Way Training Set and Old LMC GCs  \label{tab:alpha_table}}
\tablehead{
\colhead{}&\colhead{Milky Way}&\colhead{LMC ($>$10 Gyr)}}
\startdata

$[$Ca/Fe$]$	 & $+$0.35 $\pm$	0.08 	&$+$0.29 $\pm$ 0.09\\
$[$Ti/Fe$]$	& $+$0.46 $\pm$	0.15 	&$+$0.55 $\pm$ 0.37\\
$[$Si/Fe$]$	& $+$0.52 $\pm$	0.20		&$+$0.37 $\pm$ 0.40  \\

\hline
\\
$[$$\alpha$$_{\rm{CaTiSi}}$/Fe$]$	& $+$0.44 $\pm$	0.09	&$+$0.40 $\pm$ 0.13\\

\enddata

\end{deluxetable}

\end{document}